\documentclass[twocolumn,amsmath,amssymb,aps,pra,superscriptaddress,showpacs,10pt,floatfix, reprint
]{revtex4-1}
\usepackage{graphicx}\usepackage{grffile} \usepackage{array}
\usepackage{color}
\usepackage[utf8]{inputenc}
\usepackage[T1]{fontenc}
\usepackage[english]{babel}

\usepackage{units} 

\usepackage{upgreek}

\usepackage{newtxtext} 
\usepackage{newtxmath}

\usepackage{microtype}

\usepackage{dsfont}

\usepackage{xcolor} 

\definecolor{cset-aps-blueberry}{RGB}{28,128,158}
\definecolor{cset-aps-blue}{RGB}{46,44,184}
\definecolor{cset-aps-turquoise}{RGB}{0,67,88}
\definecolor{cset-aps-limegreen}{RGB}{190,219,67}
\definecolor{cset-aps-green}{RGB}{31,138,112}
\definecolor{cset-aps-yellow}{RGB}{255,225,25}
\definecolor{cset-aps-orange}{RGB}{253,116,0}
\definecolor{cset-aps-red}{RGB}{219,0,43}
\definecolor{cset-aps-violett}{RGB}{142,68,173}

\usepackage{hyperref}
\hypersetup{colorlinks=true,
	linkcolor={cset-aps-red},
	linkbordercolor={cset-aps-red},
	filecolor={cset-aps-orange},
	filebordercolor={cset-aps-orange},
	citecolor={cset-aps-blue},
	citebordercolor={cset-aps-blue},
	urlcolor={cset-aps-green},
	urlbordercolor={cset-aps-green},
	menucolor={cset-aps-limegreen},
	menubordercolor={cset-aps-limegreen},
	breaklinks=true,
	pdfborderstyle={/S/U/W 2},
pdfpagemode=UseOutlines,
	pdfstartpage={1},
}

\usepackage[showonlyrefs]{mathtools}		\mathtoolsset{showonlyrefs=true}

\renewcommand{\L}{\text{L}}
\newcommand{\W}{\text{W}}

\newcommand{\bra}[1]{\left\langle{#1}\right|}
 \newcommand{\ket}[1]{\left|{#1}\right\rangle}
 \newcommand{\braket}[2]{\langle{#1}|{#2}\rangle}

 \usepackage{braket}

\newcommand{\e}[1]{\operatorname{e}^{#1}}

\newcommand{\I}{\mathrm{i}}
\newcommand{\D}{\text{d}}

\def\ba#1\ea{\begin{align}#1\end{align}}

\usepackage{tikz}                       \usepackage{pgfplots}                   \usetikzlibrary{spy}
\usetikzlibrary{pgfplots.groupplots}

\usepgfplotslibrary{external}
\usepackage{shellesc}

\pgfplotsset{every axis legend/.append style={cells={anchor=west},
    at={(0.96,0.04)},
    anchor=south east,
    font=\scriptsize
  },
width=\linewidth,
  height=5cm, xmajorgrids=false, xminorgrids=false, minor x tick num=1
}
\pgfplotsset{
every axis plot/.append style={very thick}
}

\definecolor{matlab1}{rgb}{0, 0.4470, 0.7410}
\definecolor{matlab2}{rgb}{0.8500, 0.3250, 0.0980}
\definecolor{matlab3}{rgb}{0.9290, 0.6940, 0.1250}
\definecolor{matlab4}{rgb}{0.4940, 0.1840, 0.5560}
\definecolor{matlab5}{rgb}{0.4660, 0.6740, 0.1880}
\definecolor{matlab6}{rgb}{0.3010, 0.7450, 0.9330}
\definecolor{matlab7}{rgb}{0.6350, 0.0780, 0.1840}

\pgfplotscreateplotcyclelist{matlab}{
	{matlab1},
	{matlab2},
	{matlab3},
	{matlab4},
	{matlab5},
	{matlab6},
	{matlab7},
}

\pgfplotsset{cycle list name=matlab}

\definecolor{TolRB10}{HTML}{1965B0}
\definecolor{TolRB26}{HTML}{DC050C}
\definecolor{TolRB18}{HTML}{F7F056}
\definecolor{TolRB15}{HTML}{4EB265}

\definecolor{TolDRBlue}{HTML}{1965B0}
\definecolor{TolDRRed}{HTML}{DC050C}
\definecolor{TolDRYellow}{HTML}{F7F056}
\definecolor{TolDRGreen}{HTML}{4EB265}

\definecolor{TolHCYellow}{HTML}{DDAA33}
\definecolor{TolHCRed}{HTML}{BB5566}
\definecolor{TolHCBlue}{HTML}{004488}

\pgfplotscreateplotcyclelist{tol4HC cycle}{{black, mark=diamond*, mark size=1.5pt},
	{TolHCYellow, mark=triangle*, mark size=1.5pt},
	{TolHCRed, mark=square*, mark size=1.3pt},
	{TolHCBlue, mark=*, mark size=1.5pt},
}

\pgfplotsset{compat=newest}

\newcounter{plot}[figure]

\usepackage{xparse}
\makeatletter
\NewDocumentCommand{\SubLabel}{s O{} m}{\IfBooleanTF{#1}{\node[	anchor=north west,	fill=white!.1,	opacity=0.9] at (axis description cs:0.02,0.98){\refstepcounter{plot}\label{#3}\@refstar{#3} #2};	
	}{\node[anchor=north west] at (axis description cs:0.02,0.98){\refstepcounter{plot}\label{#3}\@refstar{#3} #2};}}
\makeatother

\def\sa{\sqrt{\alpha}}

\usepackage{physics}

\usepackage{blindtext} 
\usepackage{booktabs}

\renewcommand{\varpi}{\mathcal{W}}

\begin{document}
\title[Quantum and classical phase-space dynamics of a free-electron laser]{Quantum and classical phase-space dynamics of a free-electron laser\\[1ex]
	\normalsize\normalfont{Published in \href{https://journals.aps.org/prresearch/abstract/10.1103/PhysRevResearch.2.023027}{Phys. Rev. Research {\bfseries 2}, 023027 (2020)}}} 
\author{C.~Moritz Carmesin}
\affiliation{Institut für Quantenphysik and Center for Integrated Quantum Science and Technology 
$\left(\text{IQ}^{\text{ST}}\right)$, Universität Ulm, Albert-Einstein-Allee 11, D-89081, Germany} 
\affiliation{Helmholtz-Zentrum Dresden-Rossendorf e.V., Bautzner Landstraße 400, D-01328 Dresden, Germany}
\author{Peter Kling}
\affiliation{Institute of Quantum Technologies, German Aerospace Center (DLR), Söflinger Straße 100, D-89077 Ulm, 
Germany}
\affiliation{Institut für Quantenphysik and Center for Integrated Quantum Science and Technology $\left(\text{IQ}^{\text{ST}}\right)$, Universität Ulm, Albert-Einstein-Allee 11, D-89081, Germany} 
\author{Enno Giese}
\affiliation{Institut für Quantenphysik and Center for Integrated Quantum Science and Technology $\left(\text{IQ}^{\text{ST}}\right)$, Universität Ulm, Albert-Einstein-Allee 11, D-89081, Germany}
\author{Roland Sauerbrey}
\affiliation{Helmholtz-Zentrum Dresden-Rossendorf e.V., Bautzner Landstraße 400, D-01328 Dresden, Germany}

\author{Wolfgang P. Schleich}
\affiliation{Institut für Quantenphysik and Center for Integrated Quantum Science and Technology 
$\left(\text{IQ}^{\text{ST}}\right)$, 
Universität Ulm, Albert-Einstein-Allee 11, D-89081, Germany}
\affiliation{Institute of Quantum Technologies, German Aerospace Center (DLR), Söflinger Straße 100, D-89077 Ulm, 
	Germany}
\affiliation{Hagler Institute for Advanced Study at Texas A\&M University, Texas A\&M AgriLife Research, Institute for Quantum Science 
and Engineering (IQSE) and Department of Physics and Astronomy, Texas A\&M University, College Station, Texas 77843, USA }

\begin{abstract}
	In a quantum mechanical description of the free-electron laser (FEL) the electrons jump on discrete momentum 
	ladders, while they follow continuous trajectories according to the classical description.  In order 
	to observe the transition from  quantum to classical dynamics, it is not sufficient that 
	many momentum levels are involved. Only if additionally the initial momentum spread of the electron beam is larger than the quantum mechanical recoil, 
	caused by the emission and absorption of photons, the quantum dynamics in phase space 
	resembles the classical one.
	Beyond these criteria, quantum signatures of
	averaged quantities like 
	the FEL gain might be washed out.
\end{abstract}

\maketitle

\section{Introduction}
  \label{sec:Introduction}

  Usually, an FEL is considered as a device that can be fully described within classical physics. However, there exists 
  a “quantum regime” \cite{schroeder,*boni06,*boni17,*anisimov18,NJP2015,kling_high-gain_2019} where quantum 
  mechanics is indeed mandatory for an accurate description of 
  the FEL dynamics.
  
  In this article, we analyze the transition from quantum to classical in a low-gain FEL by contrasting the 
  dynamics of an electron in phase space with the corresponding classical description. We find that the 
  occurrence of quantum effects depends on the quantum mechanical recoil, caused by the absorption and emission of 
  photons: A small recoil energy, compared to the coupling to the fields, and a small recoil momentum, compared to the 
  initial momentum spread, are necessary to observe a classical evolution of the Wigner function.
Furthermore, we study quantum corrections to the FEL gain.
  
  \subsection{Historical overview}
  
  In his groundbreaking article~\cite{madey1} in 1971 John Madey had already formulated a quantum theory  
  for the FEL  even before the  first classical theories emerged~\cite{scully_fel,colson}.  His
  approach relies on a perturbative solution for the electron wave function and  was lateron refined
  for example in Refs.~\cite{becker79,*becker88,colson_quantum,friedman,banacloche}.  
  
  On first sight, this model perfectly describes the transition between the quantum and the classical regime of the 
  FEL: in the  former one the resonances for photon emission and absorption are well separated; by 
  taking the limit $\hbar \rightarrow 0$, however, these resonances overlap and the \textit{difference} of photon 
  emission and absorption turns into a \textit{derivative}. Since in this case  all terms with Planck's constant 
  $\hbar$ drop out, the expression for the FEL gain is purely classical~\cite{madey2}.
  
  Nevertheless,  it was soon realized~\cite{mciver} that the correct description of this transition is more 
  subtle. An electron in the FEL emits \emph{many} photons during the interaction. The first-order 
  perturbation theory in Madey's work, however, includes only \emph{single}-photon processes. Although Madey derived the 
  correct result for the FEL gain, we  strictly speaking cannot employ his method.
                 
  This puzzling result has led to a variety of different approaches towards a quantum theory for the 
  FEL~\cite{mciver,becker79strong,*becker80strong,becker80,becker82,*becker83,*becker_pstat,*becker2,*becker_lw,banacloche,banacloche_pstat,orszag_lw,gover_lw}.
   For example, in 
  Ref.~\cite{becker80} it was argued that the higher-order contributions due to multiphoton transitions cancel 
  similar to the elementary model of a classical and fixed electron current that is coupled to a quantized  radiation 
  field~\cite{glauber}.         
  
  In Ref.~\cite{banacloche}, however, the problem was considered from a more practical point of view. 
  Although the  perturbative expansion of the quantum state does not converge, the 
  corresponding expansion for the observable of interest nevertheless may converge. Hence, in such 
  a situation the results from standard perturbation theory can be used to calculate the corresponding 
  expectation value, regardless of the underlying physical mechanism.

   \subsection{Wigner function}
  \label{sec:Model}
  
  A new facet to these topics was added by formulating a quantum theory of the FEL in terms of the 
  Wigner function~\cite{boni05,*pio07_wigner,*boni_wigner,serbeto08,dattoli18}.
  Indeed, the Wigner function is a perfect choice to study the transition from classical to quantum:
  On one hand it contains all information of the quantum state~\cite{vogel89}. One the other hand, this 
  description of quantum mechanics~\cite{wigner32,carruthers83,hillery84,schleich,case08}
  is as close as one can get to classical phase space~\cite{case08}.

  The dynamics of an electron in the FEL can be interpreted as one-dimensional motion of a particle with mass $m$ in a  periodic potential~\cite{colson}.
  For convenience we choose the dimensionless representations $\theta \equiv 2kz +\text{const.}$ and $\wp \equiv 
  p/\sqrt{U_0 m}$ for the position $z$ along the wiggler axis 
  and its conjugate momentum $p$, respectively, see App.~\ref{sec:Derivation_of_model}.
Here $k$ denotes the wave number of the laser field as well as of the wiggler field in the Bambini-Renieri 
  frame~\cite{bambi,*brs}, while $U_0$ 
is the height of the periodic potential
  and includes the amplitudes of both fields.

 The quantum state  of a single electron in Wigner representation $\mathcal{W}=\mathcal{W}(\theta,\wp;\tau)$ evolves according to the Quantum Liouville 
equation~\cite{schleich,carruthers83}
  \begin{equation}\label{eq:qLiou}
   \left(\frac{\partial}{\partial \tau} +\wp\frac{\partial}{\partial \theta}\right)\mathcal{W}=\mathcal{L}^{(1)}\mathcal{W}
  \end{equation}
 from Eq.~\eqref{eq:appA_qLiou_skal}, where   $\tau \equiv 2k t\sqrt{U_0/m}$ denotes a dimensionless version of the 
 time $t$.
The left-hand side of this partial differential equation describes the free time evolution while 
   \begin{equation}\label{eq:L1}
   \mathcal{L}^{(1)} \equiv -\varepsilon(\tau)\sin\theta 
     \sum\limits_ {m=0}^\infty \frac{1}{(2m+1)!}\frac{1}{(4\alpha)^m}
     \frac{\partial^{2m+1}}{\partial\wp^{2m+1}}
  \end{equation}
 corresponds to the periodic potential. For a derivation see Appendices~\ref{sec:Derivation_of_model} and 
 \ref{sec:Perturbation_theory}. 

 Here we have introduced the parameter
$\alpha \equiv U_0/(2\hbar\omega_\text{r})$
as the ratio of potential height and recoil energy $\hbar \omega_\text{r}$. The recoil frequency $\omega_\text{r}\equiv (2\hbar k)^2/(2m\hbar)$  is associated with the  recoil $2\hbar k$ 
 the electron experiences when scattered from a laser and a wiggler photon. While the recoil itself is the origin of 
 gain in the FEL~\cite{becker80}, 
 its discrete nature is responsible for the emergence of quantum effects~\cite{NJP2015}.

 Finally, the dynamics of the normalized amplitude $\varepsilon=\varepsilon(\tau)$ of the laser field follows from Maxwell's equations resulting in a  semiclassical model for the FEL dynamics.

\subsection{Quantum vs classical}
 
In Ref.~\cite{NJP2015} we have identified $\alpha$
as the \emph{quantum parameter} that governs the transition from the classical limit 
 $\alpha \gg 1$ to the quantum regime $\alpha \ll 1$ of the FEL. 
For large values of $\alpha$ the Quantum Liouville equation
at first sight reduces~\cite{boni05,boni_wigner} to the Boltzmann equation~\cite{scully_fel}
\begin{equation}\label{eq:boltzmann}
    \left(\frac{\partial}{\partial \tau} +\wp\frac{\partial}{\partial \theta}\right)f_\text{cl}=\mathcal{L}_\text{cl}^{(1)}f_\text{cl}\equiv -\varepsilon(\tau)\sin\theta  \frac{\partial f_\text{cl}}{\partial\wp}
  \end{equation}
for a classical phase-space  distribution function $f_\text{cl}=f_\text{cl}(\theta,\wp;\tau)$. While the free part is 
the same as in the quantum model,  the potential term 
 $\mathcal{L}_\text{cl}^{(1)}$
contains only one derivative with respect to $\wp$ instead of an infinite sum. Thus, we apparently obtain $\mathcal{L}^{(1)}\rightarrow \mathcal{L}_\text{cl}^{(1)}$ in the limit $\alpha \rightarrow \infty$.

However, the situation is more involved: only increasing $\alpha$ does \emph{not necessarily} lead to the classical 
regime. In Eq.~\eqref{eq:L1} we cannot simply truncate the series
if we do not take the magnitude of the derivatives of $\mathcal W$ into account \cite{heller76,*heller77,case08}.
 
Let the Wigner function be characterized by the width 
 $\Delta \wp\equiv \Delta p/\sqrt{U_0 m} $ in momentum space. Due to the derivatives, powers of $\Delta p$  will appear in the denominator when 
  $\mathcal{L}^ {(1)}$ acts on $\mathcal{W}$. We thus infer the scaling
   \begin{equation}
     \frac{1}{(4\alpha)^m}
     \frac{\partial^{2m+1}\mathcal{W}}{\partial\wp^{2m+1}}\sim 
\left(\frac{\hbar k}{\Delta p}\right)^{2m}\frac{\mathcal{W}}{\Delta \wp}
   \end{equation}
 for the $m$-th term of the series after recalling the definition of $\alpha$. Indeed, if $\Delta p$ is of the order of the recoil $\hbar k$ 
 (that is $\Delta \wp \sim \alpha^{-1/2}$) all  higher contributions of the series are of the same order 
as the ``classical'' contribution $\mathcal{W}/\Delta \wp$ (for $m=0$) and we must not neglect them. 
 Hence, to obtain the classical limit it is not sufficient that the quantum mechanical recoil is small compared to the 
 height of the potential. In addition, it has to be small compared to the momentum spread of the electron beam.  
 
 For a pure state, for example a Gaussian wave packet~\cite{gover18, dattoli_gaussian_2019}, decreasing the momentum 
 uncertainty is inevitably connected to an increase in position uncertainty.
 In a true classical limit, however,  momentum and position uncertainties can be chosen \emph{independently} of each other~\cite{case08}. 
 This feature only emerges if the quantum state of an electron is described by a statistical mixture rather than by a pure state. 
 In other words, the classical uncertainties for momentum and position are larger than the intrinsic quantum ones from  the uncertainty principle.
 
 It is convenient to assume that the initial Wigner function for the electron beam is given by the product
   \begin{equation}\label{eq:init}
    \mathcal{W}(\theta,\wp;0)=\frac{1}{2\pi }\,\rho(\wp)
   \end{equation}
 of an arbitrary momentum distribution $\rho=\rho(\wp)$ and a uniform distribution in $\theta$-direction.
 This choice in position space is in accordance with classical FEL theory~\cite{scully_fel,schmueser,borenstein}
 since we cannot control the exact positions of each one of the electrons that are distributed over several wiggler 
 wavelengths~\cite{schmueser,meystre}. In contrast to a pure state, we can choose the width of $\rho$ without affecting 
 the distribution for $\theta$~\cite{case08}. 
 
 However, the evolution of a “classical state”  does not ensure  that we are in the classical regime \emph{for all 
 times}~\cite{case08,Katz07}. 
Local modulations of the Wigner function on the scale of $\hbar k$  may emerge  
 so that quantum effects reappear. We therefore expect that ultimately other effects enforce classicality, namely 
 sources of decoherence~\cite{Zurek2003}, for example induced by  space charge or spontaneous emission 
 \cite{debus_realizing_2018}.
 Alternatively, one can argue that a ``classical observer'' is unable to resolve these fine quantum signatures and is 
 therefore limited to an averaged measurement result \cite{berry91}.

We emphasize that the use of a statistical mixture does not imply a many-particle theory. In the low-gain 
regime~\cite{schmueser} the motions of all electrons decouple from each other and every electron interacts separately 
with the fields. Each electron in a bunch is a copy of itself, initially distributed according to 
$\mathcal{W}(\theta,\wp;0)$, and we interpret the $N$  single-particle interactions as $N$ repetitions of the same experiment. At the 
end we calculate the observable quantities, like the FEL gain, by averaging over all possible outcomes weighted by the 
time-evolved Wigner function 
$\mathcal{W}(\theta,\wp;\tau)$.   
 
  For example, we derive in App.~\ref{sec:Derivation_of_model} the semiclassical equation of motion 
\begin{equation}\label{eq:edot}
      \frac{\D\varepsilon(\tau)}{\D\tau}=-\chi \int\!\!\text{d}\theta
      \int\!\!\text{d}\wp \, \mathcal{W}(\theta,\wp;\tau)\sin\theta
   \end{equation} 
  for the dimensionless electric field $\varepsilon$ with the constant $\chi$ that includes the initial electron density $n_\text{e}$ and the initial strengths of wiggler and laser fields. The dynamics of $\varepsilon$ follows from Maxwell's equations with the electron current being determined by the Wigner function.

From Eq.~\eqref{eq:edot} it is evident why it is difficult to observe  quantum effects in the FEL radiation: even if  the Wigner function shows distinct quantum signatures,
these features might be washed out when we average over position and momentum.

\subsection{Outline}
 
 The remainder of this article is structured in the following way: In Sec.~\ref{sec:Wigner} we compare the time 
 evolution in phase space obtained from a quantum and a classical theory for the same initial state. 
 We first employ a perturbative expansion valid for the small signal-limit
before we resort to numerics in order to treat also longer interaction times. Next, we consider in Sec.~\ref{sec:gain}  the gain of the FEL.
 Again, we cover the small-signal limit as well as the evolution for longer times. 
 Finally, we discuss our results in Sec.~\ref{sec:Conclusions}.
 
 For completeness we derive in App.~\ref{sec:Derivation_of_model} the semiclassical model of the FEL used 
 throughout this article. Further, we have moved the perturbative calculation for the Wigner function to 
 App.~\ref{sec:Perturbation_theory}. In App.~\ref{app:time-evolution} we introduce energy eigenstates in terms  of 
 Mathieu functions and employ them to derive a formal expression for the Wigner function  used for the 
 numerical computations.

\section{Evolution of Wigner function}
  \label{sec:Wigner}

The Quantum Liouville equation constitutes a partial differential equation to which  we cannot find an exact, analytic  solution in general  since it contains infinitely many derivatives with respect to $\wp$. However,
before we turn to a numerical approach we asymptotically solve Eq.~\eqref{eq:qLiou} for small times. This short-time limit corresponds to the small-signal regime of an FEL.

\subsection{Small-signal limit}
\label{ssec:wigner_small-signal}  
  
In App.~\ref{sec:Perturbation_theory}  we perform the asymptotic expansion 
   \begin{equation}\label{eq:expansion}
     \varpi\cong \varpi^{(0)}+\varpi^{(1)}+\varpi^{(2)}+...
   \end{equation}  
of $\varpi$  and solve Eq.~\eqref{eq:qLiou} order by order. 
This perturbative approach resembles the procedure in Ref.~\cite{scully_fel} for the Boltzmann equation and thus differs 
from the usual perturbative solution of the Schr\"odinger equation~\cite{madey1,becker79,*becker88,colson_quantum,friedman} which is restricted to single-photon processes. 
While the former procedure is allowed as long as $\tau \ll 1 $, the latter one is only valid
for $\sqrt{\alpha}\tau\ll 1$ and thus breaks down quickly in the classical regime due to $\alpha \gg 1$.

  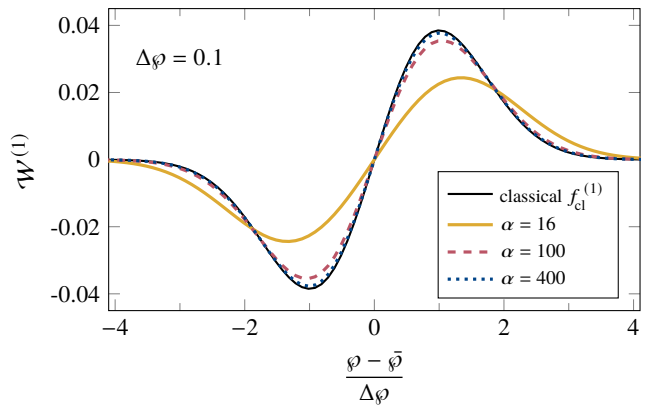
\begin{figure}
   \begin{tikzpicture}
     \begin{axis}
      [height=5.6cm,
       scaled ticks=false,
       xmin=-4.1,
       xmax=4.1,
       ymin=-0.045,
       ymax=0.045,
       xlabel={$\displaystyle\frac{\wp-\bar{\wp}}{\Delta\wp}$},
       ylabel={$\varpi^{(1)}$},
       xtick={-4,-2,0,2,4},
ytick={-0.04,-0.02,0,0.02,0.04},
       yticklabels={-0.04,-0.02,0,0.02,0.04},
]
     \addplot[black, thick] table{Daten/Abb1Tcl.dat};
      \addlegendentry{classical $f_\text{cl}^{(1)}$};
\addplot[TolHCYellow] table{Daten/Abb1T16.dat};
      \addlegendentry{$\alpha=16$};
\addplot[TolHCRed, dashed] table{Daten/Abb1T100.dat};
      \addlegendentry{$\alpha=100$};
\addplot[TolHCBlue, dotted] table{Daten/Abb1T400.dat};
            \addlegendentry{$\alpha=400$};
      
      \node at (-3,0.03) {$\Delta \wp =0.1$};
      
\end{axis}
   \end{tikzpicture}
  \caption{Small-signal contribution $\varpi^{(1)}$, Eq.~\eqref{eq:app_closed}, to the Wigner function of an electron 
   in the FEL for the fixed phase $\theta=\pi$ 
	and as a function of the relative and normalized momentum $(\wp-\bar{\wp})/\Delta\wp$. We compare the results for 
	three different values of the quantum parameter  $\alpha$, that is $\alpha=16$ (solid yellow line), $\alpha=100$ 
	(red dashed line),
	and $\alpha=400$ (blue dotted line), to the classical distribution function $f_\text{cl}^ {(1)}$, 
	Eq.~\eqref{eq:app_fcl_1} (blue line)~\cite{noteimbild}. In all cases we have chosen the values $\varepsilon=1$, 
	$\tau=0.01$, $\bar{\wp}=\pi$, and
	$\Delta\wp =0.1$ for the normalized field amplitude, the dimensionless time, the mean initial momentum, and the 
	initial momentum spread, respectively. Although $\alpha=16\gg 1$ indicates that we are close to the classical 
	regime the 
	corresponding curve for $\varpi^{(1)}$
	significantly differs from the classical result. Only after increasing $\alpha$ to $\alpha=400$ we observe an 
	agreement between the quantum and the classical theory. 
	As apparent from Eq.~\eqref{eq:Q} the important parameter that governs this transition is not $\alpha$ but rather 
	the ratio $\hbar k/\Delta p$ of recoil $\hbar k $ and momentum spread $\Delta p$.
	Indeed, for $\alpha=16 $ this parameter is  with $\hbar k/\Delta p=1.25$ of the order of unity and we cannot 
	neglect quantum corrections while $\alpha =400$ leads to the decreased value  $\hbar k/\Delta p=0.25\ll 1$. }
\label{fig:Q}        
\end{figure}

We choose the initial state from Eq.~\eqref{eq:init} and assume that the initial momentum distribution $\rho$ is  
Gaussian with mean value $\bar{\wp}$ and standard deviation $\Delta \wp$. 
In App.~\ref{sec:Perturbation_theory} we derive the expression \begin{equation}\label{eq:Q_def}
\varpi^{(1)}(\theta,\wp;\tau)\equiv f_\text{cl}^{(1)}(\theta,\wp;\tau)
\left[1+\mathcal{Q}\left(\frac{\wp-\bar{\wp}}{\sqrt{2}\Delta \wp}\right)\right]
\end{equation}      
for the first-order contribution to the unperturbed Wigner function, where $f_\text{cl}^{(1)}$ from 
Eq.~\eqref{eq:app_fcl_1}
denotes the corresponding solution for the classical Boltzmann equation~\eqref{eq:boltzmann}.

The quantum corrections $\mathcal{Q}= \mathcal{Q}(\xi)$ are given by the series
  \begin{equation}\label{eq:Q}
   \mathcal{Q}(\xi)=\sum\limits_{m=1}^\infty\frac{1}{(2m+1)!}\left(\frac{\hbar k}{\sqrt{2}\Delta p}\right)^{2m}\frac{ H_{2m+1}(\xi)}{ H_1(\xi)}\end{equation}
of Hermite polynomials which depend only on the relative momentum $\xi\equiv (\wp-\bar{\wp})/(\sqrt{2}\Delta \wp)$, but 
not on position or time. The analysis of Eq.~\eqref{eq:Q} reveals the importance of the ratio  of recoil and momentum spread: The terms of the series scale with powers of $\hbar k/\Delta p$ and 
only if this parameter is small, each term of the series decreases and the Wigner function $\varpi^{(1)}$ approaches to 
the classical distribution function $f_\text{cl}^{(1)}$.

In Fig.~\ref{fig:Q} we plot $\mathcal W^{(1)}$ as a function of the relative momentum for different values of the quantum parameter but for a fixed momentum spread and compare it to the classical result $f_\text{cl}^{(1)}$.
Although $\alpha=16$ indicates that we are close to the classical regime,
$\mathcal W^{(1)}$ does not agree with the classical curve. 
Following the discussion above the important parameter is not $\alpha$ but rather 
$\hbar k/\Delta p=1/(2\sqrt{\alpha}\Delta\wp)$ which is in this case of the order of unity, that is $\hbar k/\Delta 
p=1.25$.
Indeed, increasing the quantum parameter to $\alpha=400$ leads to $\hbar k/\Delta p=0.25$ and thus to negligible quantum corrections in Fig.~\ref{fig:Q}.

In conclusion, the phrase “small recoil” in the present context means: small compared to the initial momentum spread. 
Only in this
limit we can neglect the quantum corrections to the Wigner function.

\begin{figure*}[t!]
\begin{tikzpicture}

\begin{groupplot}[group style={
group size=4 by 3,
xticklabels at=edge bottom,
yticklabels at=edge left,
horizontal sep=1em,
vertical sep=1em,
},
height=6.5cm,
width=5.1cm,
clip=true,
axis on top,
xmin=0,
xmax=6.31491856630676,
ymin=-4.5,
ymax=4.5,
xtick={0,1.5708, 3.15159, 4.7124,  6.2832},
xticklabels={$0$, $\frac{\pi}{2}$, $\pi$, $\frac{3\pi}{2}$, \clap{$2\pi$}},
typeset ticklabels with strut,
colormap={mymap}{[1pt] rgb(0pt)=(0.2422,0.1504,0.6603); rgb(1pt)=(0.25039,0.164995,0.707614); rgb(2pt)=(0.257771,0.181781,0.751138); rgb(3pt)=(0.264729,0.197757,0.795214); rgb(4pt)=(0.270648,0.214676,0.836371); rgb(5pt)=(0.275114,0.234238,0.870986); rgb(6pt)=(0.2783,0.255871,0.899071); rgb(7pt)=(0.280333,0.278233,0.9221); rgb(8pt)=(0.281338,0.300595,0.941376); rgb(9pt)=(0.281014,0.322757,0.957886); rgb(10pt)=(0.279467,0.344671,0.971676); rgb(11pt)=(0.275971,0.366681,0.982905); rgb(12pt)=(0.269914,0.3892,0.9906); rgb(13pt)=(0.260243,0.412329,0.995157); rgb(14pt)=(0.244033,0.435833,0.998833); rgb(15pt)=(0.220643,0.460257,0.997286); rgb(16pt)=(0.196333,0.484719,0.989152); rgb(17pt)=(0.183405,0.507371,0.979795); rgb(18pt)=(0.178643,0.528857,0.968157); rgb(19pt)=(0.176438,0.549905,0.952019); rgb(20pt)=(0.168743,0.570262,0.935871); rgb(21pt)=(0.154,0.5902,0.9218); rgb(22pt)=(0.146029,0.609119,0.907857); rgb(23pt)=(0.138024,0.627629,0.89729); rgb(24pt)=(0.124814,0.645929,0.888343); rgb(25pt)=(0.111252,0.6635,0.876314); rgb(26pt)=(0.0952095,0.679829,0.859781); rgb(27pt)=(0.0688714,0.694771,0.839357); rgb(28pt)=(0.0296667,0.708167,0.816333); rgb(29pt)=(0.00357143,0.720267,0.7917); rgb(30pt)=(0.00665714,0.731214,0.766014); rgb(31pt)=(0.0433286,0.741095,0.73941); rgb(32pt)=(0.0963952,0.75,0.712038); rgb(33pt)=(0.140771,0.7584,0.684157); rgb(34pt)=(0.1717,0.766962,0.655443); rgb(35pt)=(0.193767,0.775767,0.6251); rgb(36pt)=(0.216086,0.7843,0.5923); rgb(37pt)=(0.246957,0.791795,0.556743); rgb(38pt)=(0.290614,0.79729,0.518829); rgb(39pt)=(0.340643,0.8008,0.478857); rgb(40pt)=(0.3909,0.802871,0.435448); rgb(41pt)=(0.445629,0.802419,0.390919); rgb(42pt)=(0.5044,0.7993,0.348); rgb(43pt)=(0.561562,0.794233,0.304481); rgb(44pt)=(0.617395,0.787619,0.261238); rgb(45pt)=(0.671986,0.779271,0.2227); rgb(46pt)=(0.7242,0.769843,0.191029); rgb(47pt)=(0.773833,0.759805,0.16461); rgb(48pt)=(0.820314,0.749814,0.153529); rgb(49pt)=(0.863433,0.7406,0.159633); rgb(50pt)=(0.903543,0.733029,0.177414); rgb(51pt)=(0.939257,0.728786,0.209957); rgb(52pt)=(0.972757,0.729771,0.239443); rgb(53pt)=(0.995648,0.743371,0.237148); rgb(54pt)=(0.996986,0.765857,0.219943); rgb(55pt)=(0.995205,0.789252,0.202762); rgb(56pt)=(0.9892,0.813567,0.188533); rgb(57pt)=(0.978629,0.838629,0.176557); rgb(58pt)=(0.967648,0.8639,0.16429); rgb(59pt)=(0.96101,0.889019,0.153676); rgb(60pt)=(0.959671,0.913457,0.142257); rgb(61pt)=(0.962795,0.937338,0.12651); rgb(62pt)=(0.969114,0.960629,0.106362); rgb(63pt)=(0.9769,0.9839,0.0805)},
colorbar style={width=5mm},
cycle list name = tol4HC cycle,
]
\nextgroupplot[ylabel style={align=center},
ylabel={$\Delta\wp=0.1$\\$\wp$},
title={momentum distribution},
x dir=reverse,
no marks,
xmax = 4.5,
legend style ={ at={(0.6,.05)}}
]
\addplot+ [dotted]
table[]{Daten/q_cl_comp_3x3_pdist_zero_wp_a03-1.tsv};
\addlegendentry{initial}\label{initial_dist}

\addplot+ []
table[]{Daten/q_cl_comp_3x3_pdist_zero_wp_a03-2.tsv};
\addlegendentry{$\alpha=1/3$}

\addplot+ []
table[]{Daten/q_cl_comp_3x3_pdist_zero_wp_a03-3.tsv};
\addlegendentry{$\alpha=10$}

\addplot+ []
table[]{Daten/q_cl_comp_3x3_pdist_zero_wp_a03-4.tsv};
\addlegendentry{classical}
\SubLabel{plt:qcl_comp_s01_pdist}

\nextgroupplot[title={$\varpi(\tau=\pi),\ \alpha=1/3$},
point meta min=-0.63493937253952,
point meta max=0.63493937253952,]
\addplot [forget plot] graphics [xmin=-0.00524472897093455, xmax=6.28843003615052, ymin=-4.5075, ymax=4.5075] {Daten/q_cl_comp_3x3_pdist_zero_wp_a03-1.png};
\SubLabel*{plt:qcl_comp_s01a03}

\nextgroupplot[title={$\varpi(\tau=\pi),\ \alpha=10$},
point meta min=-0.63493937253952,
point meta max=0.63493937253952,]
\addplot [forget plot] graphics [xmin=-0.00524472897093455, xmax=6.28843003615052, ymin=-4.5075, ymax=4.5075] {Daten/q_cl_comp_3x3_pdist_zero_wp_a03-2.png};
\SubLabel*{plt:qcl_comp_s01a10}

\nextgroupplot[title={$f_\text{cl}(\tau=\pi)$},
point meta min=-0.63493937253952,
point meta max=0.63493937253952,,
  colorbar,colorbar style={ scaled ticks=false,
 	width=3mm,
 	yticklabel style={
 		/pgf/number format/precision=1,
 		/pgf/number format/fixed,
 		/pgf/number format/fixed zerofill,
 	}
 }]
\addplot [forget plot] graphics [xmin=-0.00524472897093455, xmax=6.28843003615052, ymin=-4.5075, ymax=4.5075] {Daten/q_cl_comp_3x3_pdist_zero_wp_a03-3.png};
\SubLabel*{plt:qcl_comp_s01cl}

 \nextgroupplot[ylabel style={align=center},ylabel={$\Delta\wp=1.0$\\$\wp$},
x dir=reverse,
xmax=.5,
no marks]
\addplot+ [dotted]
table[]{Daten/q_cl_comp_3x3_pdist_zero_wp_a03-5.tsv};
\addplot+ 
table[]{Daten/q_cl_comp_3x3_pdist_zero_wp_a03-6.tsv};
\addplot+ 
table[]{Daten/q_cl_comp_3x3_pdist_zero_wp_a03-7.tsv};
\addplot+ 
table[]{Daten/q_cl_comp_3x3_pdist_zero_wp_a03-8.tsv};

\SubLabel*{plt:qcl_comp_s1_pdist}

 \nextgroupplot[
point meta min=-0.10719109326601,
point meta max=0.10719109326601,]
\addplot [forget plot] graphics [xmin=-0.00524472897093455, xmax=6.28843003615052, ymin=-4.5075, ymax=4.5075] {Daten/q_cl_comp_3x3_pdist_zero_wp_a03-4.png};
\SubLabel*{plt:qcl_comp_s1a03}

 \nextgroupplot[point meta min=-0.10719109326601,
 point meta max=0.10719109326601,]
\addplot [forget plot] graphics [xmin=-0.00524472897093455, xmax=6.28843003615052, ymin=-4.5075, ymax=4.5075] {Daten/q_cl_comp_3x3_pdist_zero_wp_a03-5.png};
\SubLabel*{plt:qcl_comp_s1a10}
 
 \nextgroupplot[point meta min=-0.10719109326601,
 point meta max=0.10719109326601,
 colorbar,colorbar style={ scaled ticks=false,
 	width=3mm,
 	yticklabel style={
 		/pgf/number format/precision=1,
 		/pgf/number format/fixed,
 		/pgf/number format/fixed zerofill,
 	}
 }]
\addplot [forget plot] graphics [xmin=-0.00524472897093455, xmax=6.28843003615052, ymin=-4.5075, ymax=4.5075] {Daten/q_cl_comp_3x3_pdist_zero_wp_a03-6.png};
\SubLabel*{plt:qcl_comp_s1cl}

 \nextgroupplot[ylabel style={align=center}, ylabel={$\Delta\wp=2.0$\\$\wp$},
  xlabel={$P(\wp)$ arb. units},
  xmin=0,
  xmax=0.25,
  x dir=reverse,
  no marks
  ]  
\addplot+ [dotted]
table[]{Daten/q_cl_comp_3x3_pdist_zero_wp_a03-9.tsv};
\addplot+ []
table[]{Daten/q_cl_comp_3x3_pdist_zero_wp_a03-10.tsv};
\addplot+ []
table[]{Daten/q_cl_comp_3x3_pdist_zero_wp_a03-11.tsv};
\addplot+ []
table[]{Daten/q_cl_comp_3x3_pdist_zero_wp_a03-12.tsv};
\SubLabel*{plt:qcl_comp_s2_pdist}

 \nextgroupplot[ylabel style={align=center}, 
 xlabel={$\theta$},
point meta min=-0.0427097827196121,
point meta max=0.0427097827196121,]
\addplot [forget plot] graphics [xmin=-0.00524472897093455, xmax=6.28843003615052, ymin=-4.5075, ymax=4.5075] {Daten/q_cl_comp_3x3_pdist_zero_wp_a03-7.png};
\SubLabel*{plt:qcl_comp_s2a03}

\nextgroupplot[xlabel={$\theta$},
point meta min=-0.0427097827196121,
point meta max=0.0427097827196121,]
\addplot [forget plot] graphics [xmin=-0.00524472897093455, xmax=6.28843003615052, ymin=-4.5075, ymax=4.5075] {Daten/q_cl_comp_3x3_pdist_zero_wp_a03-8.png};
\SubLabel*{plt:qcl_comp_s2a10}

 \nextgroupplot[xlabel={$\theta$},
point meta min=-0.0427097827196121,
point meta max=0.0427097827196121,
 colorbar, colorbar style={ scaled ticks=false,
 	width=3mm,
yticklabel style={
	/pgf/number format/precision=2,
	/pgf/number format/fixed,
	/pgf/number format/fixed zerofill,
}
}]
\addplot [forget plot] graphics [xmin=-0.00524472897093455, xmax=6.28843003615052, ymin=-4.5075, ymax=4.5075] {Daten/q_cl_comp_3x3_pdist_zero_wp_a03-9.png};
\SubLabel*{plt:qcl_comp_s2cl}

\end{groupplot}
\end{tikzpicture} \caption{Comparison of Wigner function $\mathcal W$ and classical phase-space distribution 	
		$f_\text{cl}$ both as functions of dimensionless position and momentum, $\theta$ and $\wp$, 
		respectively, for different values of the quantum parameter $\alpha$ and of the initial momentum spread 
		$\Delta\wp$. In all cases we have chosen the time $\tau=\pi$, which classically corresponds to half of 
		a rotation near to the center of the potential. 
		Each row corresponds to a different initial momentum width, that is $\Delta \wp = 0.1$, $1.0$, and $2.0$. The 
		marginals on the left show the initial momentum distribution (dotted) and the evolved ones for $\alpha=1/3$ and 
		$10$ as well as the corresponding classical result (solid lines). 
		The second and third column show the evolved Wigner functions for 
		$\alpha=1/3$ and $\alpha=10$, the rightmost column displays the classical phase-space distribution.	
		In the case of small $\alpha$ and $\Delta\wp$ the discrete structure in	momentum is visible in 
		\ref{plt:qcl_comp_s01a03}, where we have $\hbar k/\Delta p=8.66$. With an increased $\alpha$ more levels are 
		involved in \ref{plt:qcl_comp_s01a10} and the shape of the  corresponding classical 
		distribution in \ref{plt:qcl_comp_s01cl} starts to appear,
even though there are a lot of interference structures. This behavior is consistent with the value of  $\hbar k/\Delta p=1.58$, which is of the order of unity.
		A broader initial momentum distribution washes out the discrete levels, but still the Wigner 
		functions for $\alpha=1/3$  in \ref{plt:qcl_comp_s1a03} with $\hbar k/\Delta p=0.87$, and in 
		\ref{plt:qcl_comp_s2a03} with $\hbar k/\Delta p=0.43$ are quite different from their classical counterparts 
		in \ref{plt:qcl_comp_s1cl} and in \ref{plt:qcl_comp_s2cl}, respectively.
		For large $\alpha=10$ and $\Delta\wp=1$, that is $\hbar k/\Delta p=0.16$, there are only a few interference 
		fringes left in \ref{plt:qcl_comp_s1a10}.
		An even wider initial distribution with $\Delta\wp=2$, corresponding to $\hbar k/\Delta p=0.08$, makes them 
		nearly invisible such that Wigner function in \ref{plt:qcl_comp_s2a10} and classical distribution 
		in \ref{plt:qcl_comp_s2cl} resemble each other.} 
	\label{fig:q_cl_comparison}
\end{figure*}
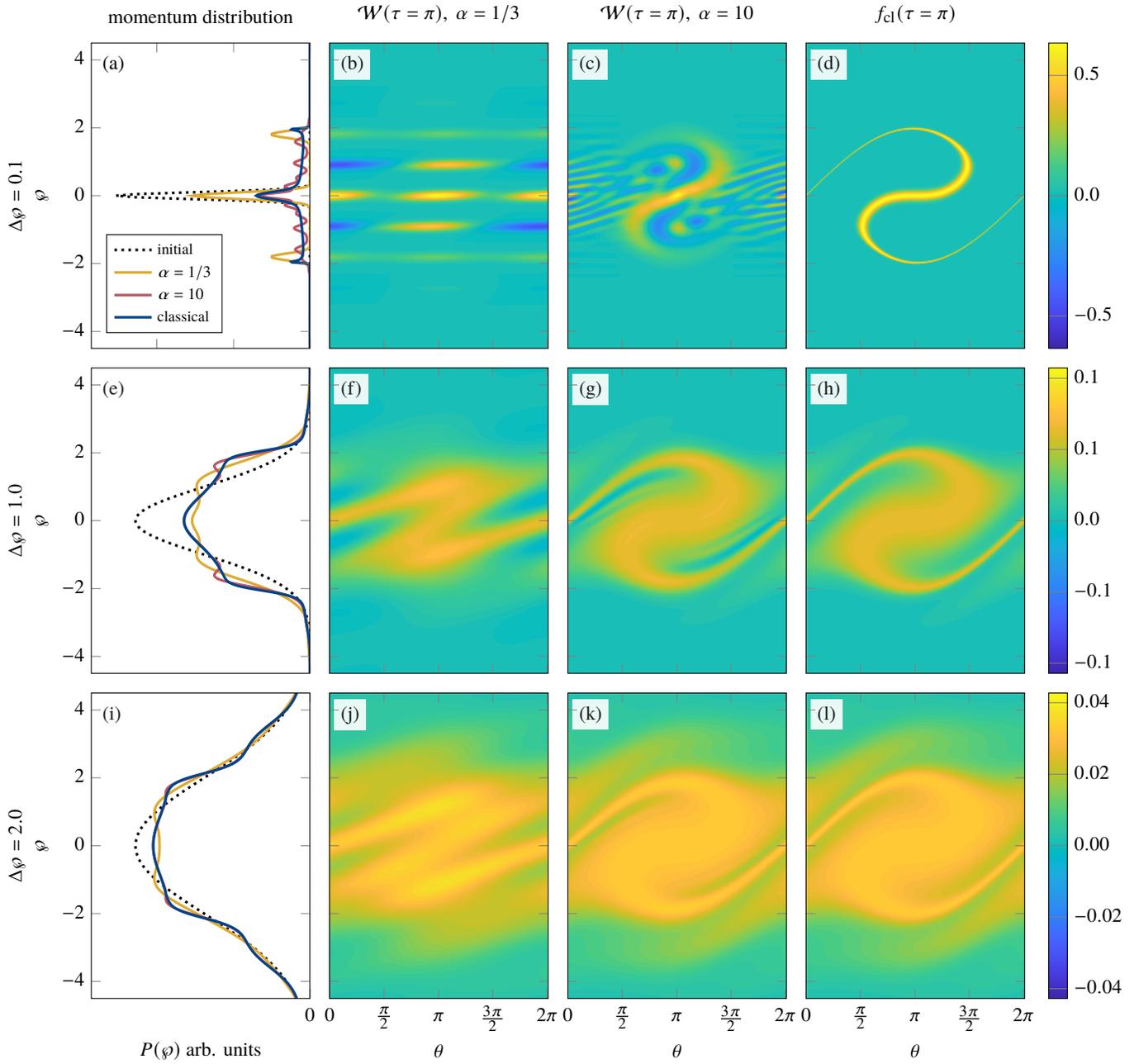

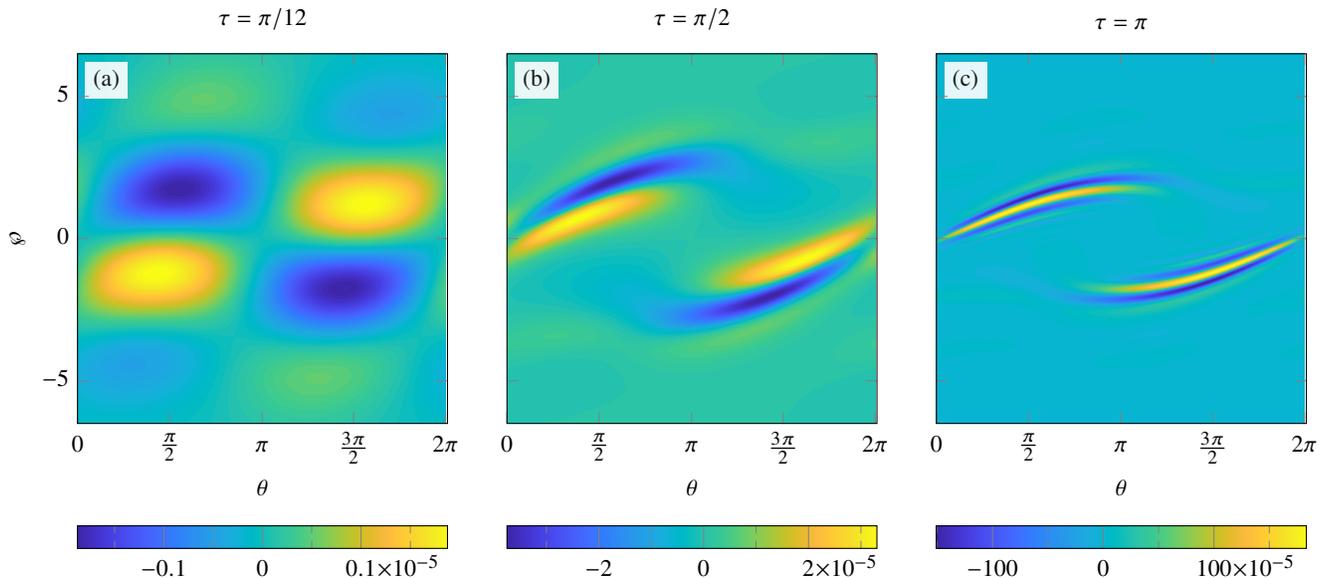
\begin{figure*}[t]
\begin{tikzpicture}
\begin{groupplot}[group style={
	group size=3 by 1,
	xticklabels at=edge bottom,
	yticklabels at=edge left,
	horizontal sep=2.5em,
	vertical sep=1em,
},
height=6.5cm,
width=6.5cm,
clip=true,
axis on top,
xmin=0,
xmax=6.31491856630676,
ymin=-6.5,
ymax=6.5,
xtick={0,1.5708, 3.15159, 4.7124,  6.2832},
xticklabels={$0$, $\frac{\pi}{2}$, $\pi$, $\frac{3\pi}{2}$, $2\pi$},
typeset ticklabels with strut,
xlabel={$\theta$},
axis on top,
colormap={mymap}{[1pt] rgb(0pt)=(0.2422,0.1504,0.6603); rgb(1pt)=(0.25039,0.164995,0.707614); 
	rgb(2pt)=(0.257771,0.181781,0.751138); rgb(3pt)=(0.264729,0.197757,0.795214); 
	rgb(4pt)=(0.270648,0.214676,0.836371); 
	rgb(5pt)=(0.275114,0.234238,0.870986); rgb(6pt)=(0.2783,0.255871,0.899071); rgb(7pt)=(0.280333,0.278233,0.9221); 
	rgb(8pt)=(0.281338,0.300595,0.941376); rgb(9pt)=(0.281014,0.322757,0.957886); 
	rgb(10pt)=(0.279467,0.344671,0.971676); 
	rgb(11pt)=(0.275971,0.366681,0.982905); rgb(12pt)=(0.269914,0.3892,0.9906); rgb(13pt)=(0.260243,0.412329,0.995157); 
	rgb(14pt)=(0.244033,0.435833,0.998833); rgb(15pt)=(0.220643,0.460257,0.997286); 
	rgb(16pt)=(0.196333,0.484719,0.989152); 
	rgb(17pt)=(0.183405,0.507371,0.979795); rgb(18pt)=(0.178643,0.528857,0.968157); 
	rgb(19pt)=(0.176438,0.549905,0.952019); 
	rgb(20pt)=(0.168743,0.570262,0.935871); rgb(21pt)=(0.154,0.5902,0.9218); rgb(22pt)=(0.146029,0.609119,0.907857); 
	rgb(23pt)=(0.138024,0.627629,0.89729); rgb(24pt)=(0.124814,0.645929,0.888343); 
	rgb(25pt)=(0.111252,0.6635,0.876314); 
	rgb(26pt)=(0.0952095,0.679829,0.859781); rgb(27pt)=(0.0688714,0.694771,0.839357); 
	rgb(28pt)=(0.0296667,0.708167,0.816333); rgb(29pt)=(0.00357143,0.720267,0.7917); 
	rgb(30pt)=(0.00665714,0.731214,0.766014); rgb(31pt)=(0.0433286,0.741095,0.73941); 
	rgb(32pt)=(0.0963952,0.75,0.712038); 
	rgb(33pt)=(0.140771,0.7584,0.684157); rgb(34pt)=(0.1717,0.766962,0.655443); rgb(35pt)=(0.193767,0.775767,0.6251); 
	rgb(36pt)=(0.216086,0.7843,0.5923); rgb(37pt)=(0.246957,0.791795,0.556743); rgb(38pt)=(0.290614,0.79729,0.518829); 
	rgb(39pt)=(0.340643,0.8008,0.478857); rgb(40pt)=(0.3909,0.802871,0.435448); rgb(41pt)=(0.445629,0.802419,0.390919); 
	rgb(42pt)=(0.5044,0.7993,0.348); rgb(43pt)=(0.561562,0.794233,0.304481); rgb(44pt)=(0.617395,0.787619,0.261238); 
	rgb(45pt)=(0.671986,0.779271,0.2227); rgb(46pt)=(0.7242,0.769843,0.191029); rgb(47pt)=(0.773833,0.759805,0.16461); 
	rgb(48pt)=(0.820314,0.749814,0.153529); rgb(49pt)=(0.863433,0.7406,0.159633); 
	rgb(50pt)=(0.903543,0.733029,0.177414); 
	rgb(51pt)=(0.939257,0.728786,0.209957); rgb(52pt)=(0.972757,0.729771,0.239443); 
	rgb(53pt)=(0.995648,0.743371,0.237148); 
	rgb(54pt)=(0.996986,0.765857,0.219943); rgb(55pt)=(0.995205,0.789252,0.202762); 
	rgb(56pt)=(0.9892,0.813567,0.188533); 
	rgb(57pt)=(0.978629,0.838629,0.176557); rgb(58pt)=(0.967648,0.8639,0.16429); rgb(59pt)=(0.96101,0.889019,0.153676); 
	rgb(60pt)=(0.959671,0.913457,0.142257); rgb(61pt)=(0.962795,0.937338,0.12651); 
	rgb(62pt)=(0.969114,0.960629,0.106362); 
	rgb(63pt)=(0.9769,0.9839,0.0805)},
colorbar horizontal,
colorbar style={height=3mm,
	scaled x ticks=base 10:5,
every x tick label/.append style={alias=XTick,inner xsep=0pt},
	every x tick scale label/.style={at=(XTick.base east),anchor=base west},
	tick scale binop=\times
}
]

\nextgroupplot[title={$\tau=\pi/12$},
ylabel={$\wp$},
point meta min=-1.87627909998666e-06,
point meta max=1.87698963304526e-06,
]
\addplot [forget plot] graphics [xmin=-0.000787366579847066, xmax=6.28397267375943, ymin=-10.1242295713828, 
ymax=10.1242295713828] {Daten/wigner-diff-evolution-zero-wp-sp=2-1.png};
\SubLabel*{plt:w_diff-pi/12}

\nextgroupplot[title={$\tau=\pi/2$},
point meta min=-3.75127342322028e-05,
point meta max=3.30566491735847e-05,
]
\addplot [forget plot] graphics [xmin=-0.000787366579847066, xmax=6.28397267375943, ymin=-10.1242295713828, 
ymax=10.1242295713828] {Daten/wigner-diff-evolution-zero-wp-sp=2-2.png};
\SubLabel*{plt:w_diff-pi/2}

\nextgroupplot[title={$\tau=\pi$},
point meta min=-0.00150658540201804,
point meta max=0.00183517744529021,
]
\addplot [forget plot] graphics [xmin=-0.000787366579847066, xmax=6.28397267375943, ymin=-10.1242295713828, 
ymax=10.1242295713828] {Daten/wigner-diff-evolution-zero-wp-sp=2-3.png};
\SubLabel*{plt:w_diff-pi}
\end{groupplot}
\end{tikzpicture} 	
	\caption{Deviation $\varpi-f_\text{cl}$ of the Wigner function from the classical distribution function 
	depending on dimensionless position and momentum, $\theta$ and $\wp$, respectively, at three different times. In 
	all plots we have chosen the values  $\alpha=10$ and $\Delta\wp=2$, that are the 
    configurations of Fig.~\ref{fig:q_cl_comparison}\ref{plt:qcl_comp_s2a10} and \ref{plt:qcl_comp_s2cl}.
    For a small time $\tau=\pi/12$, the interference fringes have 
	a width proportional to the initial distribution, but a low amplitude. In the course of time, the 
	amplitude grows (note the different scaling of the colormap for each plot) due to the increasing magnitude of the 
	higher-order derivatives of the Wigner function. Those, in turn, are a consequence of the narrow width in momentum 
	space that appears first in the proximity to the separatrix due to dispersion. The appearing interference fringes 
	are also narrow, leading to a self-reinforcing effect. 
However, the amplitude of the deviation remains in the depicted case about one order of 
	magnitude smaller than the maximal values of $\mathcal W$ in Fig.~\ref{fig:q_cl_comparison}\ref{plt:qcl_comp_s2a10}.
} \label{fig:difference_time}
\end{figure*}

\subsection{Longer times}
\label{ssec:wigner_longer_times}

The small-signal regime allows for a perturbative treatment, which is not possible for longer times.
Hence, we have to determine the time evolution of the Wigner function numerically.
In contrast, the solution of the classical equation of motion can be given in a closed form \footnote{We apply the 
method of 
characteristics \protect\cite{courant_methods_2008}, which reduces the problem to solving the equation of motion for 
the classical trajectories \protect\cite{louisell_exact_1979}. These can be written in terms of Jacobi elliptic 
functions \cite{Lawden}.}.
In Fig.~\ref{fig:q_cl_comparison} we compare our results for the Wigner function to the classical distribution function 
for different initial momentum widths $\Delta\wp$ and for different values of the quantum parameter $\alpha$.

The relevant—classical and quantum mechanical—phase-space dynamics takes place inside the 
classical separatrix, which separates open and closed trajectories, because here the largest changes of momentum  
appear.  The classical evolution of a phase-space distribution is a rotation inside the separatrix.
Since we consider an anharmonic oscillator, the angular velocity depends on position and momentum and decreases going 
from the center to the separatrix. As a consequence, the phase-space distribution is stretched during the 
evolution, since the inner parts move faster than the outer ones, see the right column of 
Fig.~\ref{fig:q_cl_comparison}.

The time evolution of a Wigner function that is initially uniform in space can be expressed as
\begin{equation}
\mathcal W(\theta,\wp;\tau)=\frac{\sa}{\pi}
\sum_{s=-\infty}^\infty
w_{s}(\theta,\wp;\tau)
\rho\left(\wp+\frac{s/2}\sa\right)\label{eqn:Wigner_sum},
\end{equation}
for details see App.~\ref{app:Wigner_formal}. This expression can be understood in the following way:
The momentum of the electrons changes by integer multiples $s$  of the discrete recoil $2\hbar k$.
The counter-intuitive occurrence of  half-integer multiples $s/2$ of $2\hbar k$  in Eq.~\eqref{eqn:Wigner_sum}
comes from the interference between two adjacent momentum levels
\footnote{
This effect is common to any Wigner function consisting of a coherent superposition of two or more 
individual states. Consider the superposition state $\ket{\psi}=\alpha \ket\phi +\beta\ket\chi$. Then the total Wigner 
function is given by $W(x,p)=\abs{\alpha}^2 W_{\ket\phi}(x,p)+ \abs{\beta}^2 W_{\ket{\chi}}(x,p) + 
2\Re{\alpha\beta^*\int\dd{y}\mathrm e^{\mathrm i py/\hbar}\phi(x-y/2)\chi^*(x+y/2)}$,
where $W_{\ket\phi}$ and $W_{\ket{\chi}}$, respectively, denote the Wigner functions for the 
states $\ket\phi$ and $\ket\chi$, while $\phi(x) =\braket{x}{\phi}$ and $\chi(x)=\braket{x}{\chi}$ are the position 
representations of their wave function. The last contribution is the interference term. See also 
Ref.~{\protect\cite{BUZEK19951}}.
}. Note, that the momentum changes of $s2\hbar k$ correspond to the emission and absorption of $|s|$ 
photons each. Therefore, multiple scattering events are  included in our low-gain model.

We interpret the prefactors $w_s$ as scattering amplitudes for the shifted parts of the Wigner function.
They depend not only on time and position but also on the momentum itself. As a consequence of this momentum 
dependence, only a fraction of the distribution is selected to participate in the interaction. 
This effect is also known as “velocity selectivity” in atomic Bragg diffraction \cite{giese_double_2013}.

In the quantum regime \cite{NJP2015}, that is $\alpha \ll 1$ and $\Delta p/2\hbar k\ll1$, the summation in Eq.\,\eqref{eqn:Wigner_sum} breaks down to a few terms, as  only few momentum levels are 
involved. In this extreme limit, single-photon processes dominate the dynamics. 
Since the initial momentum spread is 
sufficiently small, the shifted distributions do not overlap and hence appear as discrete lines in phase space.

In Fig.~\ref{fig:q_cl_comparison}\ref{plt:qcl_comp_s01a03}, where the time-evolved  Wigner function for $\alpha=1/3$ 
and $\Delta\wp=0.1$ is shown, we indeed observe only three momentum levels, that is the initial momentum  $p=0$ and the 
levels $p=2\hbar k$ and $p=-2\hbar k$, as well as two interference patterns between these levels. From the marginal 
distribution in \ref{fig:q_cl_comparison}\ref{plt:qcl_comp_s01_pdist} we recognize that these intermediate momenta do 
not contribute to the momentum distribution.  As expected from the value $\hbar k/\Delta p=8.66$, there is no agreement 
with the classical distribution in Fig.~\ref{fig:q_cl_comparison} \ref{plt:qcl_comp_s01cl}.

With increasing $\alpha$,  more and more momentum levels intersect with the separatrix and therefore contribute to the 
sum in  Eq.~\eqref{eqn:Wigner_sum}, that is, multi-photon processes occur. The phase-space structure starts to resemble 
the classical dynamics, but with 
an additional fine structure due to the interference between many levels, see 
Fig.~\ref{fig:q_cl_comparison}\ref{plt:qcl_comp_s01a10}. Here we have $\hbar k/\Delta p=1.58$, which is still 
outside the range for a classical evolution. From the momentum distribution in 
Fig.~\ref{fig:q_cl_comparison}\ref{plt:qcl_comp_s01_pdist}, we recognize several single peaks corresponding 
to the distinct momentum levels, while the classical distribution is spread out rather homogeneously over the area 
enclosed by the separatrix.

We infer from these plots that a large $\alpha$ by itself is not sufficient to obtain a classical time evolution, 
which underlines the results from the preceding section. A broad initial momentum distribution is also not enough, as 
we can see in Fig.~\ref{fig:q_cl_comparison}\ref{plt:qcl_comp_s1a03} ($\hbar k/\Delta p=0.87$) and 
\ref{plt:qcl_comp_s2a03} ($\hbar k/\Delta p=0.43$) for $\alpha=1/3$. Here the discreteness of the few momentum levels 
is washed out due to the overlap of the shifted contributions \cite{NJP2015}. The shape of each distribution is 
however quite different from the classical counterpart in Fig.~\ref{fig:q_cl_comparison}\ref{plt:qcl_comp_s1cl} and  in 
\ref{plt:qcl_comp_s2cl}, respectively, since only few momentum levels are involved.

The combination of a broad initial distribution and the inclusion of many momentum levels, that is a large $\Delta\wp$ 
and a large $\alpha$, eventually leads to a Wigner function which seems to resemble the classical distribution 
function. By comparing Fig.~\ref{fig:q_cl_comparison}\ref{plt:qcl_comp_s2cl} with \ref{plt:qcl_comp_s2a10}, where the 
value $\hbar k/\Delta p=0.08$  matches the condition for a classical evolution, we can hardly recognize any difference 
between these two distribution functions. The momentum distributions agree even better, see 
Fig.~\ref{fig:q_cl_comparison}\ref{plt:qcl_comp_s2_pdist}.

At first sight, Fig.~\ref{fig:q_cl_comparison} completely confirms our expectations: For a small value of $\hbar 
k/\Delta p$ 
the Wigner function resembles its classical counterpart. However, if we go beyond this purely visual comparison, we 
indeed observe 
significant differences. For this purpose, we consider the numerical 
deviation $\mathcal{W}(\theta,\wp,\tau)-f_\text{cl}(\theta,\wp,\tau)$ 
of both distributions.
From Fig.~\ref{fig:difference_time} we notice that the differences increase in the course of time, even if the
conditions for a classical evolution are satisfied, that is $\alpha=10$ and $\Delta\wp = 2$ leading to $\hbar 
k/\Delta p=0.08$. 

For short times the deviations have a low amplitude, see the colorbar of 
\ref{fig:difference_time}\ref{plt:w_diff-pi/12}. The structure in momentum direction can be identified as the Hermite 
polynomials of the approximated analytical solution from Eq.~\eqref{eq:Q}. 
For longer times, the anharmonic dynamics leading to dispersion of the distribution in proximity to the separatrix 
becomes more prominent.
At the points in phase space where the distribution is narrow, the deviation increases by orders of magnitude, see
Fig.~\ref{fig:difference_time}\ref{plt:w_diff-pi/2} and \ref{plt:w_diff-pi}. 
This observation can be explained in the following way: Where the distribution is narrow, it has large 
higher-order derivatives. Those in turn lead to a significant difference between the potential term of the classical 
Boltzmann equation and the one of the Quantum Liouville equation.
Only if the terms involving the higher-order derivatives are suppressed, both equations are approximately the same. 
A small value of $\hbar k /\Delta p$ compensates the increasing contributions of the higher-order derivatives only for some time.
Consequently, every initial distribution that evolves coherently sooner or later becomes non-classical.  
Decoherence effects like spontaneous emission or space charge may impede or even remove the appearance of the quantum 
features in the distribution and hence might lead to a classical evolution \cite{Zurek2003}.

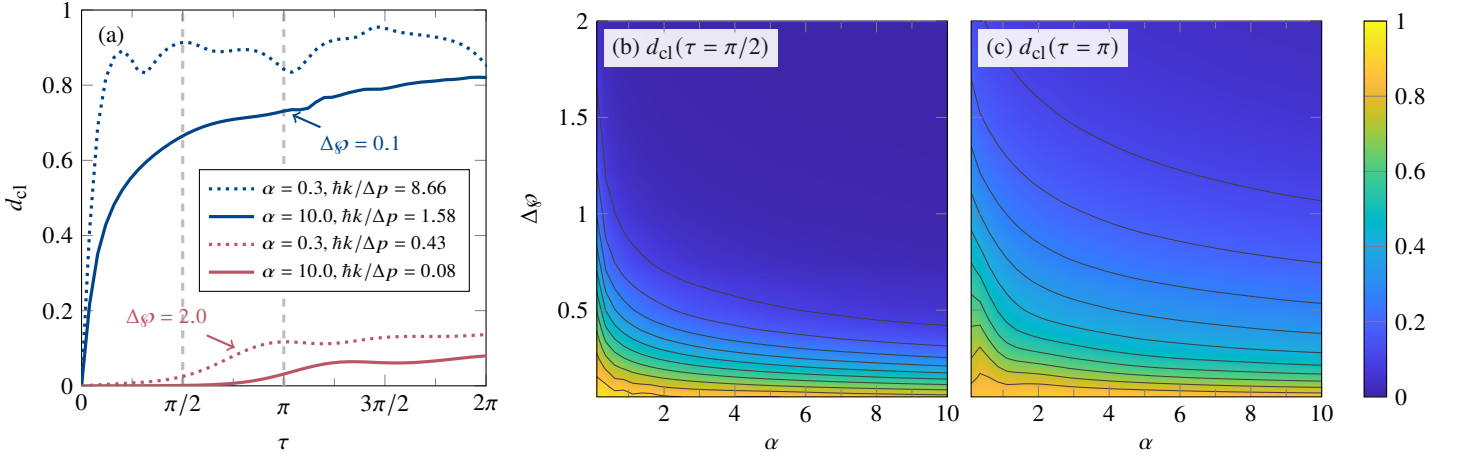
\begin{figure*}[t]
	\centerline{\begin{tikzpicture}\begin{axis}[width=.3\linewidth,
height=5cm,
scale only axis,
xtick = { 0, .5*3.1415, 1*3.1415, 1.5*3.1415, 2*3.1415},
xticklabels = {$0$,$\pi/2$, $\strut\pi$, $3\pi/2$, $2\pi$},
xmin=0,	
xmax=3.14159*2,
xlabel={$\tau$},
ymin=0,
ymax=1,
ylabel={$d_{\text{cl}}$},
legend style={legend cell align=left, align=left, at={(0.625,0.25)}, anchor=south},
no marks,
extra x ticks={3.14159/2,3.14159},
extra x tick labels={},
extra tick style={grid=major, grid style={dashed, very thick}},
]
\addplot+ [TolHCBlue, dotted]
  table[]{Daten/wigner-dist-evolution-2x2-1.tsv};
\addlegendentry{$\alpha=0.3$,\ $\hbar k/\Delta p=8.66$}
\addplot+ [TolHCBlue]
table[]{Daten/wigner-dist-evolution-2x2-3.tsv};
\addlegendentry{$\alpha=10.0$,\ $\hbar k/\Delta p=1.58$}
\addplot+ [TolHCRed, dotted]
  table[]{Daten/wigner-dist-evolution-2x2-2.tsv};
\addlegendentry{$\alpha=0.3$,\ $\hbar k/\Delta p=0.43$}
\addplot+ [TolHCRed]
  table[]{Daten/wigner-dist-evolution-2x2-4.tsv};
\addlegendentry{$\alpha=10.0$,\ $\hbar k/\Delta p=0.08$}
\node[pin={[pin distance=3mm,pin edge={TolHCBlue, 
thick,<-}]-30:{\footnotesize\textcolor{TolHCBlue}{$\Delta\wp=0.1$}}}] 
at 
(3.1415,0.7311){};
\node[pin={[pin distance=3mm,pin edge={TolHCRed, 
thick,<-}]150:{\footnotesize\textcolor{TolHCRed}{$\Delta\wp=2.0$}}}] 
at 
(2.513,0.09141){};
\SubLabel{plt:dist-te}
\end{axis}
\end{tikzpicture} \hfill
		\begin{tikzpicture}
\setcounter{plot}{1}
\begin{groupplot}[group style={
	group size=2 by 1,
	xticklabels at=edge bottom,
	yticklabels at=edge left,
	horizontal sep=1em,
	vertical sep=1em,
},
height=5cm,
width=.26\linewidth,
scale only axis,
clip=true,
axis on top,
xmin=0.1,
xmax=10,
ymin=0.05,
ymax=2,
point meta min=0,
point meta max=1,
colormap={mymap}{[1pt] rgb(0pt)=(0.2422,0.1504,0.6603); rgb(1pt)=(0.25039,0.164995,0.707614); 
	rgb(2pt)=(0.257771,0.181781,0.751138); rgb(3pt)=(0.264729,0.197757,0.795214); 
	rgb(4pt)=(0.270648,0.214676,0.836371); 
	rgb(5pt)=(0.275114,0.234238,0.870986); rgb(6pt)=(0.2783,0.255871,0.899071); rgb(7pt)=(0.280333,0.278233,0.9221); 
	rgb(8pt)=(0.281338,0.300595,0.941376); rgb(9pt)=(0.281014,0.322757,0.957886); 
	rgb(10pt)=(0.279467,0.344671,0.971676); 
	rgb(11pt)=(0.275971,0.366681,0.982905); rgb(12pt)=(0.269914,0.3892,0.9906); rgb(13pt)=(0.260243,0.412329,0.995157); 
	rgb(14pt)=(0.244033,0.435833,0.998833); rgb(15pt)=(0.220643,0.460257,0.997286); 
	rgb(16pt)=(0.196333,0.484719,0.989152); 
	rgb(17pt)=(0.183405,0.507371,0.979795); rgb(18pt)=(0.178643,0.528857,0.968157); 
	rgb(19pt)=(0.176438,0.549905,0.952019); 
	rgb(20pt)=(0.168743,0.570262,0.935871); rgb(21pt)=(0.154,0.5902,0.9218); rgb(22pt)=(0.146029,0.609119,0.907857); 
	rgb(23pt)=(0.138024,0.627629,0.89729); rgb(24pt)=(0.124814,0.645929,0.888343); 
	rgb(25pt)=(0.111252,0.6635,0.876314); 
	rgb(26pt)=(0.0952095,0.679829,0.859781); rgb(27pt)=(0.0688714,0.694771,0.839357); 
	rgb(28pt)=(0.0296667,0.708167,0.816333); rgb(29pt)=(0.00357143,0.720267,0.7917); 
	rgb(30pt)=(0.00665714,0.731214,0.766014); rgb(31pt)=(0.0433286,0.741095,0.73941); 
	rgb(32pt)=(0.0963952,0.75,0.712038); 
	rgb(33pt)=(0.140771,0.7584,0.684157); rgb(34pt)=(0.1717,0.766962,0.655443); rgb(35pt)=(0.193767,0.775767,0.6251); 
	rgb(36pt)=(0.216086,0.7843,0.5923); rgb(37pt)=(0.246957,0.791795,0.556743); rgb(38pt)=(0.290614,0.79729,0.518829); 
	rgb(39pt)=(0.340643,0.8008,0.478857); rgb(40pt)=(0.3909,0.802871,0.435448); rgb(41pt)=(0.445629,0.802419,0.390919); 
	rgb(42pt)=(0.5044,0.7993,0.348); rgb(43pt)=(0.561562,0.794233,0.304481); rgb(44pt)=(0.617395,0.787619,0.261238); 
	rgb(45pt)=(0.671986,0.779271,0.2227); rgb(46pt)=(0.7242,0.769843,0.191029); rgb(47pt)=(0.773833,0.759805,0.16461); 
	rgb(48pt)=(0.820314,0.749814,0.153529); rgb(49pt)=(0.863433,0.7406,0.159633); 
	rgb(50pt)=(0.903543,0.733029,0.177414); 
	rgb(51pt)=(0.939257,0.728786,0.209957); rgb(52pt)=(0.972757,0.729771,0.239443); 
	rgb(53pt)=(0.995648,0.743371,0.237148); 
	rgb(54pt)=(0.996986,0.765857,0.219943); rgb(55pt)=(0.995205,0.789252,0.202762); 
	rgb(56pt)=(0.9892,0.813567,0.188533); 
	rgb(57pt)=(0.978629,0.838629,0.176557); rgb(58pt)=(0.967648,0.8639,0.16429); rgb(59pt)=(0.96101,0.889019,0.153676); 
	rgb(60pt)=(0.959671,0.913457,0.142257); rgb(61pt)=(0.962795,0.937338,0.12651); 
	rgb(62pt)=(0.969114,0.960629,0.106362); 
	rgb(63pt)=(0.9769,0.9839,0.0805)},
colorbar style={width=3mm},
every axis plot/.append style={thin}
]
\nextgroupplot[
xlabel={$\alpha$},
ylabel=$\Delta\wp$,
]
\addplot[surf,
shader=interp, mesh/rows=17]
table[point meta=\thisrow{c}] {Daten/wigner-norm-dist-dep-2t-1.tsv};
\addplot[contour prepared, contour prepared format=matlab, contour/labels=false, contour/draw color=darkgray] table[] {Daten/wigner-norm-dist-dep-2t-2.tsv};
\SubLabel*[$d_\text{cl}(\tau=\pi/2)$]{plt:dcl_pi2}
\nextgroupplot[
xlabel={$\alpha$},
colorbar
]
\addplot[surf,
shader=interp, mesh/rows=17]
table[point meta=\thisrow{c}] {Daten/wigner-norm-dist-dep-2t-3.tsv};
\addplot[contour prepared, contour prepared format=matlab, contour/labels=false, contour/draw color=darkgray] table[] {Daten/wigner-norm-dist-dep-2t-4.tsv};
\SubLabel*[$d_\text{cl}(\tau=\pi)$]{plt:dcl_pi}
\end{groupplot}
\end{tikzpicture} }
\caption{Distance $d_\text{cl}$, Eq.~\eqref{eqn:dist}, between Wigner and classical phase-space distribution. A 
		large distance (close to unity) means that the distributions are very different, while a small value (close to 
		zero) indicates that they are similar. 
		\ref{plt:dist-te} Distance  $d_\text{cl}$ as a function of the dimensionless time $\tau$ for the 
		values $\alpha=0.3$ (dotted lines) and $\alpha=10$ (solid lines) of the quantum parameter  as well as for small 
		(blue)  and large (red) initial momentum spread $\Delta\wp$. Note that these are the same configurations 
		as for the first and third row of Fig.~\ref{fig:q_cl_comparison}.
		The distance increases with increasing values of $\hbar k/\Delta p=1/(2\sqrt{\alpha}\Delta\wp)$. For large 
		values of this parameter the distance rises quickly, while it remains small at short times for small values of 
		$\hbar k/\Delta p$.  
		After a longer time the distance suddenly increases even for small $\hbar k/\Delta p$ but saturates afterwards.
In \ref{plt:dcl_pi2} and \ref{plt:dcl_pi} we draw $d_\text{cl}$ as a function of the initial momentum width 
		$\Delta \wp$ and of the quantum parameter $\alpha$ at times $\tau=\pi/2$ and $\tau=\pi$, respectively, 
		corresponding to the grey vertical lines in plot \ref{plt:dist-te}.
		For small values of $\alpha$ and $\Delta\wp$, that is the lower left corner, $d_\text{cl}$ is maximized and 
		close to unity, while it decreases for large values of $\alpha$ and $\Delta\wp$ towards the upper right 
		corner, that is, the evolution becomes more classical. At the earlier time $\tau=\pi/2$ the decrease is steep, 
		while it is more gradual at the later time $\tau=\pi$. 
Even if the distance is very small at an earlier time, this doesn't mean that it will be small also for later 
		times.
	}\label{fig:distance}  
\end{figure*}

Even though all existing x-ray FELs operate in the high-gain regime, we rely on their parameters for an 
estimation of the decoherence time scales, since they are to date the only FEL devices with large recoils that approach 
the quantum regime. We emphasize that the high-gain regime is strictly-speaking not covered by our single-electron 
model.

For the European XFEL \cite{altarelli_xfel_2006} we obtain $\hbar k / \Delta p=0.014$,   
a value that is still much smaller than unity but at least corrections in the Wigner function due to the discrete 
momentum steps might become conceivable.
The interference pattern in Fig.~\ref{fig:difference_time}\ref{plt:w_diff-pi} emerges at $\tau\cong\pi$, 
that is the typical time for which the FEL gain saturates \cite{fedorov_rev,*fedorov_book}\nobreak. For the European 
XFEL the saturation length amounts to $L_\text{sat}=133\,\text{m}$ corresponding to a saturation time 
$T_\text{sat}=440\,\text{ns}$.

The corresponding time scales of possible decoherence mechanisms \cite{debus_realizing_2018} are calculated 
\footnote{The time scale for space charge effects $T_\text{sc}=1/\omega_\text{p}$ depends on the  
	relativistic plasma frequency  defined (in laboratory frame) as 
	$\omega_\text{p}=\sqrt{e^2n_\text{el}/(\varepsilon_0\gamma^3m_\text{el})}$, 
	where $n_\text{el}$ is the electron density, $\gamma$ the relativistic factor, and $\varepsilon_0$ the vacuum 
	permittivity while 	$m_\text{el}$ and $e$ are the electron mass and charge, respectively. For the timescale of 
	spontaneous emission, we use the formula for a classical electron in a wiggler
	\protect\cite[p. 692]{jackson_classical_1998} $T_\text{se}=3\lambda_\text{W}/(2\pi\alpha_\text{f}ca_0^2)$, 
	where $\alpha_\text{f}$ is the fine structure constant, $a_0$ the wiggler parameter and $\lambda_\text{W}$ the 
	wiggler	wavelength in the laboratory frame. See also Ref.~\cite{debus_realizing_2018}}  to 
$T_\text{se}=0.7\,\text{ns}$ (corresponding to the length $L_\text{se}=0.2\,\text{m}$) for spontaneous emission 
and $T_\text{p}=1\,\text{µs}$ (corresponding to the length $L_\text{p}=300\,\text{m}$) for space charge,
respectively.  The comparison of these three time scales reveals that, while we can neglect space charge effects, 
spontaneous emission occurs long before an interference pattern can emerge. Hence, we expect inerference effects to be 
suppressed in state-of-the-art machines.

In order to quantify the deviation of quantum and classical phase-space evolution we introduce the quantity
\begin{equation}
d_\text{cl}(\tau)=\left\{\frac{\int_{-\infty}^{\infty}\dd{\wp}\int_0^{2\pi}\dd{\theta}\left[\mathcal 
W(\theta,\wp;\tau)-f_\text{cl}(\theta,\wp;\tau)\right]^2}
{\int_{-\infty}^{\infty}\dd{\wp}\int_0^{2\pi}\dd{\theta}\left[\mathcal 
	W(\theta,\wp;\tau)^2 +f_\text{cl}(\theta,\wp;\tau)^2\right]}
\right\}^{\frac12},\label{eqn:dist}
\end{equation}
which defines a Hilbert-Schmidt-like distance \cite{dodonov_hilbert-schmidt_2000} of the Wigner function to the 
classical distribution, 
normalized to values between zero and unity. 
The smaller the value of $d_\text{cl}$, the closer is the Wigner 
function to the classical distribution.  
Due to the integration over the whole phase space, $d_\text{cl}$ constitutes a global measure in contrast to the local 
modulations that we observe in Fig.~\ref{fig:difference_time}.

The introduction of $d_\text{cl}$ gives us a convenient tool to study the influence of the different parameters on the 
similarity of quantum and classical evolution. 
In Fig.~\ref{fig:distance}\ref{plt:dist-te} we plot the time evolution $d_\text{cl}=d_\text{cl}(\tau)$ for different 
values of $\alpha$ and $\Delta\wp$ used in Fig.~\ref{fig:q_cl_comparison}.
We observe that the distance $d_\text{cl}$ increases rapidly for larger values of $\hbar k/\Delta p$, which matches the results 
from the perturbative treatment.
Moreover, we notice that for small values of $\hbar k/\Delta p$, the distance is very small for short times, see the 
curves corresponding to $\Delta\wp=2$.
This behavior fulfills our expectations of a nearly classical dynamics. However, after some time the quantum features 
in the Wigner function become more prominent and the distance suddenly increases before it saturates. The
saturation value is still about an order of magnitude below the possible maximum.

In  Fig.~\ref{fig:distance}\ref{plt:dcl_pi2} and \ref{plt:dcl_pi} we plot $d_\text{cl}$ as a function of $\alpha$ and 
$\Delta\wp$ for two different times $\tau$. In both panels we observe the behavior that small values of $\alpha$ and 
$\Delta\wp$, see the lower left corner, lead to a large distance while it decreases when these parameters are 
increased, corresponding to a more classical evolution in phase space. The contour lines approximately correspond to 
hyperbolae of constant $\hbar k/\Delta p$.

At the earlier time $\tau=\pi/2$, shown in Fig.~\ref{fig:distance}\ref{plt:dcl_pi2}, the distance quickly falls off when
approaching the upper right corner. For the later time $\tau=\pi$ depicted in Fig.~\ref{fig:distance}\ref{plt:dcl_pi} 
the distance from a classical evolution declines much slower, since the Wigner function and the classical distribution 
function differ after some time, as seen in Fig.~\ref{fig:distance}\ref{plt:dist-te}.

\section{FEL gain}
\label{sec:gain}

So far, we have only discussed the motion of the electron. However, in an experiment we are mainly interested in the 
dynamics of the laser field, which is why we investigate in the following quantum effects of the FEL gain.

\subsection{Small-signal limit}
\label{ssec:gain_small-signal}

Inserting the expression from Eq.~\eqref{eq:app_w1_zwischen} for $\mathcal W^{(1)}$ into the equation of motion 
\eqref{eq:edot} for the laser field, and averaging over the phase $\theta$  yields a linear differential equation 
for $\varepsilon$ which can be straightforwardly solved. 
This solution reads~\footnote{We note that the linearization of the exponential is allowed since we are in the low-gain regime, where the relative change of $\varepsilon$  during one passage of electrons
is small, that is $G\ll 1$.}
   \begin{equation}\label{eq:lin}
     \varepsilon(\tau)=\exp{\left[G(\tau)\right]}\cong1+G(\tau)\,,
   \end{equation}
where we have introduced the gain
   \begin{equation}\label{eq:gain_int}
      G(\tau)\equiv -\frac{\chi\tau^ 2}{4}\!\sum\limits_{m=0}^\infty\!\frac{(4\alpha)^ {-m}}{(2m+1)!}\!\int\!\!\text{d}\wp\,\,\text{sinc}^2\!\!\left(\frac{\wp\tau}{2}\right)\frac{\partial^{2m+1}\rho(\wp)}{\partial \wp^ {2m+1}}
   \end{equation}
of the laser field in the small-signal limit.

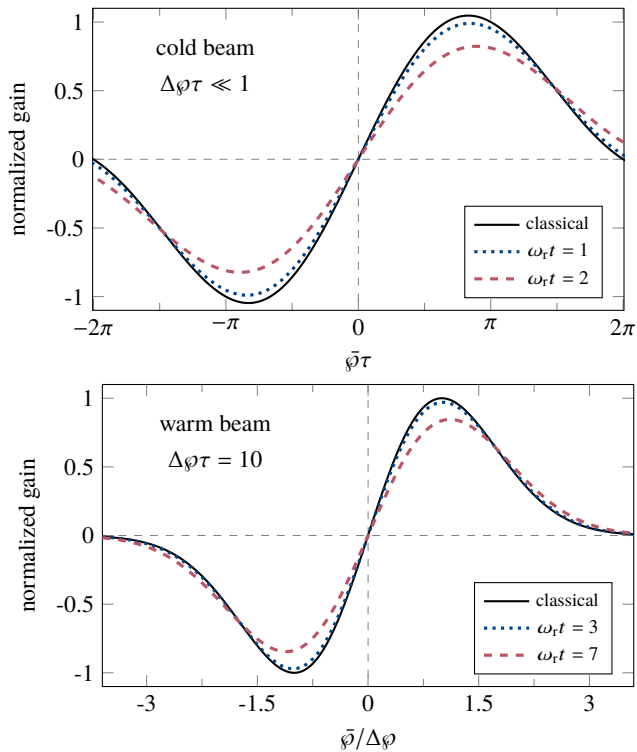
\begin{figure}
   \begin{tikzpicture}
     \begin{axis}
      [height=5.6cm,
       xmin=-6.3,
       xmax=6.3,
       ymin=-1.1,
       ymax=1.1,
       xlabel={$\bar{\wp}\tau$},
       ylabel={normalized gain},
       xtick={-6.28,-3.14,0,3.14,6.28},
       xticklabels={$-2\pi$,$-\pi$,$0$,$\pi$,$2\pi$}, 
       ytick={-1,-0.5,0,0.5,1},
       yticklabels={-1,-0.5,0,0.5,1}]
      
      \draw[help lines,dashed] (current axis.above origin) -- (current axis.below origin);
      \draw[help lines,dashed] (current axis.left of origin) -- (current axis.right of origin);

     \addplot[black, thick] table{Daten/Cold0.dat};
     \addlegendentry{classical};
\addplot[TolHCBlue,dotted] table{Daten/Cold1.dat};
     \addlegendentry{$\omega_\text{r}t=1$};
\addplot[TolHCRed,dashed] table{Daten/Cold2.dat};
     \addlegendentry{$\omega_\text{r}t=2$};
     \end{axis}
     
     \node at (1.5,3.5) {cold beam};
     \node at (1.5,3) {$\Delta \wp\tau \ll 1$};
     
   \end{tikzpicture}
    \begin{tikzpicture}
     \begin{axis}
      [height=5.6cm,
       xmin=-3.6,
       xmax=3.6,
       ymin=-1.1,
       ymax=1.1,
       xlabel={$\bar{\wp}/\Delta \wp$},
       ylabel={normalized gain},
       xtick={-3,-1.5,0,1.5,3},
       xticklabels={-3,-1.5,0,1.5,3}, 
       ytick={-1,-0.5,0,0.5,1},
       yticklabels={-1,-0.5,0,0.5,1}]
      
      \draw[help lines,dashed] (current axis.above origin) -- (current axis.below origin);
      \draw[help lines,dashed] (current axis.left of origin) -- (current axis.right of origin); 
      
     \addplot[black,thick] table{Daten/Warm0.dat};
     \addlegendentry{classical};
\addplot[TolHCBlue,dotted] table{Daten/Warm3.dat};
     \addlegendentry{$\omega_\text{r}t=3$};
\addplot[TolHCRed,dashed] table{Daten/Warm7.dat};
     \addlegendentry{$\omega_\text{r}t=7$};
     \end{axis}
     
     \node at (1.5,3.5) {warm beam};
     \node at (1.5,3) {$\Delta \wp\tau =10$};
     
   \end{tikzpicture}
  \caption{Quantum corrections to the FEL gain: we have drawn the small-signal gain for a cold (above), Eq.~\eqref{eq:gain_cold}, and a warm electron beam (below), Eq.~\eqref{eq:gain_warm}, as functions of 
  $\bar{\wp}\tau=2k\bar{p}t/m$ and $\bar{\wp}/\Delta \wp=\bar{p}/\Delta p$, respectively. In the cold-beam case, 
  $\Delta \wp \tau \ll 1$, we compare the classical gain (black line) to gain curves including the lowest-order quantum 
  corrections
  ($m=1$ in Eq.~\eqref{eq:gain_cold}) for the values $\omega_\text{r}t=1$ (blue, dotted line) and $\omega_\text{r}t=2$ 
  (red, dashed line), respectively, of the recoil parameter $\omega_\text{r}t$.
  We observe that for increasing values of $\omega_\text{r}t$ the positions of  minimum and maximum are slightly shifted to the left and to the right, respectively, while their magnitude is decreased. The behavior in the warm beam case with
  $\Delta \wp \tau\gg 1$, is qualitatively similar. However, for our -- already moderate -- choice of $\Delta \wp \tau= 10$ visible quantum effects (again in lowest order, that is $m=1$ in Eq.~\eqref{eq:gain_warm}) 
  occur not until $\omega_\text{r}t=3$ (blue, dotted line) and are still quite small for $\omega_\text{r}t=7$ (red, 
  dashed line).
  Hence, quantum corrections in the warm beam case are small in comparison to a cold beam in accordance to Eq.~\eqref{eq:connect}. We conclude that a large momentum spread suppresses quantum effects in the FEL gain. 
  } 
  \label{fig:gain_small_signal}
  \end{figure}

To derive analytical expressions we restrict ourselves to the extreme cases of a cold and a warm electron beam~\cite{fedorov_book} defined by the limits $\Delta\wp\tau\ll 1$ and
$\Delta\wp\tau\gg 1$, respectively. Making the approximation
$\rho(\wp)\cong \updelta(\wp-\bar{\wp})$
leads us from Eq.~\eqref{eq:gain_int} to the FEL gain
    \begin{equation}\label{eq:gain_cold}
      G_\text{cold}(\tau)\cong-\frac{\chi\tau^3}{4}\!\sum\limits_{m=0}^\infty\!\frac{(\omega_\text{r}t)^ {2m}}{(2m+1)!}\left.\frac{\partial^{2m+1}}{\partial x^{2m+1}}\text{sinc}^2\!\left(\frac{x}{2}\right)\right|_{x=\bar{\wp}\tau}
    \end{equation}
for a cold beam, while the assumption $\text{sinc}^2(\wp\tau/2)\cong 2\pi\updelta(\wp)/\tau$
yields the corresponding quantity  
    \begin{equation}\label{eq:gain_warm}
      \begin{aligned}
         G_\text{warm}(\tau)\cong\frac{\pi\chi\tau\rho(0)}{2\sqrt{2}\Delta\wp}\!\sum\limits_{m=0}^\infty\!\frac{\left(\frac{\hbar k}{\sqrt{2}\Delta p}\right)^{2m}}{(2m+1)!} H_{2m+1}\left(\frac{\bar{\wp}}{\sqrt{2}\Delta\wp}\right)
       \end{aligned}
    \end{equation}
in the warm-beam limit. 

In contrast to the Wigner function and the warm-beam case, the quantum corrections for a cold beam do not scale with  powers of $\hbar k/\Delta p$ but instead  
with even powers of the recoil parameter, defined as $\omega_\text{r}t=\tau/\sqrt{4\alpha}$.
We note that the relation
\begin{equation}\label{eq:connect}
 \frac{\hbar k}{\Delta p}=\frac{1}{\Delta \wp\tau}\omega_\text{r}t
\end{equation}
connects the three relevant parameters.

For a cold beam with $\Delta \wp \tau\ll 1$ the recoil parameter $\omega_\text{r}t$ is always smaller than the ratio of recoil and momentum spread, that is 
$\omega_\text{r}t\ll \hbar k /\Delta p$. Hence, quantum effects apparent in the Wigner function are washed out due to the averaging for the FEL gain.
In the opposing limit of a warm beam, Eq.~\eqref{eq:connect} predicts that $\hbar k/\Delta p$ in turn  is smaller 
than the recoil parameter $\omega_\text{r}t $ dominating the cold-beam limit, that is   $\hbar k/ \Delta p \ll 
\omega_\text{r}t$ due to $\Delta \wp\tau \gg 1$.

In Fig.~\ref{fig:gain_small_signal} we compare the classical gain function to the one including quantum corrections for a cold and a warm electron beam, respectively. 
We observe the same qualitative behavior in both cases, that is a slight shift of the positions for maximum and minimum 
gain and a decrease of its magnitude for increasing values of the recoil parameter
$\omega_\text{r}t$. However, while moderate quantum corrections in the cold-beam case emerge already for 
$\omega_\text{r}t=1$, it is not until $\omega_\text{r}t=3$ that they appear for a warm beam and they remain small even 
for $\omega_\text{r}t=7$. Hence, quantum effects are suppressed for a warm beam in comparison to a cold beam in 
accordance with Eq.~\eqref{eq:connect}.

We conclude that the averaging over all momenta as well as a large momentum spread prevents the appearance of quantum effects in the FEL gain, at least in the small-signal limit.

The expression in Eq.~\eqref{eq:gain_cold} is not a new result. It is in principle the same formula derived by Madey 
and many others~\cite{madey1,scully_fel,colson,becker79,*becker88,colson_quantum,friedman,banacloche}.
For $m=0$ we indeed recover from Eq.~\eqref{eq:gain_cold} the classical result---Madey's famous gain formula. In 
contrast to these earlier works, our approach is not restricted to single-photon processes but instead generalizes the 
procedure~\cite{scully_fel} for the classical Boltzmann equation~\eqref{eq:boltzmann}.    
  
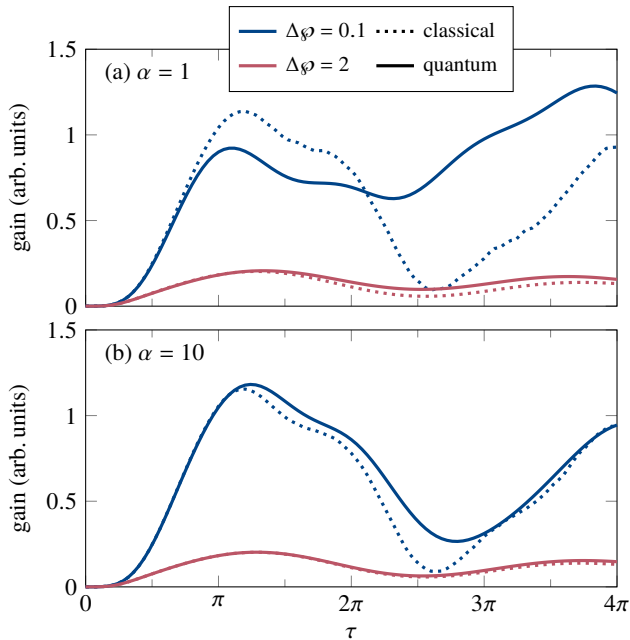
\begin{figure}[t]
	\begin{tikzpicture}

\begin{groupplot}[group style={
	group size=1 by 2,
xticklabels at=edge bottom,
	yticklabels at=edge left,
	horizontal sep=1em,
	vertical sep=1em,
},
width=\linewidth,
	xmin =0,
	xmax = 4,
	ymin =0,
	ymax = 1.5,
	xtick = { 0, 1, 2, 3, 4},
	xticklabels = {$0$,$\pi$, $2\pi$, $3\pi$, $4\pi$ },
	ylabel style={align=center},
	clip=false,
		ylabel={gain (arb.\,units)},
]

\nextgroupplot[]
\node[matrix anchor=south, matrix,anchor=base west, font=\footnotesize, draw, fill=white]  at (2.15,1.25){
	\draw[TolHCBlue, very thick] (0,.5ex) -- ++(5mm,0);\pgfmatrixnextcell \node{$\Delta\wp=0.1$};\pgfmatrixnextcell
	\draw[dotted, very thick] (0,.5ex) -- ++(5mm,0);\pgfmatrixnextcell \node{classical};\pgfmatrixnextcell\\
	\draw[TolHCRed, very thick] (0,.5ex) -- ++(5mm,0);\pgfmatrixnextcell \node{$\Delta\wp=2$};\pgfmatrixnextcell
	\draw[very thick] (0,.5ex) -- ++(5mm,0);\pgfmatrixnextcell \node{quantum};\pgfmatrixnextcell\\
};
\addplot+ [TolHCBlue, dotted]
  table[]{Daten/gain-evolution-5.tsv};

\addplot+ [TolHCRed, dotted]
  table[]{Daten/gain-evolution-6.tsv};

\addplot+ [TolHCBlue]
  table[]{Daten/gain-evolution-7.tsv};

\addplot+ [TolHCRed]
  table[]{Daten/gain-evolution-8.tsv};
\SubLabel[$\alpha=1$]{plt:gain_te_a1}

\nextgroupplot[
	xlabel = {$\tau$},
]
\addplot+ [TolHCBlue, dotted]
  table[]{Daten/gain-evolution-9.tsv};

\addplot+[TolHCRed, dotted]
  table[]{Daten/gain-evolution-10.tsv};

\addplot+[TolHCBlue]
table[]{Daten/gain-evolution-11.tsv};

\addplot+[TolHCRed]
table[]{Daten/gain-evolution-12.tsv};
\SubLabel[$\alpha=10$]{plt:gain_te_a10}
\end{groupplot}

\end{tikzpicture} 	\caption{Comparison of the FEL gain $G$ as a function of the dimensionless time $\tau$ 
	computed with initial mean momentum $\bar\wp=1.6$ and for different values of the quantum parameter $\alpha$ 
	as well as for narrow ($\Delta\wp=0.1$) and  broad 	($\Delta\wp=2$) initial momentum distributions.
	The solid	lines correspond to the result obtained from the Wigner function, while the dotted ones 
	are calculated from the classical distribution function with the same initial state. 
The latter curves feature oscillations  around a saturation value with a frequency 
	based on the rotation period of	the bounded trajectories and are independent of $\alpha$. 
	For $\alpha=1$, as shown in \ref{plt:gain_te_a1}, the results from both theories are similar for short 
	times, especially for the wide momentum spread with $\hbar k/\Delta p=0.25$, in contrast the narrow momentum 
	spread with $\hbar k/\Delta p=5.00$.
	For $\alpha=10$, as shown in \ref{plt:gain_te_a10}, the quantum and classical behavior of the gain  is 
	qualitatively very 	similar and agrees very well at least until the first maximum is reached.
	Note the large value of $\hbar k/\Delta p=1.58$ for $\Delta\wp=0.1$ in contrast to $\hbar k/\Delta p=0.08$ for 
	$\Delta\wp=2$.
	The good agreement comes from the fact that even substantial quantum features of the Wigner function, see 
	for instance Fig.~\ref{fig:q_cl_comparison}\ref{plt:qcl_comp_s01a10}, average out in the gain, since it is obtained 
	from the change of mean momentum. Further, we observe that the point in time where the curves diverge appears 
	later, when $\alpha$ is increased. 
}
	\label{fig:gain-time-evolution}
\end{figure}

\subsection{Longer times}
\label{ssec:gain_longer_times}

It is possible to relate the gain of the laser field to the change of the mean momentum of the electrons.
Inserting the Quantum Liouville equation \eqref{eq:qLiou}  into the time derivative of the  expectation value of the 
electron momentum and making use of Eq.~\eqref{eq:edot} we find
\begin{equation}
\dv{\ev{\hat \wp}}{\tau}=
\int\!\dd{\theta}\int\! \dd{\wp}\wp \pdv{\tau} \mathcal W(\theta,\wp;\tau)=-\frac{1}{\chi}\dv{\varepsilon(\tau)}{\tau},
\end{equation} 
after assuming $\varepsilon(\tau)\cong 1$. Identifying the gain as the relative change of the field in accordance with 
Eq.~\eqref{eq:lin}, we arrive at the expression
\begin{align}\label{eqn:gain_p_diff}
G(\tau)&=-\chi\ev{\hat\wp(\tau)-\hat\wp(0)}.
\end{align} 
Having already calculated the Wigner function, this expression for the gain can easily 
be evaluated. The classical gain can also be inferred from Eq.~\eqref{eqn:gain_p_diff} with the help of the classical 
distribution function $f_\text{cl}$.

In Fig.~\ref{fig:gain-time-evolution} we plot the time evolution of the FEL gain, obtained from 
Eq.~\eqref{eqn:gain_p_diff} and the numerical solution of the Wigner function. 
Because the gain vanishes for $\bar\wp=0$, we chose a nonzero initial mean momentum ($\bar\wp =1.6$ in the 
plots). 

For longer times, the gain does not increase monotonically, but exhibits oscillations in a saturation 
regime. The classical time evolution of the phase-space distribution yields a rotation inside the separatrix, leading 
to an oscillating change of the mean momentum and consequently also to the oscillations of the gain.
This behavior was for example discussed in Ref.~\cite{fedorov_rev,*fedorov_book}.

In Fig.~\ref{fig:gain-time-evolution} we do not present the case $\alpha \ll1$, since here the gain shows a behavior 
completely different 
from the classical 
description~\footnote{The reason for this is the different underlying process, namely a Rabi oscillation 
\cite{NJP2015} between the involved momentum levels with a different time-scale and not the rotation inside the 
separatrix.}.
With increasing $\alpha$ the frequency of oscillation approaches the classical one such that the quantitative 
differences between quantum and classical evolution become less prominent and only appear at relatively large times, 
such that both theories lead to a similar behavior of the gain.
This convergence becomes apparent in Fig.~\ref{fig:gain-time-evolution}\ref{plt:gain_te_a1} and \ref{plt:gain_te_a10}.
A larger $\Delta\wp$ further reduces the difference between the two gain curves.

In contrast to the phase-space distance $d_\text{cl}$, which increases rapidly in the beginning for an initially small 
momentum spread, the classical and quantum gain resemble each other very well for short times. We have seen this 
behavior already in the section above, where the quantum corrections to the gain scaled asymptotically with $\omega_\text{r}t=\tau/\sqrt{4\alpha}$.
The good agreement between quantum and classical theory arises from
the fact that the gain depends on the change of mean momentum, that is a quantity where 
the differences between classical and Wigner distribution are averaged out, even if they are as 
substantial as in Fig.~\ref{fig:q_cl_comparison}\ref{plt:qcl_comp_s01a10}, while the distance takes all 
deviations into account.
Further, we observe that the quantum corrections become smaller for decreasing values of   
$1/(2\sqrt{\alpha}\Delta\wp)=\hbar k/\Delta p$.

\section{Conclusions}
  \label{sec:Conclusions}
  
By comparing the time evolution of the Wigner function and of the classical phase-space distribution  in the low-gain 
regime, we have analyzed the effects of the initial momentum width $\Delta\wp$ and the quantum parameter $\alpha$. The 
asymptotic treatment valid
for the small-signal regime, that is short times, allowed us to identify the critical quantity $\hbar k/\Delta 
p=1/(2\sqrt\alpha\Delta\wp)$ governing the transition from quantum to classical. The numerical solution for longer 
interaction times confirmed that neither the variation of $\Delta\wp$ nor $\alpha$ by themselves are sufficient to 
obtain  matching dynamics of the Wigner function and the classical distribution function, but only the combination of 
both. Furthermore, we observe always that, after sufficient time, both distributions differ due to nonlinear effects 
imposed by the anharmonic potential.

For the FEL gain, the quantum effects are much less prominent and hence hard to observe in comparison to the Wigner 
function. At the highly relativistic electron energies of an FEL we cannot determine the Wigner function sufficiently 
precisely from experiment. However, our theory is not limited to this specific energy range and can also be applied to 
electrons at lower energies. Other experimental situations, for example Kapitza-Dirac \cite{Efremov1999}, provide a 
more favorable basis for the obvervation of the quantum to classical transition, since here the momentum level 
structure and the oscillations between the levels are experimentally resolvable 
\cite{feist_quantum_2015,kozak_inelastic_2018,*kozak_ponderomotive_2018,vanacore_attosecond_2018}. 
There is even the possibility to reconstruct the Wigner function from measurement data 
\cite{priebe_attosecond_2017}.

To extend our model to the high-gain regime  \cite{pellegrini_physics_2016}, a many-electron 
description including a varying laser field becomes necessary. However, the quantum limit of the high-gain regime 
also reduces to a system of two momenta for each electron, such that the process is dominated by single-photon 
transitions. The classical regime, for low- as well as for high-gain FELs, differs from the quantum limit by the 
emission of multiple photons per electron.
In the high-gain regime many electrons simultaneously interact with the light fields, and thus they may become 
entangled \cite{kling_high-gain_2019}. However, if we neglect these quantum correlations between the individual 
electrons, the high-gain regime mainly differs through an rapidly increasing laser field from the low-gain limit. 
Hence, we expect that the parameter $\hbar k / \Delta p$ plays a similar role in both regimes.

\begin{acknowledgments}
	We thank P. Anisimov,  A. Debus,  M. Efremov, M. Freyberger,  A. Gover, Y. Pan, and  K. Steiniger
	for many fruitful discussions.  
	W.\,P.\,S. is grateful to Texas A\&M University for a Faculty Fellowship at the 
	Hagler Institute of Advanced Study at Texas A\&M University, and Texas A\&M AgriLife 
	Research for the support of his work. The Research of $\mathrm{IQ}^{\mathrm{ST}}$ is 
	financially supported by the Ministry of Science, Research and Arts Baden-Württemberg.
\end{acknowledgments}

 \begin{appendix}
   \section{Derivation of model}
  \label{sec:Derivation_of_model}
In this appendix we derive our model for the FEL dynamics in terms of the Wigner function.
While we describe the motion of the electron quantum mechanically, the laser field is treated as a classical quantity 
in analogy to semiclassical laser theory~\cite{scullylamb}.

   \subsection{Pendulum Hamiltonian}  
     \label{sec:Quantum_mechanical_description_for_electron}
  
The one-dimensional motion of an electron with mass $m$ and charge $e$ in the FEL is dictated by the Hamiltonian~\cite{meystre}
    \begin{equation}\label{eq:H_ALAW}
        \hat{H}=\frac{\hat{p}^2}{2m}+\frac{e^2}{m}A_\L(\hat{z},t)A_\W(\hat{z},t)
    \end{equation}    
with $\hat{z}$ and  $\hat{p}$ being the position and the momentum operators along the wiggler axis.

We assume that the vector potentials $\boldsymbol{A}_\L=A_\L \boldsymbol{e}_x$ and $\boldsymbol{A}_\W=A_\W\boldsymbol{e}_x$ of the laser and the wiggler field, respectively, are linearly polarized perpendicular to the longitudinal axis. Moreover, we model their amplitudes as plane waves~\cite{schmueser}     
      \begin{align}
        A_\L(z,t)&=\frac{E_\L(t)}{\omega}\sin{(\omega t +\phi_\L(t)-kz)}, \ \ \ \text{and}  \label{eq:app_AL}\\
        A_\W(z,t)&=\frac{B_0}{k}\cos{(\omega t +kz)}\,,\label{eq:app_AW}
      \end{align}   
where $\omega$ and $k=\omega/c$ denote the frequency and wave number, respectively, of the two counterpropagating fields, while $c$ gives 
the velocity of light. We consider the reference frame~\cite{bambi,*brs}  where the frequencies of the two modes coincide and thus the 
electron interacts with a standing light field. For resonant interaction the electron motion in this frame is always 
nonrelativistic~\cite{NJP2015} justifying the Hamiltonian in Eq.~\eqref{eq:H_ALAW}.  
  
The amplitude and phase corresponding to the electric field of the laser mode, $E_\L=E_\L(t)$ and $\phi_\L=\phi_\L(t)$, respectively, 
are slowly varying with time. In contrast, we assume that the corresponding quantities for the strong magnetic field of the wiggler, 
$B_0$ and $\phi_\W\equiv 0$, are constant.

By inserting the fields from Eqs.~\eqref{eq:app_AL} and~\eqref{eq:app_AW} into Eq.~\eqref{eq:H_ALAW} and neglecting 
rapid oscillations with $2\omega$ in a rotating-wave-like approximation~\cite{fedorov_book}, we arrive at the pendulum 
Hamiltonian~\cite{colson,brs}
    \begin{equation}\label{eq:appA_H_fin}
      \hat{H}=\frac{\hat{p}^2}{2m}+U_0\,\varepsilon(t)\sin\left[2k\hat{z}-\phi_\L(t)
      \right]
    \end{equation}    
for the electron dynamics in the FEL. Here we have introduced the amplitude   
    \begin{equation}\label{eq:appA_U0}
    U_0\equiv \frac{e^2E_0B_0}{mck^2}
    \end{equation}         
 of the potential as well as the dimensionless electric field $\varepsilon(t)\equiv E_\L(t)/E_0$  normalized with respect to the initial 
 field amplitude $E_0\equiv E_\L(0)$ before the electrons enter the wiggler. In the low-gain regime of FEL operation the value of 
 $\varepsilon$ is always close to unity.

 \subsection{Laser field couples to electron current}

In contrast to the quantized electron, we describe the laser field as a classical quantity that evolves according to Maxwell's equations. Hereby we strongly follow the lines of semiclassical laser theory~\cite{scullylamb}. 
The classical wave equation~\cite{boni90}
\begin{equation}\label{eq:maxwell}
\left(\frac{\partial^2}{\partial z^2}-\frac{1}{c^2}\frac{\partial^2}{\partial t^2}\right)A_\L(z,t)=
\mu_0 \,j_\text{el}(z,t)
\end{equation}    
couples $A_\L$ to the $x$-component  $j_\text{el}$ of the electron current with $\mu_0$ denoting the vacuum permeability.

Inserting Eq.~\eqref{eq:app_AL} into Eq.~\eqref{eq:maxwell}, projecting on the laser mode,
performing  the slowly-varying phase and amplitude approximation~\cite{scullylamb,meystre},
and taking the imaginary part leads to the equation of motion
       \begin{equation}\label{eq:edot1}
            \frac{\D}{\D t}E_\L(t)=-\frac{1}{2\varepsilon_0}\,
            \text{Re}{\mathcal{J}_\L(t)}
       \end{equation}             
for the laser amplitude that includes the vacuum permittivity $\varepsilon_0=1/(c^2\mu_0)$. We observe  that only the Fourier component
       \begin{equation}\label{eq:IL}
        \mathcal{J}_\L(t)=\e{\I[\omega t +\phi_\L(t)]}\frac{2}{L_z}\int\limits_0^{L_z}\!\!\text{d}z\,
        j_\text{el}(z,t)\e{-\I k z}
      \end{equation} 
of the current that corresponds to the laser mode appears in the equation of motion, with $L_z$ denoting the longitudinal extend of the 
quantization volume. In the course of the slowly-varying phase and amplitude approximation we have neglected terms with $\ddot{E}_\L$, 
$\ddot{\phi}_\L$, $\dot{\phi}_\L \dot{E}_\L$, $\dot{\phi}_\L^2$, $\dot{\phi}_\L \mathcal{J}_\L$, and $\dot{\mathcal{J}}_\L$ in  
accordance with Ref.~\cite{scullylamb}.

Since the gradient of the electron wave function $\psi=\psi(z,t)$ in $x$-direction vanishes we deduce that   
      \begin{equation}\label{eq:jel_AW}
        j_\text{el}(z,t)\cong -\frac{e^2N }{m L_x L_y} A_\W(z,t)|\psi(z,t)|^2
      \end{equation}        
is the $x$-component of a current that satisfies a continuity equation following from the Schr\"odinger equation of a charged particle in the electromagnetic field. 
Here $N$ gives the number of electrons in the bunch while $L_x$ and $L_y$ describe normalization lengths. We note that  we have neglected a contribution with $A_\L \ll A_\W$.

By inserting Eq.~\eqref{eq:jel_AW} into Eq.~\eqref{eq:IL} and neglecting a rapidly oscillating term,  we finally arrive 
from Eq.~\eqref{eq:edot1} at the equation of motion     
      \begin{equation}\label{eq:edot_cos}
         \frac{\D}{\D t}E_\L(t)=
         \frac{e^2B_0 n_\text{el}}{2 \varepsilon_0 m k}
         \left\langle{\cos{\left[2k\hat{z}-\phi_\L(t)\right]}}\right\rangle
      \end{equation}
for the field amplitude coupling to the expectation value of $\cos{2k\hat{z}}$.           
Here we have  introduced the initial density  $n_\text{el}\equiv N/V$ of the electrons after identifying the quantization volume  $V\equiv L_x L_y L_z$ with the volume of the electron bunch.

\subsection{Formulation with Wigner function}

With the pendulum Hamiltonian for the dynamics of the electron and the equation of motion
for the laser amplitude we have obtained the two important relations of our semiclassical theory for the FEL.
In order to illuminate the transition to a purely classical theory~\cite{scully_fel}, we introduce the Wigner 
representation~\cite{schleich}
   \begin{equation}\label{eq:app_W_def}
      W(z,p;t)\equiv \frac{1}{2\pi \hbar}\int\!\!\text{d}\zeta\, \bra{p-\zeta/2}\hat{\rho}(t)\ket{p+\zeta/2}\e{\I \zeta z/
      \hbar}
   \end{equation}
for the density operator $\hat{\rho}$ of an electron. This function depends on the position $z$ and its conjugate momentum $p$ which  together form the Wigner phase space.

 It is convenient to use the dimensionless variables       
   \begin{equation}\label{eq:dimless}
    \begin{aligned}
       \theta &\equiv 2kz -\phi_\L(t)-\frac{\pi}{2},\\
       \wp &\equiv \frac{p}{\sqrt{U_0 m}}, \ \ \ \ \ \  \text{and}\\
       \tau &\equiv 2k \sqrt{\frac{U_0}{m}}\,t \\
\end{aligned} 
   \end{equation}
which correspond to position, momentum, and time, respectively. 
Accordingly, we  write the Wigner function $\varpi=\varpi(\theta,\wp;\tau)$ in its dimensionless form, which, for the correct normalization, includes
the factor $\sqrt{U_0m}/(2k)$ in comparison to $W$, that is $\varpi=\sqrt{U_0m}\,W/(2k)$.
   
With the help of the von Neumann equation
\begin{equation}
\I\hbar\frac{\D}{\D t}\hat{\rho}(t)=\left[\hat{H},\hat{\rho}(t)\right]
\end{equation}
and the Hamiltonian from Eq.~\eqref{eq:appA_H_fin} we derive the equation of motion~\cite{pio07_wigner}
   \begin{equation}\label{eq:qloiu_diff}
     \begin{aligned}
       \left(\frac{\partial}{\partial\tau}+\right.&\left.\wp \frac{\partial}{\partial \theta}\right)\varpi(\theta,\wp;
       \tau)=-\varepsilon(\tau)\sqrt{\alpha}\sin\theta \\
       &\times \left[\varpi\left(\theta,\wp+\frac{1}{2\sqrt{\alpha}};
       \tau\right)-\varpi\left(\theta,\wp-\frac{1}{2\sqrt{\alpha}};\tau\right)\right]
     \end{aligned}  
   \end{equation}
for the Wigner function also known as Quantum Liouville equation~\cite{schleich}. While the left-hand side of this equation corresponds to the free motion of the electron, the right-hand side emerges due to the periodic potential. 

In the Wigner description of quantum mechanics the expectation value of an observable is formed by weighting the Weyl 
representation of the corresponding operator with the Wigner function and integrating over phase 
space~\cite{schleich,case08}.
For the expectation value in Eq.~\eqref{eq:edot_cos} this procedure yields the relation 
   \begin{equation}\label{eq:app_edot_wigner}
      \frac{\D}{\D \tau}\varepsilon(\tau)=-\chi \int\!\!\text{d}\theta
      \int\!\!\text{d}\wp \, \mathcal W(\theta,\wp;\tau)\sin\theta
   \end{equation}
for the dynamics of the dimensionless field amplitude $\varepsilon$ with 
   \begin{equation}\label{eq:chi}
       \chi \equiv \frac{n_\text{el}}{4k \varepsilon_0}\frac{\sqrt{cB_0}}{E_0^{3/2}}
   \end{equation}
characterizing the coupling between field and electrons.

    \section{Perturbation theory}
  \label{sec:Perturbation_theory}
  
This appendix is devoted to the  solution of the Quantum Liouville equation by means of perturbation theory~\cite{carruthers83} which is valid in the small-signal limit of FEL operation.
  
\subsection{Structure of equation}
  
 We first cast Eq.~\eqref{eq:qloiu_diff}  into  the form 
    \begin{equation}\label{eq:appA_qLiou_skal}
     \mathcal{L}^{(0)}\mathcal W=\mathcal{L}^{(1)} \mathcal W\,, 
    \end{equation}    
 where we have defined the operators
    \begin{align}
     \mathcal{L}^{(0)} &\equiv \frac{\partial}{\partial \tau} +\wp\frac{\partial}{\partial \theta} \ \ \ \ \text{and}\\
      \mathcal{L}^{(1)} &\equiv -\varepsilon(\tau)\sin\theta 
       \sum\limits_ {m=0}^\infty \frac{1}{(2m+1)!}\frac{1}{(4\alpha)^m}
       \frac{\partial^{2m+1}}{\partial\wp^{2m+1}}
    \end{align}
for the free motion and the potential, respectively. By performing a Taylor expansion of the shifted Wigner functions in Eq.~\eqref{eq:qloiu_diff} in powers of $1/(2\sqrt{\alpha})$  we identify $\mathcal{L}^{(1)}$ as differential operator containing infinitely many derivatives  with respect to momentum.

  We interpret the right-hand side of Eq.~\eqref{eq:appA_qLiou_skal} as a perturbation to the free motion and  make the asymptotic expansion  
    \begin{equation}
       \mathcal{W} \cong \mathcal W^{(0)}+ \mathcal W^{(1)} + \mathcal W^{(2)}+...
    \end{equation}
  in analogy to Ref.~\cite{carruthers83}.   
By inserting this expansion into Eq.~\eqref{eq:appA_qLiou_skal}  we obtain the equations  
    \begin{align}
        \mathcal{L}^{(0)}\mathcal W^{(0)}&=0\, , \ \ \ \text{and} \label{eq:pert_w0}\\
        \mathcal{L}^{(0)}\mathcal W^{(n)}&=\mathcal{L}^{(1)}\mathcal W^{(n-1)}\label{eq:pert_wn}
    \end{align}   
 for zeroth order and  for higher orders ($n>0$), respectively, which we solve iteratively.

\subsection{Zeroth order}
   
The equation in lowest order  corresponds to a free particle and is solved by
     \begin{equation}\label{eq:app_w0}
        \mathcal W^{(0)}(\theta,\wp;\tau)=\mathcal W(\theta-\wp\tau,\wp;0)=\mathcal W(\theta,\wp;0)\,.
     \end{equation}          
In our case, the lowest-order contribution of the asymptotic expansion  equals the initial distribution because it is independent of $\theta$ according to Eq.~\eqref{eq:init}.   

\subsection{First order}
    
We now turn to the first-order calculations. The formal solution of Eq.~\eqref{eq:pert_wn} for $n=1$ reads
      \begin{equation}\label{eq:app_formal}
        \begin{aligned}
          \mathcal W^{(1)}(\theta,\wp;\tau)=\!\!\int\limits_0^\tau \!\!\D\tau'\!\!\int\!\!\D\theta'\,
           G(\theta,\tau|\theta',\tau')\mathcal{L}^{(1)}\mathcal W^{(0)}(\theta',\wp;\tau'),
        \end{aligned}  
      \end{equation}  
     where we have recalled from  Ref.~\cite{carruthers83} the Green's function
      \begin{equation}
        G(\theta,\tau|\theta',\tau')\equiv \updelta\left[\theta-\theta'-\wp (\tau-\tau')\right]
      \end{equation}
     for the unperturbed Quantum Liouville equation.

Inserting the lowest-order solution $\mathcal{W}^{(0)}$ from Eq.~\eqref{eq:app_w0} into Eq.~\eqref{eq:app_formal} and 
evaluating the operators and integrals yields the result
      \begin{equation}\label{eq:app_w1_zwischen}
       \begin{aligned}
         \mathcal W^{(1)}(\theta,\wp;\tau)&=-\varepsilon(\tau)\,\tau\frac{\cos{(\theta-\wp\tau)}-\cos{\theta}}
         {\wp\tau}\\
         &\times
         \sum\limits_ {m=0}^\infty \frac{1}{(2m+1)!}\frac{1}{(4\alpha)^m}\frac{1}{2\pi}
         \frac{\partial^{2m+1}\rho(\wp)}{\partial\wp^{2m+1}}
       \end{aligned} 
      \end{equation}         
    for the first-order term $ \mathcal{W}^{(1)}$, which still contains infinitely many derivatives of $\rho$. We note 
    that we have treated the relative field change $\varepsilon$ as a constant in this procedure, that is 
    $\varepsilon(\tau')\cong \varepsilon(\tau)$, in accordance with our low-gain approach.

The assumption that the initial momentum distribution $\rho$ is  Gaussian with mean value 
$\bar{\wp}$ and standard deviation $\Delta \wp$ leads to  a series of Hermite polynomials of odd order. From this series we finally arrive at the closed expression 
     \begin{equation}\label{eq:app_closed}
      \begin{aligned}  
         \mathcal W^{(1)}&(\theta,\wp;\tau)=-\varepsilon(\tau)\,\tau\frac{\cos{(\theta-\wp\tau)}-\cos{\theta}}
         {\wp\tau} \\
         \times& 2\sqrt{\alpha}\e{-\frac{1}{2}\left(\frac{\hbar k}{\Delta p}\right)^2}\sinh{\left[\frac{\hbar k}{ \Delta p}\frac{\wp-\bar{\wp}}{\Delta\wp}\right]} 
         \frac{1}{2\pi}\rho(\wp)
      \end{aligned} 
     \end{equation}           
for $\mathcal{W}^{(1)}$ after using a generating function for Hermite polynomials~\cite{koekoek} as well as $\left(4\alpha 
\,\Delta\wp^2\right)^{-1/2}=\hbar k/\Delta p$.
      
  \subsection{Connection to classical theory}    
     
     We briefly compare our results for the Wigner function to classical FEL theory~\cite{scully_fel} in terms of the classical distribution function $f_\text{cl}=f_\text{cl}(\theta,\wp;\tau)$ in phase space. In the equation of motion for $f_\text{cl}$ we simply replace $\mathcal{L}^{(1)}$ in the Quantum  Liouville equation by $\mathcal{L}_\text{cl}^{(1)} \equiv -\varepsilon(\tau)\sin\theta\frac{\partial}{\partial\wp}$    
and arrive at the colissonless Boltzmann equation~\eqref{eq:boltzmann}.  
      
Consequently, the procedure for finding a perturbative solution of $f_\text{cl}$ is analogous to the approach for 
$\mathcal{W}$ in the preceding section. 
     For the same initial distribution we thus  obtain the first-order contribution~\footnote{Indeed, the small 
     parameter for the perturbative expansion of $f_\text{cl}$ (or $\mathcal{W}$ close to the classical regime) is not 
     $\tau$ but rather $\tau/\Delta \wp$.
     Hence, we require $\tau \ll \Delta \wp$ for this asymptotic expansion to converge. We emphasize that $\tau/\Delta \wp$ is a purely classical quantity, where no term with $\hbar$ enters.}  
     \begin{equation}\label{eq:app_fcl_1}
       \begin{aligned}
         f_\text{cl}^{(1)}(\theta,\wp;\tau)=-\varepsilon(\tau)\,\tau\frac{\cos{(\theta-\wp\tau)}-
         \cos{\theta}}
         {\wp\tau} 
         \frac{H_1(\xi)}{\sqrt{2}\Delta \wp} \frac{\rho(\wp)}
         {2\pi}
       \end{aligned} 
     \end{equation} 
    for the classical distribution function with $H_1$ denoting the Hermite polynomial of first order evaluated at $\xi \equiv (\wp-\bar{\wp})/\break(\sqrt{2}\Delta\wp)$. 
      
    By comparing Eq.~\eqref{eq:app_closed} to Eq.~\eqref{eq:app_fcl_1} we realize  that the Wigner function
     \begin{equation}\label{eq:app_w1_tocl}
      \mathcal{W}^{(1)}(\theta,\wp;\tau)=f_\text{cl}^{(1)}(\theta,\wp;\tau)
      \left[1+\mathcal{Q}\left(\frac{\wp-\bar{\wp}}{\sqrt{2}\Delta \wp}\right)\right]
     \end{equation}      
     and the classical distribution just differ by a momentum-dependent contribution $\mathcal{Q}$. 
     These `quantum corrections' are given by the series  
     \begin{equation}
       \mathcal{Q}(\xi)\equiv\sum\limits_{m=1}^\infty \frac{1}{(2m+1)!}\left(\frac{\hbar k}{\sqrt{2}\Delta 
       p}\right)^{2m} \frac{H_{2m+1}(\xi)}{H_1(\xi)}
     \end{equation}      
     of  Hermite polyniomials $H_{2m+1}$ of odd order that arise from the derivatives with respect 
     to $\wp$ in Eq.~\eqref{eq:app_w1_zwischen}.

\section{Exact time evolution of the Wigner function}
\label{app:time-evolution}
Since the infinitely many derivatives in the Quantum Liouville equation \eqref{eq:qLiou}
impede the solution in a closed form, we resort to solving the Schrödinger equation.
From the resulting solutions we can construct the time-evolution operator. We then use this operator to 
obtain the time evolution of the density operator from which the Wigner function is computed. 

\subsection{Schrödinger equation and Mathieu functions}
  \label{sec:Mathieu_equation}

At first, we determine the eigenfunctions and energy eigenvalues of the 
Hamiltonian 
Eq.~\eqref{eq:appA_H_fin}. They directly lead to the time evolution of an arbitrary quantum state subject to the 
periodic potential $U_0\cos\theta$.

We write the stationary Schrödinger equation in position space with the dimensionless position $\theta$ and the quantum 
parameter $\alpha$ and arrive at
\begin{equation}
-\frac{}{}\dv[2]{\theta}u(\theta) +2 \alpha \cos(\theta)
u(\theta) = \mathcal E u(\theta),  \label{eqn:Mathieu}
\end{equation}
where we have introduced the dimensionless energy $\mathcal E$ scaled in units of $\hbar\omega_\text{r}$. This equation is known as Mathieu equation \cite{meixner_mathieusche_1954, NIST:DLMF}. Due to the periodicity of the 
potential, we restrict the range of $\theta$ to the interval $[0,2\pi]$.

We use here only a subset of the solutions of Eq.~\eqref{eqn:Mathieu}, namely the bounded Mathieu 
functions $u(\alpha,\theta)\equiv \mathrm{me}_\nu(\alpha, \theta)$, since the other solutions diverge for 
$\theta \to \pm \infty$ and hence do not describe a physical quantum state.
The functions $\mathrm{me}_\nu$ are, in general, not expressible in terms of standard functions and are 
$2\pi/\nu$-pseudoperiodic. That is, they obey the relation
\begin{equation}
\mathrm{me}_\nu(\alpha,\theta +2\pi)= \mathrm e ^{\I 2\pi \nu}\mathrm{me}_\nu(\alpha,\theta).
\end{equation} 
The parameter $\nu \in \mathbb R$ is hence determined by the particular choice of periodic boundary conditions
\footnote{There is still an ambiguity in the choice of $\nu$, as $\nu'=\nu+n2\pi$, $n\in\mathbb{N}$ yields the same 
periodic boundary condition. We circumvent this by demanding $\lim_{\alpha\to 0} \mathrm{me}_{\nu}(\theta)=\mathrm e 
^{\I \nu\theta}$ for $\nu\not \in\mathbb{Z}$. This approach is different form the convention in solid state 
physics, where $\nu$ is reduced to the first Brillouin zone, i.\,e. the interval $[-\pi,\pi]$, and an additional band 
index $n$ is introduced.}.
Connected to each solution is the characteristic value $\mathcal E_\nu(\alpha)$, that is the eigenenergy of that state.
By solving the time-dependent Schrödinger equation we find
\begin{equation}
 u_\nu(\alpha, \theta, \tau) = \mathrm e ^{-\I\mathcal{E}_\nu(\alpha)\tau/(2\sqrt{\alpha})} 
 \mathrm{me}_\nu(\alpha, \theta),
 \label{eqn:mathieu_time-evolution}
\end{equation}
describing the time evolution of an energy eigenstate.

In order to attribute a physical meaning to the parameter $\nu$, we recall that the periodicity of a momentum 
eigenstate 
is given by the De Brogli wavelength $\lambda_\mathrm{DB}=2\pi\hbar/p$, which translates in the dimensionless 
coordinates to $2\pi\sqrt\alpha/\wp=2\pi/\nu$. Hence, we identify the parameter $\nu$ of the Mathieu functions as 
the initial momentum. Thus, the initial momentum by itself  determines the boundary conditions 
we have to impose.

In order to explicitly find the solutions of Eq.~\eqref{eqn:Mathieu}, we exploit the periodicity of the potential with 
the help of Bloch theory. Accordingly, the solutions can be represented in a special form of Fourier series, given by 
\begin{equation}
\mathrm{me}_\nu(\alpha,\theta) = \mathrm e^{\I \nu\theta}\sum_{r=-\infty}^{\infty}c_r^\nu(\alpha)\mathrm e^{\I r\theta},
\label{eqn:mathieu-fourier}
\end{equation}
where the $c^\nu_r(\alpha)$ denote the Fourier coefficients.
Here the series itself has the periodicity of the potential, while the pseudoperiodicity is achieved only through the 
pre\-factor $\mathrm e ^{\I\nu\theta}$. 
For the sake of readability we will from now on suppress the parameter $\alpha$. 

By inserting Eq.~\eqref{eqn:mathieu-fourier} into the Schrödinger equation, Eq.~\eqref{eqn:Mathieu}, we obtain
\begin{equation}
\left[(\nu+r)^2 -\mathcal E_\nu\right] c_r^\nu +\alpha \left(c_{r+1}^\nu + c_{r-1}^\nu\right)=0, \label{eqn:SEQ_coeffs}
\end{equation}
which is an infinite set of coupled algebraic equations for the coefficients $c^\nu_r$.
This representation is equivalent to an infinite-dimensional matrix eigenvalue problem with the eigenvalues $\mathcal 
E_\nu$ and eigenvectors $\mathbf{c}^\nu$. After truncation to finite dimension, we solve the problem by numerical 
means \cite{coisson_mathieu_2009}. Recursive insertion of Eq.~\eqref{eqn:SEQ_coeffs} into itself leads to a continued 
fraction expansion for the coefficients \cite{meixner_mathieusche_1954, shirts_computation_1993}. Not only can this be 
used for the numerical computation, but it also is particularly useful to derive asymptotic expressions for the limit 
$\alpha \ll 1$.

Furthermore, the set $\left\{\mathrm{me}_{\nu +r}(\alpha,\theta)\right\}_{r=-\infty}^\infty$ forms a complete basis of 
the Hilbert space. This allows us to introduce the concise Dirac notation $\ket{r_\nu}$, such that we have
\begin{equation}
\braket{\theta}{r_\nu} = \frac{1}{\sqrt{2\pi}}\mathrm{me}_{\nu+r}(\theta)
\end{equation}
in position representation. It holds the orthonormality relation $\braket{r_\nu}{l_\nu}=\delta_{rl}$,
where $\delta_{rl}$ denotes the Kronecker delta,  and the completeness relation
$\sum_{r=-\infty}^\infty \ketbra{r_\nu}{r_\nu}=\mathds{1}.
$These properties allow us to expand other functions in terms of Mathieu functions.

In order to calculate the time evolution of a momentum eigenstate, that is a plane  wave with initial momentum 
$\nu_0=\wp_0/\sa$, we expand such a state in terms of the Mathieu functions. In this basis, we read off the time 
evolution from 
Eq.~\eqref{eqn:mathieu_time-evolution}. A 
transformation back 
into the momentum representation yields
\begin{align}
\ket{\psi_{\nu_0}(t)}=\sum_{s=-\infty}^\infty S_{s}^{\nu_0}(\tau)\ket{\nu_0+s}
,\label{eqn:mathieu-momentum-time-evolution}
\end{align}
where we have introduced the scattering amplitude
\begin{equation}
S_{s}^{\nu_0}(\tau) \equiv \sum_{n=-\infty}^\infty c_{-n}^{{\nu_0}+n}c_{s-n}^{{\nu_0}+n}\mathrm e^{-\I 
\mathcal{E}_{{\nu_0}+n}\tau/(2\sa)}.\label{eqn:scattering-amplitude}
\end{equation}
This representation highlights that an initial momentum $p_0 = 2\hbar k \nu_0 $ is coupled only to other 
momenta $p_0 + s (2\hbar k)$, which are separated by multiples of the recoil momentum $2\hbar k$.
In the limit $\alpha \ll 1$ most coefficients (except $c^{\nu_0}_0$ and $c_{-n}^{\nu_0}$ for 
$\nu_0\approx \frac{n}{2}\in\mathbb{Z}$) vanish and the sums in Eqs.~\eqref{eqn:mathieu-momentum-time-evolution} 
and \eqref{eqn:scattering-amplitude} can be reduced to only a few terms. This procedure leads to Rabi oscillations 
between momentum levels that dominate in the quantum regime of the FEL \cite{NJP2015}.

In the next step we generalize Eq.\,\eqref{eqn:mathieu-momentum-time-evolution} by integrating over all possible 
momenta as initial states.
Thus, we obtain the time-evolution operator in momentum representation
\begin{align}
	\hat{\mathcal{U}}(\tau)=\int_{-\infty}^\infty&\dd{\nu}
	\sum_{s=-\infty}^\infty S^{\nu}_s(\tau)\ketbra{\nu+s}{\nu},
\label{eqn:mathieu-te-operator}
\end{align}
which can be applied to an arbitrary initial state.

\subsection{Wigner function}\label{app:Wigner_formal}
We are now in the position to calculate the time evolution of a quantum state in Wigner representation. Here we 
restrict ourselves to an initial 
state which is initially uniform in space and hence fully determined by the initial momentum distribution $\rho(\wp)$, 
see Eq.~\eqref{eq:init}. 

A similar approach to calculate the Wigner function has been persued in Ref.~\cite{pio07_wigner}, but not for 
mixed states.
Other attempts to use the Mathieu functions to obtain the Wigner function of an electron in a periodic potential 
\cite{kandemir_wigner_1998} do not cover dynamics nor arbitrary momenta. In the context of the FEL, Mathieu functions 
have been used to derive asymptotic expressions for the FEL gain \cite{becker79strong,*becker80strong}, but not in 
Wigner phase space.

With the help of the time-evolution operator in Eq.~\eqref{eqn:mathieu-te-operator}, we obtain the time-evolved density 
operator
\begin{equation}
\hat\rho(\tau) = \int \dd{\wp}\rho(\wp)\,\hat{\mathcal{U}}(\tau)\ketbra{\wp}{\wp}\hat{\mathcal{U}}^\dagger(\tau).
\end{equation}
By inserting this expression into the definition for the Wigner representation of a density operator 
Eq.~\eqref{eq:app_W_def}
we find
\begin{gather}
\mathcal W(\theta,\wp;\tau)=\frac{\sa}{\pi}
\sum_{ s=-\infty}^\infty
w_{s}(\theta,\wp;\tau)
\rho\left(\wp+\frac{s/2}{\sa}\right)\label{eqn:Wigner_compact}
\end{gather}
after performing all integrations. 
Here we absorbed one summation as well as the position dependence into the weighting factors
\begin{equation}
w_{s}(\theta,\wp;\tau)=\!\!\!\sum_{s'=-\infty}^\infty\!\!\!
\eval{S_{s'}^{\nu}(\tau) \left[S_{ s -s'}^{\nu}(\tau)\right]^*
	\mathrm e^{\I2\theta(s'- s/2)}}_{\nu=\sa\wp+ s/2},
\end{equation}
which we interpret as scattering amplitudes of the Wigner function.

 \end{appendix}
 
\bibliography{references}

\begin{thebibliography}{85}%
\makeatletter
\providecommand \@ifxundefined [1]{%
 \@ifx{#1\undefined}
}%
\providecommand \@ifnum [1]{%
 \ifnum #1\expandafter \@firstoftwo
 \else \expandafter \@secondoftwo
 \fi
}%
\providecommand \@ifx [1]{%
 \ifx #1\expandafter \@firstoftwo
 \else \expandafter \@secondoftwo
 \fi
}%
\providecommand \natexlab [1]{#1}%
\providecommand \enquote  [1]{``#1''}%
\providecommand \bibnamefont  [1]{#1}%
\providecommand \bibfnamefont [1]{#1}%
\providecommand \citenamefont [1]{#1}%
\providecommand \href@noop [0]{\@secondoftwo}%
\providecommand \href [0]{\begingroup \@sanitize@url \@href}%
\providecommand \@href[1]{\@@startlink{#1}\@@href}%
\providecommand \@@href[1]{\endgroup#1\@@endlink}%
\providecommand \@sanitize@url [0]{\catcode `\\12\catcode `\$12\catcode
  `\&12\catcode `\#12\catcode `\^12\catcode `\_12\catcode `\%12\relax}%
\providecommand \@@startlink[1]{}%
\providecommand \@@endlink[0]{}%
\providecommand \url  [0]{\begingroup\@sanitize@url \@url }%
\providecommand \@url [1]{\endgroup\@href {#1}{\urlprefix }}%
\providecommand \urlprefix  [0]{URL }%
\providecommand \Eprint [0]{\href }%
\providecommand \doibase [0]{http://dx.doi.org/}%
\providecommand \selectlanguage [0]{\@gobble}%
\providecommand \bibinfo  [0]{\@secondoftwo}%
\providecommand \bibfield  [0]{\@secondoftwo}%
\providecommand \translation [1]{[#1]}%
\providecommand \BibitemOpen [0]{}%
\providecommand \bibitemStop [0]{}%
\providecommand \bibitemNoStop [0]{.\EOS\space}%
\providecommand \EOS [0]{\spacefactor3000\relax}%
\providecommand \BibitemShut  [1]{\csname bibitem#1\endcsname}%
\let\auto@bib@innerbib\@empty
%</preamble>
\bibitem [{\citenamefont {Schroeder}\ \emph {et~al.}(2001)\citenamefont
  {Schroeder}, \citenamefont {Pellegrini},\ and\ \citenamefont
  {Chen}}]{schroeder}%
  \BibitemOpen
  \bibfield  {author} {\bibinfo {author} {\bibfnamefont {C.~B.}\ \bibnamefont
  {Schroeder}}, \bibinfo {author} {\bibfnamefont {C.}~\bibnamefont
  {Pellegrini}}, \ and\ \bibinfo {author} {\bibfnamefont {P.}~\bibnamefont
  {Chen}},\ }\href {http://link.aps.org/doi/10.1103/PhysRevE.64.056502}
  {\bibfield  {journal} {\bibinfo  {journal} {Phys. Rev. E}\ }\textbf {\bibinfo
  {volume} {64}},\ \bibinfo {pages} {056502} (\bibinfo {year}
  {2001})}\BibitemShut {NoStop}%
\bibitem [{\citenamefont {Bonifacio}\ \emph {et~al.}(2006)\citenamefont
  {Bonifacio}, \citenamefont {Piovella}, \citenamefont {Robb},\ and\
  \citenamefont {Schiavi}}]{boni06}%
  \BibitemOpen
  \bibfield  {author} {\bibinfo {author} {\bibfnamefont {R.}~\bibnamefont
  {Bonifacio}}, \bibinfo {author} {\bibfnamefont {N.}~\bibnamefont {Piovella}},
  \bibinfo {author} {\bibfnamefont {G.~R.~M.}\ \bibnamefont {Robb}}, \ and\
  \bibinfo {author} {\bibfnamefont {A.}~\bibnamefont {Schiavi}},\ }\href@noop
  {} {\bibfield  {journal} {\bibinfo  {journal} {Phys. Rev. Spec. Top.--Accel.
  Beams}\ }\textbf {\bibinfo {volume} {9}},\ \bibinfo {pages} {090701}
  (\bibinfo {year} {2006})}\BibitemShut {NoStop}%
\bibitem [{\citenamefont {Bonifacio}\ \emph {et~al.}(2017)\citenamefont
  {Bonifacio}, \citenamefont {Fares}, \citenamefont {Ferrario}, \citenamefont
  {McNeil},\ and\ \citenamefont {Robb}}]{boni17}%
  \BibitemOpen
  \bibfield  {author} {\bibinfo {author} {\bibfnamefont {R.}~\bibnamefont
  {Bonifacio}}, \bibinfo {author} {\bibfnamefont {H.}~\bibnamefont {Fares}},
  \bibinfo {author} {\bibfnamefont {M.}~\bibnamefont {Ferrario}}, \bibinfo
  {author} {\bibfnamefont {B.~W.~J.}\ \bibnamefont {McNeil}}, \ and\ \bibinfo
  {author} {\bibfnamefont {G.~R.~M.}\ \bibnamefont {Robb}},\ }\href@noop {}
  {\bibfield  {journal} {\bibinfo  {journal} {Opt. Commun.}\ }\textbf {\bibinfo
  {volume} {382}},\ \bibinfo {pages} {58} (\bibinfo {year} {2017})}\BibitemShut
  {NoStop}%
\bibitem [{\citenamefont {Anisimov}(2018)}]{anisimov18}%
  \BibitemOpen
  \bibfield  {author} {\bibinfo {author} {\bibfnamefont {P.~M.}\ \bibnamefont
  {Anisimov}},\ }\href {https://doi.org/10.1080/09500340.2017.1375567}
  {\bibfield  {journal} {\bibinfo  {journal} {J. Mod. Opt.}\ }\textbf {\bibinfo
  {volume} {65}},\ \bibinfo {pages} {1370} (\bibinfo {year}
  {2018})}\BibitemShut {NoStop}%
\bibitem [{\citenamefont {Kling}\ \emph {et~al.}(2015)\citenamefont {Kling},
  \citenamefont {Giese}, \citenamefont {Endrich}, \citenamefont {Preiss},
  \citenamefont {Sauerbrey},\ and\ \citenamefont {Schleich}}]{NJP2015}%
  \BibitemOpen
  \bibfield  {author} {\bibinfo {author} {\bibfnamefont {P.}~\bibnamefont
  {Kling}}, \bibinfo {author} {\bibfnamefont {E.}~\bibnamefont {Giese}},
  \bibinfo {author} {\bibfnamefont {R.}~\bibnamefont {Endrich}}, \bibinfo
  {author} {\bibfnamefont {P.}~\bibnamefont {Preiss}}, \bibinfo {author}
  {\bibfnamefont {R.}~\bibnamefont {Sauerbrey}}, \ and\ \bibinfo {author}
  {\bibfnamefont {W.~P.}\ \bibnamefont {Schleich}},\ }\href {\doibase
  10.1088/1367-2630/17/12/123019} {\bibfield  {journal} {\bibinfo  {journal}
  {New J. Phys.}\ }\textbf {\bibinfo {volume} {17}},\ \bibinfo {pages} {123019}
  (\bibinfo {year} {2015})}\BibitemShut {NoStop}%
\bibitem [{\citenamefont {Kling}\ \emph {et~al.}(2019)\citenamefont {Kling},
  \citenamefont {Giese}, \citenamefont {Carmesin}, \citenamefont {Sauerbrey},\
  and\ \citenamefont {Schleich}}]{kling_high-gain_2019}%
  \BibitemOpen
  \bibfield  {author} {\bibinfo {author} {\bibfnamefont {P.}~\bibnamefont
  {Kling}}, \bibinfo {author} {\bibfnamefont {E.}~\bibnamefont {Giese}},
  \bibinfo {author} {\bibfnamefont {C.~M.}\ \bibnamefont {Carmesin}}, \bibinfo
  {author} {\bibfnamefont {R.}~\bibnamefont {Sauerbrey}}, \ and\ \bibinfo
  {author} {\bibfnamefont {W.~P.}\ \bibnamefont {Schleich}},\ }\href {\doibase
  10.1103/PhysRevA.99.053823} {\bibfield  {journal} {\bibinfo  {journal} {Phys.
  Rev. A}\ }\textbf {\bibinfo {volume} {99}},\ \bibinfo {pages} {053823}
  (\bibinfo {year} {2019})}\BibitemShut {NoStop}%
\bibitem [{\citenamefont {Madey}(1971)}]{madey1}%
  \BibitemOpen
  \bibfield  {author} {\bibinfo {author} {\bibfnamefont {J.~M.~J.}\
  \bibnamefont {Madey}},\ }\href@noop {} {\bibfield  {journal} {\bibinfo
  {journal} {J. Appl. Phys.}\ }\textbf {\bibinfo {volume} {42}},\ \bibinfo
  {pages} {1906} (\bibinfo {year} {1971})}\BibitemShut {NoStop}%
\bibitem [{\citenamefont {Hopf}\ \emph {et~al.}(1976)\citenamefont {Hopf},
  \citenamefont {Meystre}, \citenamefont {Scully},\ and\ \citenamefont
  {Louisell}}]{scully_fel}%
  \BibitemOpen
  \bibfield  {author} {\bibinfo {author} {\bibfnamefont {F.~A.}\ \bibnamefont
  {Hopf}}, \bibinfo {author} {\bibfnamefont {P.}~\bibnamefont {Meystre}},
  \bibinfo {author} {\bibfnamefont {M.~O.}\ \bibnamefont {Scully}}, \ and\
  \bibinfo {author} {\bibfnamefont {W.~H.}\ \bibnamefont {Louisell}},\
  }\href@noop {} {\bibfield  {journal} {\bibinfo  {journal} {Opt. Commun.}\
  }\textbf {\bibinfo {volume} {18}},\ \bibinfo {pages} {413 } (\bibinfo {year}
  {1976})}\BibitemShut {NoStop}%
\bibitem [{\citenamefont {Colson}(1977)}]{colson}%
  \BibitemOpen
  \bibfield  {author} {\bibinfo {author} {\bibfnamefont {W.~B.}\ \bibnamefont
  {Colson}},\ }\href@noop {} {\bibfield  {journal} {\bibinfo  {journal} {Phys.
  Lett. A}\ }\textbf {\bibinfo {volume} {64}},\ \bibinfo {pages} {190 }
  (\bibinfo {year} {1977})}\BibitemShut {NoStop}%
\bibitem [{\citenamefont {Becker}\ and\ \citenamefont
  {Mitter}(1979)}]{becker79}%
  \BibitemOpen
  \bibfield  {author} {\bibinfo {author} {\bibfnamefont {W.}~\bibnamefont
  {Becker}}\ and\ \bibinfo {author} {\bibfnamefont {H.}~\bibnamefont
  {Mitter}},\ }\href@noop {} {\bibfield  {journal} {\bibinfo  {journal} {Z.
  Phys. B: Condens. Matter}\ }\textbf {\bibinfo {volume} {35}},\ \bibinfo
  {pages} {399} (\bibinfo {year} {1979})}\BibitemShut {NoStop}%
\bibitem [{\citenamefont {Becker}\ and\ \citenamefont
  {McIver}(1988)}]{becker88}%
  \BibitemOpen
  \bibfield  {author} {\bibinfo {author} {\bibfnamefont {W.}~\bibnamefont
  {Becker}}\ and\ \bibinfo {author} {\bibfnamefont {J.~K.}\ \bibnamefont
  {McIver}},\ }\href {\doibase 10.1007/BF01439805} {\bibfield  {journal}
  {\bibinfo  {journal} {Z. Phys. D: At., Mol. Clusters}\ }\textbf {\bibinfo
  {volume} {7}},\ \bibinfo {pages} {353} (\bibinfo {year} {1988})}\BibitemShut
  {NoStop}%
\bibitem [{\citenamefont {Bosco}\ \emph {et~al.}(1983)\citenamefont {Bosco},
  \citenamefont {Colson},\ and\ \citenamefont {Freedman}}]{colson_quantum}%
  \BibitemOpen
  \bibfield  {author} {\bibinfo {author} {\bibfnamefont {P.}~\bibnamefont
  {Bosco}}, \bibinfo {author} {\bibfnamefont {W.}~\bibnamefont {Colson}}, \
  and\ \bibinfo {author} {\bibfnamefont {R.}~\bibnamefont {Freedman}},\
  }\href@noop {} {\bibfield  {journal} {\bibinfo  {journal} {IEEE J. Quantum
  Electron.}\ }\textbf {\bibinfo {volume} {19}},\ \bibinfo {pages} {272}
  (\bibinfo {year} {1983})}\BibitemShut {NoStop}%
\bibitem [{\citenamefont {Friedman}\ \emph {et~al.}(1988)\citenamefont
  {Friedman}, \citenamefont {Gover}, \citenamefont {Kurizki}, \citenamefont
  {Ruschin},\ and\ \citenamefont {Yariv}}]{friedman}%
  \BibitemOpen
  \bibfield  {author} {\bibinfo {author} {\bibfnamefont {A.}~\bibnamefont
  {Friedman}}, \bibinfo {author} {\bibfnamefont {A.}~\bibnamefont {Gover}},
  \bibinfo {author} {\bibfnamefont {G.}~\bibnamefont {Kurizki}}, \bibinfo
  {author} {\bibfnamefont {S.}~\bibnamefont {Ruschin}}, \ and\ \bibinfo
  {author} {\bibfnamefont {A.}~\bibnamefont {Yariv}},\ }\href@noop {}
  {\bibfield  {journal} {\bibinfo  {journal} {Rev. Mod. Phys.}\ }\textbf
  {\bibinfo {volume} {60}},\ \bibinfo {pages} {471} (\bibinfo {year}
  {1988})}\BibitemShut {NoStop}%
\bibitem [{\citenamefont {Gea-Banacloche}(1985)}]{banacloche}%
  \BibitemOpen
  \bibfield  {author} {\bibinfo {author} {\bibfnamefont {J.}~\bibnamefont
  {Gea-Banacloche}},\ }\href@noop {} {\bibfield  {journal} {\bibinfo  {journal}
  {Phys. Rev. A}\ }\textbf {\bibinfo {volume} {31}},\ \bibinfo {pages} {1607}
  (\bibinfo {year} {1985})}\BibitemShut {NoStop}%
\bibitem [{\citenamefont {Madey}\ \emph {et~al.}(1973)\citenamefont {Madey},
  \citenamefont {Schwettman},\ and\ \citenamefont {Fairbank}}]{madey2}%
  \BibitemOpen
  \bibfield  {author} {\bibinfo {author} {\bibfnamefont {J.~M.~J.}\
  \bibnamefont {Madey}}, \bibinfo {author} {\bibfnamefont {H.~A.}\ \bibnamefont
  {Schwettman}}, \ and\ \bibinfo {author} {\bibfnamefont {W.~M.}\ \bibnamefont
  {Fairbank}},\ }\href@noop {} {\bibfield  {journal} {\bibinfo  {journal} {IEEE
  Trans. Nuc. Sci.}\ }\textbf {\bibinfo {volume} {20}},\ \bibinfo {pages} {980}
  (\bibinfo {year} {1973})}\BibitemShut {NoStop}%
\bibitem [{\citenamefont {McIver}\ and\ \citenamefont
  {Fedorov}(1979)}]{mciver}%
  \BibitemOpen
  \bibfield  {author} {\bibinfo {author} {\bibfnamefont {J.~K.}\ \bibnamefont
  {McIver}}\ and\ \bibinfo {author} {\bibfnamefont {M.~V.}\ \bibnamefont
  {Fedorov}},\ }\href@noop {} {\bibfield  {journal} {\bibinfo  {journal} {J.
  Exp. Theor. Phys.}\ }\textbf {\bibinfo {volume} {49}},\ \bibinfo {pages}
  {1012 } (\bibinfo {year} {1979})}\BibitemShut {NoStop}%
\bibitem [{\citenamefont {Becker}(1979)}]{becker79strong}%
  \BibitemOpen
  \bibfield  {author} {\bibinfo {author} {\bibfnamefont {W.}~\bibnamefont
  {Becker}},\ }\href {\doibase 10.1016/0375-9601(79)90585-1} {\bibfield
  {journal} {\bibinfo  {journal} {Phys. Lett. A}\ }\textbf {\bibinfo {volume}
  {74}},\ \bibinfo {pages} {66} (\bibinfo {year} {1979})}\BibitemShut {NoStop}%
\bibitem [{\citenamefont {Becker}(1980{\natexlab{a}})}]{becker80strong}%
  \BibitemOpen
  \bibfield  {author} {\bibinfo {author} {\bibfnamefont {W.}~\bibnamefont
  {Becker}},\ }\href@noop {} {\bibfield  {journal} {\bibinfo  {journal} {Z.
  Phys. B: Condens. Matter}\ }\textbf {\bibinfo {volume} {38}},\ \bibinfo
  {pages} {287} (\bibinfo {year} {1980}{\natexlab{a}})}\BibitemShut {NoStop}%
\bibitem [{\citenamefont {Becker}(1980{\natexlab{b}})}]{becker80}%
  \BibitemOpen
  \bibfield  {author} {\bibinfo {author} {\bibfnamefont {W.}~\bibnamefont
  {Becker}},\ }\href@noop {} {\bibfield  {journal} {\bibinfo  {journal} {Opt.
  Commun.}\ }\textbf {\bibinfo {volume} {33}},\ \bibinfo {pages} {69} (\bibinfo
  {year} {1980}{\natexlab{b}})}\BibitemShut {NoStop}%
\bibitem [{\citenamefont {Becker}\ \emph {et~al.}(1982)\citenamefont {Becker},
  \citenamefont {Scully},\ and\ \citenamefont {Zubairy}}]{becker82}%
  \BibitemOpen
  \bibfield  {author} {\bibinfo {author} {\bibfnamefont {W.}~\bibnamefont
  {Becker}}, \bibinfo {author} {\bibfnamefont {M.~O.}\ \bibnamefont {Scully}},
  \ and\ \bibinfo {author} {\bibfnamefont {M.~S.}\ \bibnamefont {Zubairy}},\
  }\href@noop {} {\bibfield  {journal} {\bibinfo  {journal} {Phys. Rev. Lett.}\
  }\textbf {\bibinfo {volume} {48}},\ \bibinfo {pages} {475} (\bibinfo {year}
  {1982})}\BibitemShut {NoStop}%
\bibitem [{\citenamefont {Becker}\ and\ \citenamefont
  {McIver}(1983)}]{becker83}%
  \BibitemOpen
  \bibfield  {author} {\bibinfo {author} {\bibfnamefont {W.}~\bibnamefont
  {Becker}}\ and\ \bibinfo {author} {\bibfnamefont {J.~K.}\ \bibnamefont
  {McIver}},\ }\href@noop {} {\bibfield  {journal} {\bibinfo  {journal} {Phys.
  Rev. A}\ }\textbf {\bibinfo {volume} {27}},\ \bibinfo {pages} {1030}
  (\bibinfo {year} {1983})}\BibitemShut {NoStop}%
\bibitem [{\citenamefont {Becker}\ and\ \citenamefont
  {Zubairy}(1982)}]{becker_pstat}%
  \BibitemOpen
  \bibfield  {author} {\bibinfo {author} {\bibfnamefont {W.}~\bibnamefont
  {Becker}}\ and\ \bibinfo {author} {\bibfnamefont {M.~S.}\ \bibnamefont
  {Zubairy}},\ }\href@noop {} {\bibfield  {journal} {\bibinfo  {journal} {Phys.
  Rev. A}\ }\textbf {\bibinfo {volume} {25}},\ \bibinfo {pages} {2200}
  (\bibinfo {year} {1982})}\BibitemShut {NoStop}%
\bibitem [{\citenamefont {Becker}\ \emph {et~al.}(1984)\citenamefont {Becker},
  \citenamefont {Scully},\ and\ \citenamefont {Zubairy}}]{becker2}%
  \BibitemOpen
  \bibfield  {author} {\bibinfo {author} {\bibfnamefont {W.}~\bibnamefont
  {Becker}}, \bibinfo {author} {\bibfnamefont {M.~O.}\ \bibnamefont {Scully}},
  \ and\ \bibinfo {author} {\bibfnamefont {M.~S.}\ \bibnamefont {Zubairy}},\
  }in\ \href@noop {} {\emph {\bibinfo {booktitle} {Coherence and Quantum Optics
  V}}},\ \bibinfo {editor} {edited by\ \bibinfo {editor} {\bibfnamefont
  {L.}~\bibnamefont {Mandel}}\ and\ \bibinfo {editor} {\bibfnamefont
  {E.}~\bibnamefont {Wolf}}}\ (\bibinfo  {publisher} {Springer US},\ \bibinfo
  {address} {Boston, MA},\ \bibinfo {year} {1984})\ pp.\ \bibinfo {pages}
  {811--818}\BibitemShut {NoStop}%
\bibitem [{\citenamefont {Becker}\ \emph {et~al.}(1986)\citenamefont {Becker},
  \citenamefont {Gea-Banacloche},\ and\ \citenamefont {Scully}}]{becker_lw}%
  \BibitemOpen
  \bibfield  {author} {\bibinfo {author} {\bibfnamefont {W.}~\bibnamefont
  {Becker}}, \bibinfo {author} {\bibfnamefont {J.}~\bibnamefont
  {Gea-Banacloche}}, \ and\ \bibinfo {author} {\bibfnamefont {M.~O.}\
  \bibnamefont {Scully}},\ }\href@noop {} {\bibfield  {journal} {\bibinfo
  {journal} {Phys. Rev. A}\ }\textbf {\bibinfo {volume} {33}},\ \bibinfo
  {pages} {2174} (\bibinfo {year} {1986})}\BibitemShut {NoStop}%
\bibitem [{\citenamefont {Gea-Banacloche}(1986)}]{banacloche_pstat}%
  \BibitemOpen
  \bibfield  {author} {\bibinfo {author} {\bibfnamefont {J.}~\bibnamefont
  {Gea-Banacloche}},\ }\href@noop {} {\bibfield  {journal} {\bibinfo  {journal}
  {Phys. Rev. A}\ }\textbf {\bibinfo {volume} {33}},\ \bibinfo {pages} {1448}
  (\bibinfo {year} {1986})}\BibitemShut {NoStop}%
\bibitem [{\citenamefont {Orszag}(1987)}]{orszag_lw}%
  \BibitemOpen
  \bibfield  {author} {\bibinfo {author} {\bibfnamefont {M.}~\bibnamefont
  {Orszag}},\ }\href@noop {} {\bibfield  {journal} {\bibinfo  {journal} {Phys.
  Rev. A}\ }\textbf {\bibinfo {volume} {36}},\ \bibinfo {pages} {189} (\bibinfo
  {year} {1987})}\BibitemShut {NoStop}%
\bibitem [{\citenamefont {Gover}\ \emph {et~al.}(1987)\citenamefont {Gover},
  \citenamefont {Amir},\ and\ \citenamefont {Elias}}]{gover_lw}%
  \BibitemOpen
  \bibfield  {author} {\bibinfo {author} {\bibfnamefont {A.}~\bibnamefont
  {Gover}}, \bibinfo {author} {\bibfnamefont {A.}~\bibnamefont {Amir}}, \ and\
  \bibinfo {author} {\bibfnamefont {L.~R.}\ \bibnamefont {Elias}},\ }\href@noop
  {} {\bibfield  {journal} {\bibinfo  {journal} {Phys. Rev. A}\ }\textbf
  {\bibinfo {volume} {35}},\ \bibinfo {pages} {164} (\bibinfo {year}
  {1987})}\BibitemShut {NoStop}%
\bibitem [{\citenamefont {Glauber}(1951)}]{glauber}%
  \BibitemOpen
  \bibfield  {author} {\bibinfo {author} {\bibfnamefont {R.~J.}\ \bibnamefont
  {Glauber}},\ }\href@noop {} {\bibfield  {journal} {\bibinfo  {journal} {Phys.
  Rev.}\ }\textbf {\bibinfo {volume} {84}},\ \bibinfo {pages} {395} (\bibinfo
  {year} {1951})}\BibitemShut {NoStop}%
\bibitem [{\citenamefont {Bonifacio}\ \emph {et~al.}(2005)\citenamefont
  {Bonifacio}, \citenamefont {Cola}, \citenamefont {Piovella},\ and\
  \citenamefont {Robb}}]{boni05}%
  \BibitemOpen
  \bibfield  {author} {\bibinfo {author} {\bibfnamefont {R.}~\bibnamefont
  {Bonifacio}}, \bibinfo {author} {\bibfnamefont {M.~M.}\ \bibnamefont {Cola}},
  \bibinfo {author} {\bibfnamefont {N.}~\bibnamefont {Piovella}}, \ and\
  \bibinfo {author} {\bibfnamefont {G.~R.~M.}\ \bibnamefont {Robb}},\
  }\href@noop {} {\bibfield  {journal} {\bibinfo  {journal} {Europhys. Lett.}\
  }\textbf {\bibinfo {volume} {69}},\ \bibinfo {pages} {55} (\bibinfo {year}
  {2005})}\BibitemShut {NoStop}%
\bibitem [{\citenamefont {Piovella}\ \emph {et~al.}(2007)\citenamefont
  {Piovella}, \citenamefont {Cola}, \citenamefont {Volpe}, \citenamefont
  {Gaiba}, \citenamefont {Schiavi},\ and\ \citenamefont
  {Bonifacio}}]{pio07_wigner}%
  \BibitemOpen
  \bibfield  {author} {\bibinfo {author} {\bibfnamefont {N.}~\bibnamefont
  {Piovella}}, \bibinfo {author} {\bibfnamefont {M.~M.}\ \bibnamefont {Cola}},
  \bibinfo {author} {\bibfnamefont {L.}~\bibnamefont {Volpe}}, \bibinfo
  {author} {\bibfnamefont {R.}~\bibnamefont {Gaiba}}, \bibinfo {author}
  {\bibfnamefont {A.}~\bibnamefont {Schiavi}}, \ and\ \bibinfo {author}
  {\bibfnamefont {R.}~\bibnamefont {Bonifacio}},\ }\href {\doibase
  https://doi.org/10.1016/j.optcom.2007.02.061} {\bibfield  {journal} {\bibinfo
   {journal} {Opt. Commun.}\ }\textbf {\bibinfo {volume} {274}},\ \bibinfo
  {pages} {347 } (\bibinfo {year} {2007})}\BibitemShut {NoStop}%
\bibitem [{\citenamefont {Piovella}\ \emph {et~al.}(2008)\citenamefont
  {Piovella}, \citenamefont {Cola}, \citenamefont {Volpe}, \citenamefont
  {Schiavi},\ and\ \citenamefont {Bonifacio}}]{boni_wigner}%
  \BibitemOpen
  \bibfield  {author} {\bibinfo {author} {\bibfnamefont {N.}~\bibnamefont
  {Piovella}}, \bibinfo {author} {\bibfnamefont {M.~M.}\ \bibnamefont {Cola}},
  \bibinfo {author} {\bibfnamefont {L.}~\bibnamefont {Volpe}}, \bibinfo
  {author} {\bibfnamefont {A.}~\bibnamefont {Schiavi}}, \ and\ \bibinfo
  {author} {\bibfnamefont {R.}~\bibnamefont {Bonifacio}},\ }\href@noop {}
  {\bibfield  {journal} {\bibinfo  {journal} {Phys. Rev. Lett.}\ }\textbf
  {\bibinfo {volume} {100}},\ \bibinfo {pages} {044801} (\bibinfo {year}
  {2008})}\BibitemShut {NoStop}%
\bibitem [{\citenamefont {Serbeto}\ \emph {et~al.}(2008)\citenamefont
  {Serbeto}, \citenamefont {Mendonça}, \citenamefont {Tsui},\ and\
  \citenamefont {Bonifacio}}]{serbeto08}%
  \BibitemOpen
  \bibfield  {author} {\bibinfo {author} {\bibfnamefont {A.}~\bibnamefont
  {Serbeto}}, \bibinfo {author} {\bibfnamefont {J.~T.}\ \bibnamefont
  {Mendonça}}, \bibinfo {author} {\bibfnamefont {K.~H.}\ \bibnamefont {Tsui}},
  \ and\ \bibinfo {author} {\bibfnamefont {R.}~\bibnamefont {Bonifacio}},\
  }\href {\doibase 10.1063/1.2833591} {\bibfield  {journal} {\bibinfo
  {journal} {Phys. Plasmas}\ }\textbf {\bibinfo {volume} {15}},\ \bibinfo
  {pages} {013110} (\bibinfo {year} {2008})}\BibitemShut {NoStop}%
\bibitem [{\citenamefont {Dattoli}\ \emph {et~al.}(2018)\citenamefont
  {Dattoli}, \citenamefont {Di Palma}, \citenamefont {Pagnutti},\ and\
  \citenamefont {Sabia}}]{dattoli18}%
  \BibitemOpen
  \bibfield  {author} {\bibinfo {author} {\bibfnamefont {G.}~\bibnamefont
  {Dattoli}}, \bibinfo {author} {\bibfnamefont {E.}~\bibnamefont {Di Palma}},
  \bibinfo {author} {\bibfnamefont {S.}~\bibnamefont {Pagnutti}}, \ and\
  \bibinfo {author} {\bibfnamefont {E.}~\bibnamefont {Sabia}},\ }\href
  {\doibase https://doi.org/10.1016/j.physrep.2018.02.005} {\bibfield
  {journal} {\bibinfo  {journal} {Phys. Rep.}\ }\textbf {\bibinfo {volume}
  {739}},\ \bibinfo {pages} {1 } (\bibinfo {year} {2018})}\BibitemShut
  {NoStop}%
\bibitem [{\citenamefont {Vogel}\ and\ \citenamefont {Risken}(1989)}]{vogel89}%
  \BibitemOpen
  \bibfield  {author} {\bibinfo {author} {\bibfnamefont {K.}~\bibnamefont
  {Vogel}}\ and\ \bibinfo {author} {\bibfnamefont {H.}~\bibnamefont {Risken}},\
  }\href@noop {} {\bibfield  {journal} {\bibinfo  {journal} {Phys. Rev. A}\
  }\textbf {\bibinfo {volume} {40}},\ \bibinfo {pages} {2847} (\bibinfo {year}
  {1989})}\BibitemShut {NoStop}%
\bibitem [{\citenamefont {Wigner}(1932)}]{wigner32}%
  \BibitemOpen
  \bibfield  {author} {\bibinfo {author} {\bibfnamefont {E.~P.}\ \bibnamefont
  {Wigner}},\ }\href@noop {} {\bibfield  {journal} {\bibinfo  {journal} {Phys.
  Rev.}\ }\textbf {\bibinfo {volume} {40}},\ \bibinfo {pages} {749} (\bibinfo
  {year} {1932})}\BibitemShut {NoStop}%
\bibitem [{\citenamefont {Carruthers}\ and\ \citenamefont
  {Zachariasen}(1983)}]{carruthers83}%
  \BibitemOpen
  \bibfield  {author} {\bibinfo {author} {\bibfnamefont {P.}~\bibnamefont
  {Carruthers}}\ and\ \bibinfo {author} {\bibfnamefont {F.}~\bibnamefont
  {Zachariasen}},\ }\href@noop {} {\bibfield  {journal} {\bibinfo  {journal}
  {Rev. Mod. Phys.}\ }\textbf {\bibinfo {volume} {55}},\ \bibinfo {pages} {245}
  (\bibinfo {year} {1983})}\BibitemShut {NoStop}%
\bibitem [{\citenamefont {Hillery}\ \emph {et~al.}(1984)\citenamefont
  {Hillery}, \citenamefont {O'Connell}, \citenamefont {Scully},\ and\
  \citenamefont {Wigner}}]{hillery84}%
  \BibitemOpen
  \bibfield  {author} {\bibinfo {author} {\bibfnamefont {M.}~\bibnamefont
  {Hillery}}, \bibinfo {author} {\bibfnamefont {R.~F.}\ \bibnamefont
  {O'Connell}}, \bibinfo {author} {\bibfnamefont {M.~O.}\ \bibnamefont
  {Scully}}, \ and\ \bibinfo {author} {\bibfnamefont {E.~P.}\ \bibnamefont
  {Wigner}},\ }\href@noop {} {\bibfield  {journal} {\bibinfo  {journal} {Phys.
  Rep.}\ }\textbf {\bibinfo {volume} {106}},\ \bibinfo {pages} {121} (\bibinfo
  {year} {1984})}\BibitemShut {NoStop}%
\bibitem [{\citenamefont {Schleich}(2001)}]{schleich}%
  \BibitemOpen
  \bibfield  {author} {\bibinfo {author} {\bibfnamefont {W.~P.}\ \bibnamefont
  {Schleich}},\ }\href@noop {} {\emph {\bibinfo {title} {{Quantum Optics in
  Phase Space}}}}\ (\bibinfo  {publisher} {Wiley-VCH, Weinheim},\ \bibinfo
  {year} {2001})\BibitemShut {NoStop}%
\bibitem [{\citenamefont {Case}(2008)}]{case08}%
  \BibitemOpen
  \bibfield  {author} {\bibinfo {author} {\bibfnamefont {W.~B.}\ \bibnamefont
  {Case}},\ }\href@noop {} {\bibfield  {journal} {\bibinfo  {journal} {Am. J.
  Phys.}\ }\textbf {\bibinfo {volume} {76}},\ \bibinfo {pages} {937} (\bibinfo
  {year} {2008})}\BibitemShut {NoStop}%
\bibitem [{\citenamefont {Bambini}\ and\ \citenamefont
  {Renieri}(1978)}]{bambi}%
  \BibitemOpen
  \bibfield  {author} {\bibinfo {author} {\bibfnamefont {A.}~\bibnamefont
  {Bambini}}\ and\ \bibinfo {author} {\bibfnamefont {A.}~\bibnamefont
  {Renieri}},\ }\href {http://dx.doi.org/10.1007/BF02762613} {\bibfield
  {journal} {\bibinfo  {journal} {Lett. Nuovo Cimento}\ }\textbf {\bibinfo
  {volume} {21}},\ \bibinfo {pages} {399} (\bibinfo {year} {1978})}\BibitemShut
  {NoStop}%
\bibitem [{\citenamefont {Bambini}\ \emph {et~al.}(1979)\citenamefont
  {Bambini}, \citenamefont {Renieri},\ and\ \citenamefont {Stenholm}}]{brs}%
  \BibitemOpen
  \bibfield  {author} {\bibinfo {author} {\bibfnamefont {A.}~\bibnamefont
  {Bambini}}, \bibinfo {author} {\bibfnamefont {A.}~\bibnamefont {Renieri}}, \
  and\ \bibinfo {author} {\bibfnamefont {S.}~\bibnamefont {Stenholm}},\
  }\href@noop {} {\bibfield  {journal} {\bibinfo  {journal} {Phys. Rev. A}\
  }\textbf {\bibinfo {volume} {19}},\ \bibinfo {pages} {2013 } (\bibinfo {year}
  {1979})}\BibitemShut {NoStop}%
\bibitem [{\citenamefont {Heller}(1976)}]{heller76}%
  \BibitemOpen
  \bibfield  {author} {\bibinfo {author} {\bibfnamefont {E.~J.}\ \bibnamefont
  {Heller}},\ }\href@noop {} {\bibfield  {journal} {\bibinfo  {journal} {J.
  Chem. Phys.}\ }\textbf {\bibinfo {volume} {65}},\ \bibinfo {pages} {1289}
  (\bibinfo {year} {1976})}\BibitemShut {NoStop}%
\bibitem [{\citenamefont {Heller}(1977)}]{heller77}%
  \BibitemOpen
  \bibfield  {author} {\bibinfo {author} {\bibfnamefont {E.~J.}\ \bibnamefont
  {Heller}},\ }\href@noop {} {\bibfield  {journal} {\bibinfo  {journal} {J.
  Chem. Phys.}\ }\textbf {\bibinfo {volume} {67}},\ \bibinfo {pages} {3339}
  (\bibinfo {year} {1977})}\BibitemShut {NoStop}%
\bibitem [{\citenamefont {Gover}\ and\ \citenamefont {Pan}(2018)}]{gover18}%
  \BibitemOpen
  \bibfield  {author} {\bibinfo {author} {\bibfnamefont {A.}~\bibnamefont
  {Gover}}\ and\ \bibinfo {author} {\bibfnamefont {Y.}~\bibnamefont {Pan}},\
  }\href@noop {} {\bibfield  {journal} {\bibinfo  {journal} {Phys. Lett. A}\
  }\textbf {\bibinfo {volume} {382}},\ \bibinfo {pages} {1550} (\bibinfo {year}
  {2018})}\BibitemShut {NoStop}%
\bibitem [{\citenamefont {Dattoli}\ and\ \citenamefont
  {Fares}(2019)}]{dattoli_gaussian_2019}%
  \BibitemOpen
  \bibfield  {author} {\bibinfo {author} {\bibfnamefont {G.}~\bibnamefont
  {Dattoli}}\ and\ \bibinfo {author} {\bibfnamefont {H.}~\bibnamefont
  {Fares}},\ }\href {\doibase 10.1063/1.5040925} {\bibfield  {journal}
  {\bibinfo  {journal} {Journal of Mathematical Physics}\ }\textbf {\bibinfo
  {volume} {60}},\ \bibinfo {pages} {042101} (\bibinfo {year}
  {2019})}\BibitemShut {NoStop}%
\bibitem [{\citenamefont {Schm\"{u}ser}\ \emph {et~al.}(2008)\citenamefont
  {Schm\"{u}ser}, \citenamefont {Dohlus},\ and\ \citenamefont
  {Rossbach}}]{schmueser}%
  \BibitemOpen
  \bibfield  {author} {\bibinfo {author} {\bibfnamefont {P.}~\bibnamefont
  {Schm\"{u}ser}}, \bibinfo {author} {\bibfnamefont {M.}~\bibnamefont
  {Dohlus}}, \ and\ \bibinfo {author} {\bibfnamefont {J.}~\bibnamefont
  {Rossbach}},\ }\href@noop {} {\emph {\bibinfo {title} {Ultraviolet and Soft
  X-Ray Free-Electron Lasers}}}\ (\bibinfo  {publisher} {Springer,
  Heidelberg},\ \bibinfo {year} {2008})\BibitemShut {NoStop}%
\bibitem [{\citenamefont {Borenstein}\ and\ \citenamefont
  {Lamb}(1972)}]{borenstein}%
  \BibitemOpen
  \bibfield  {author} {\bibinfo {author} {\bibfnamefont {M.}~\bibnamefont
  {Borenstein}}\ and\ \bibinfo {author} {\bibfnamefont {W.~E.}\ \bibnamefont
  {Lamb}},\ }\href@noop {} {\bibfield  {journal} {\bibinfo  {journal} {Phys.
  Rev. A}\ }\textbf {\bibinfo {volume} {5}},\ \bibinfo {pages} {1298} (\bibinfo
  {year} {1972})}\BibitemShut {NoStop}%
\bibitem [{\citenamefont {Meystre}\ and\ \citenamefont
  {Sargent~III}(1999)}]{meystre}%
  \BibitemOpen
  \bibfield  {author} {\bibinfo {author} {\bibfnamefont {P.}~\bibnamefont
  {Meystre}}\ and\ \bibinfo {author} {\bibfnamefont {M.}~\bibnamefont
  {Sargent~III}},\ }\href@noop {} {\emph {\bibinfo {title} {Elements of Quantum
  Optics}}}\ (\bibinfo  {publisher} {Springer, Berlin Heidelberg},\ \bibinfo
  {year} {1999})\BibitemShut {NoStop}%
\bibitem [{\citenamefont {Katz}\ \emph {et~al.}(2007)\citenamefont {Katz},
  \citenamefont {Retzker}, \citenamefont {Straub},\ and\ \citenamefont
  {Lifshitz}}]{Katz07}%
  \BibitemOpen
  \bibfield  {author} {\bibinfo {author} {\bibfnamefont {I.}~\bibnamefont
  {Katz}}, \bibinfo {author} {\bibfnamefont {A.}~\bibnamefont {Retzker}},
  \bibinfo {author} {\bibfnamefont {R.}~\bibnamefont {Straub}}, \ and\ \bibinfo
  {author} {\bibfnamefont {R.}~\bibnamefont {Lifshitz}},\ }\href {\doibase
  10.1103/PhysRevLett.99.040404} {\bibfield  {journal} {\bibinfo  {journal}
  {Phys. Rev. Lett.}\ }\textbf {\bibinfo {volume} {99}},\ \bibinfo {pages}
  {040404} (\bibinfo {year} {2007})}\BibitemShut {NoStop}%
\bibitem [{\citenamefont {Zurek}(2003)}]{Zurek2003}%
  \BibitemOpen
  \bibfield  {author} {\bibinfo {author} {\bibfnamefont {W.~H.}\ \bibnamefont
  {Zurek}},\ }\href {\doibase 10.1103/RevModPhys.75.715} {\bibfield  {journal}
  {\bibinfo  {journal} {Rev. Mod. Phys.}\ }\textbf {\bibinfo {volume} {75}},\
  \bibinfo {pages} {715} (\bibinfo {year} {2003})}\BibitemShut {NoStop}%
\bibitem [{\citenamefont {Debus}\ \emph {et~al.}(2019)\citenamefont {Debus},
  \citenamefont {Steiniger}, \citenamefont {Kling}, \citenamefont {Carmesin},\
  and\ \citenamefont {Sauerbrey}}]{debus_realizing_2018}%
  \BibitemOpen
  \bibfield  {author} {\bibinfo {author} {\bibfnamefont {A.}~\bibnamefont
  {Debus}}, \bibinfo {author} {\bibfnamefont {K.}~\bibnamefont {Steiniger}},
  \bibinfo {author} {\bibfnamefont {P.}~\bibnamefont {Kling}}, \bibinfo
  {author} {\bibfnamefont {C.~M.}\ \bibnamefont {Carmesin}}, \ and\ \bibinfo
  {author} {\bibfnamefont {R.}~\bibnamefont {Sauerbrey}},\ }\href {\doibase
  10.1088/1402-4896/aaf951} {\bibfield  {journal} {\bibinfo  {journal} {Phys.
  Scr.}\ }\textbf {\bibinfo {volume} {94}},\ \bibinfo {pages} {074001}
  (\bibinfo {year} {2019})}\BibitemShut {NoStop}%
\bibitem [{\citenamefont {Berry}(1991)}]{berry91}%
  \BibitemOpen
  \bibfield  {author} {\bibinfo {author} {\bibfnamefont {M.~V.}\ \bibnamefont
  {Berry}},\ }in\ \href@noop {} {\emph {\bibinfo {booktitle} {Les Houches LII,
  Chaos and Quantum Physics}}},\ \bibinfo {editor} {edited by\ \bibinfo
  {editor} {\bibfnamefont {M.-J.}\ \bibnamefont {Giannoni}}, \bibinfo {editor}
  {\bibfnamefont {A.}~\bibnamefont {Voros}}, \ and\ \bibinfo {editor}
  {\bibfnamefont {J.}~\bibnamefont {Zinn-Justin}}}\ (\bibinfo  {publisher}
  {Elsevier, New York},\ \bibinfo {year} {1991})\BibitemShut {NoStop}%
\bibitem [{not()}]{noteimbild}%
  \BibitemOpen
  \href@noop {} {}\bibinfo {note} {Negative values for the first-order
  contribution $f_\text{cl}^{(1)}$ do not mean that the full distribution
  function $f_\text{cl}\cong f_\text{cl}^{(0)} + f_\text{cl}^{(1)}$ becomes
  negative. The latter quantity is still positive}\BibitemShut {NoStop}%
\bibitem [{Note1()}]{Note1}%
  \BibitemOpen
  \bibinfo {note} {We apply the method of characteristics \protect \cite
  {courant_methods_2008}, which reduces the problem to solving the equation of
  motion for the classical trajectories \protect \cite {louisell_exact_1979}.
  These can be written in terms of Jacobi elliptic functions \cite
  {Lawden}.}\BibitemShut {Stop}%
\bibitem [{Note2()}]{Note2}%
  \BibitemOpen
  \bibinfo {note} {This effect is common to any Wigner function consisting of a
  coherent superposition of two or more individual states. Consider the
  superposition state $\ket {\psi }=\alpha \ket \phi +\beta \ket \chi $. Then
  the total Wigner function is given by $W(x,p)=\abs {\alpha }^2 W_{\ket \phi
  }(x,p)+ \abs {\beta }^2 W_{\ket {\chi }}(x,p) + 2\Re {\alpha \beta ^*\DOTSI
  \intop \ilimits@ \dd {y}\protect \mathrm e^{\protect \mathrm i py/\hbar }\phi
  (x-y/2)\chi ^*(x+y/2)}$, where $W_{\ket \phi }$ and $W_{\ket {\chi }}$,
  respectively, denote the Wigner functions for the states $\ket \phi $ and
  $\ket \chi $, while $\phi (x) =\braket {x}{\phi }$ and $\chi (x)=\braket
  {x}{\chi }$ are the position representations of their wave function. The last
  contribution is the interference term. See also Ref.~{\protect \cite
  {BUZEK19951}}.}\BibitemShut {Stop}%
\bibitem [{\citenamefont {Giese}\ \emph {et~al.}(2013)\citenamefont {Giese},
  \citenamefont {Roura}, \citenamefont {Tackmann}, \citenamefont {Rasel},\ and\
  \citenamefont {Schleich}}]{giese_double_2013}%
  \BibitemOpen
  \bibfield  {author} {\bibinfo {author} {\bibfnamefont {E.}~\bibnamefont
  {Giese}}, \bibinfo {author} {\bibfnamefont {A.}~\bibnamefont {Roura}},
  \bibinfo {author} {\bibfnamefont {G.}~\bibnamefont {Tackmann}}, \bibinfo
  {author} {\bibfnamefont {E.~M.}\ \bibnamefont {Rasel}}, \ and\ \bibinfo
  {author} {\bibfnamefont {W.~P.}\ \bibnamefont {Schleich}},\ }\href {\doibase
  10.1103/PhysRevA.88.053608} {\bibfield  {journal} {\bibinfo  {journal} {Phys.
  Rev. A}\ }\textbf {\bibinfo {volume} {88}},\ \bibinfo {pages} {053608}
  (\bibinfo {year} {2013})}\BibitemShut {NoStop}%
\bibitem [{\citenamefont {Altarelli}\ \emph {et~al.}(2006)\citenamefont
  {Altarelli} \emph {et~al.}}]{altarelli_xfel_2006}%
  \BibitemOpen
  \bibfield  {author} {\bibinfo {author} {\bibfnamefont {M.}~\bibnamefont
  {Altarelli}} \emph {et~al.},\ }\href
  {https://bib-pubdb1.desy.de/record/77248/files/european-xfel-tdr.pdf} {\emph
  {\bibinfo {title} {XFEL: The European X-Ray Free-Electron Laser -- Technical
  Design Report}}}\ (\bibinfo  {publisher} {DESY},\ \bibinfo {address}
  {Hamburg},\ \bibinfo {year} {2006})\BibitemShut {NoStop}%
\bibitem [{\citenamefont {Fedorov}(1981)}]{fedorov_rev}%
  \BibitemOpen
  \bibfield  {author} {\bibinfo {author} {\bibfnamefont {M.~V.}\ \bibnamefont
  {Fedorov}},\ }\href@noop {} {\bibfield  {journal} {\bibinfo  {journal} {Prog.
  Quantum Electron.}\ }\textbf {\bibinfo {volume} {7}},\ \bibinfo {pages} {73}
  (\bibinfo {year} {1981})}\BibitemShut {NoStop}%
\bibitem [{\citenamefont {Fedorov}(1997)}]{fedorov_book}%
  \BibitemOpen
  \bibfield  {author} {\bibinfo {author} {\bibfnamefont {M.~V.}\ \bibnamefont
  {Fedorov}},\ }\href@noop {} {\emph {\bibinfo {title} {Atomic and Free
  Electrons in a Strong Light Field}}}\ (\bibinfo  {publisher} {World
  Scientific, Singapore},\ \bibinfo {year} {1997})\BibitemShut {NoStop}%
\bibitem [{Note3()}]{Note3}%
  \BibitemOpen
  \bibinfo {note} {The time scale for space charge effects $T_\protect \text
  {sc}=1/\omega _\protect \text {p}$ depends on the relativistic plasma
  frequency defined (in laboratory frame) as $\omega _\protect \text
  {p}=\protect \sqrt {e^2n_\protect \text {el}/(\varepsilon _0\gamma
  ^3m_\protect \text {el})}$, where $n_\protect \text {el}$ is the electron
  density, $\gamma $ the relativistic factor, and $\varepsilon _0$ the vacuum
  permittivity while $m_\protect \text {el}$ and $e$ are the electron mass and
  charge, respectively. For the timescale of spontaneous emission, we use the
  formula for a classical electron in a wiggler \protect \cite [p.
  692]{jackson_classical_1998} $T_\protect \text {se}=3\lambda _\protect \text
  {W}/(2\pi \alpha _\protect \text {f}ca_0^2)$, where $\alpha _\protect \text
  {f}$ is the fine structure constant, $a_0$ the wiggler parameter and $\lambda
  _\protect \text {W}$ the wiggler wavelength in the laboratory frame. See also
  Ref.~\cite {debus_realizing_2018}}\BibitemShut {NoStop}%
\bibitem [{\citenamefont {Dodonov}\ \emph {et~al.}(200)\citenamefont {Dodonov},
  \citenamefont {Man'ko}, \citenamefont {Man'ko},\ and\ \citenamefont
  {Wünsche}}]{dodonov_hilbert-schmidt_2000}%
  \BibitemOpen
  \bibfield  {author} {\bibinfo {author} {\bibfnamefont {V.~V.}\ \bibnamefont
  {Dodonov}}, \bibinfo {author} {\bibfnamefont {O.~V.}\ \bibnamefont {Man'ko}},
  \bibinfo {author} {\bibfnamefont {V.~I.}\ \bibnamefont {Man'ko}}, \ and\
  \bibinfo {author} {\bibfnamefont {A.}~\bibnamefont {Wünsche}},\ }\href
  {\doibase 10.1080/09500340008233385} {\bibfield  {journal} {\bibinfo
  {journal} {J. Mod. Opt.}\ }\textbf {\bibinfo {volume} {47}},\ \bibinfo
  {pages} {633} (\bibinfo {year} {200})}\BibitemShut {NoStop}%
\bibitem [{Note4()}]{Note4}%
  \BibitemOpen
  \bibinfo {note} {We note that the linearization of the exponential is allowed
  since we are in the low-gain regime, where the relative change of
  $\varepsilon $ during one passage of electrons is small, that is $G\ll
  1$.}\BibitemShut {Stop}%
\bibitem [{Note5()}]{Note5}%
  \BibitemOpen
  \bibinfo {note} {The reason for this is the different underlying process,
  namely a Rabi oscillation \cite {NJP2015} between the involved momentum
  levels with a different time-scale and not the rotation inside the
  separatrix.}\BibitemShut {Stop}%
\bibitem [{\citenamefont {Efremov}\ and\ \citenamefont
  {Fedorov}(1999)}]{Efremov1999}%
  \BibitemOpen
  \bibfield  {author} {\bibinfo {author} {\bibfnamefont {M.~A.}\ \bibnamefont
  {Efremov}}\ and\ \bibinfo {author} {\bibfnamefont {M.~V.}\ \bibnamefont
  {Fedorov}},\ }\href {\doibase 10.1134/1.559004} {\bibfield  {journal}
  {\bibinfo  {journal} {J. Exp. Theor. Phys.}\ }\textbf {\bibinfo {volume}
  {89}},\ \bibinfo {pages} {460} (\bibinfo {year} {1999})}\BibitemShut
  {NoStop}%
\bibitem [{\citenamefont {Feist}\ \emph {et~al.}(2015)\citenamefont {Feist},
  \citenamefont {Echternkamp}, \citenamefont {Schauss}, \citenamefont
  {Yalunin}, \citenamefont {Schäfer},\ and\ \citenamefont
  {Ropers}}]{feist_quantum_2015}%
  \BibitemOpen
  \bibfield  {author} {\bibinfo {author} {\bibfnamefont {A.}~\bibnamefont
  {Feist}}, \bibinfo {author} {\bibfnamefont {K.~E.}\ \bibnamefont
  {Echternkamp}}, \bibinfo {author} {\bibfnamefont {J.}~\bibnamefont
  {Schauss}}, \bibinfo {author} {\bibfnamefont {S.~V.}\ \bibnamefont
  {Yalunin}}, \bibinfo {author} {\bibfnamefont {S.}~\bibnamefont {Schäfer}}, \
  and\ \bibinfo {author} {\bibfnamefont {C.}~\bibnamefont {Ropers}},\ }\href
  {\doibase 10.1038/nature14463} {\bibfield  {journal} {\bibinfo  {journal}
  {Nature}\ }\textbf {\bibinfo {volume} {521}},\ \bibinfo {pages} {200}
  (\bibinfo {year} {2015})}\BibitemShut {NoStop}%
\bibitem [{\citenamefont {Kozák}\ \emph {et~al.}(2018)\citenamefont {Kozák},
  \citenamefont {Eckstein}, \citenamefont {Schönenberger},\ and\ \citenamefont
  {Hommelhoff}}]{kozak_inelastic_2018}%
  \BibitemOpen
  \bibfield  {author} {\bibinfo {author} {\bibfnamefont {M.}~\bibnamefont
  {Kozák}}, \bibinfo {author} {\bibfnamefont {T.}~\bibnamefont {Eckstein}},
  \bibinfo {author} {\bibfnamefont {N.}~\bibnamefont {Schönenberger}}, \ and\
  \bibinfo {author} {\bibfnamefont {P.}~\bibnamefont {Hommelhoff}},\ }\href
  {\doibase 10.1038/nphys4282} {\bibfield  {journal} {\bibinfo  {journal} {Nat.
  Phys.}\ }\textbf {\bibinfo {volume} {14}},\ \bibinfo {pages} {121} (\bibinfo
  {year} {2018})}\BibitemShut {NoStop}%
\bibitem [{\citenamefont {Koz{\'a}k}\ \emph {et~al.}(2018)\citenamefont
  {Koz{\'a}k}, \citenamefont {Sch{\"o}nenberger},\ and\ \citenamefont
  {Hommelhoff}}]{kozak_ponderomotive_2018}%
  \BibitemOpen
  \bibfield  {author} {\bibinfo {author} {\bibfnamefont {M.}~\bibnamefont
  {Koz{\'a}k}}, \bibinfo {author} {\bibfnamefont {N.}~\bibnamefont
  {Sch{\"o}nenberger}}, \ and\ \bibinfo {author} {\bibfnamefont
  {P.}~\bibnamefont {Hommelhoff}},\ }\href {\doibase
  10.1103/PhysRevLett.120.103203} {\bibfield  {journal} {\bibinfo  {journal}
  {Phys. Rev. Lett.}\ }\textbf {\bibinfo {volume} {120}},\ \bibinfo {pages}
  {103203} (\bibinfo {year} {2018})}\BibitemShut {NoStop}%
\bibitem [{\citenamefont {Vanacore}\ \emph {et~al.}(2018)\citenamefont
  {Vanacore}, \citenamefont {Madan}, \citenamefont {Berruto}, \citenamefont
  {Wang}, \citenamefont {Pomarico}, \citenamefont {Lamb}, \citenamefont
  {McGrouther}, \citenamefont {Kaminer}, \citenamefont {Barwick}, \citenamefont
  {García~de Abajo},\ and\ \citenamefont
  {Carbone}}]{vanacore_attosecond_2018}%
  \BibitemOpen
  \bibfield  {author} {\bibinfo {author} {\bibfnamefont {G.~M.}\ \bibnamefont
  {Vanacore}}, \bibinfo {author} {\bibfnamefont {I.}~\bibnamefont {Madan}},
  \bibinfo {author} {\bibfnamefont {G.}~\bibnamefont {Berruto}}, \bibinfo
  {author} {\bibfnamefont {K.}~\bibnamefont {Wang}}, \bibinfo {author}
  {\bibfnamefont {E.}~\bibnamefont {Pomarico}}, \bibinfo {author}
  {\bibfnamefont {R.~J.}\ \bibnamefont {Lamb}}, \bibinfo {author}
  {\bibfnamefont {D.}~\bibnamefont {McGrouther}}, \bibinfo {author}
  {\bibfnamefont {I.}~\bibnamefont {Kaminer}}, \bibinfo {author} {\bibfnamefont
  {B.}~\bibnamefont {Barwick}}, \bibinfo {author} {\bibfnamefont {F.~J.}\
  \bibnamefont {García~de Abajo}}, \ and\ \bibinfo {author} {\bibfnamefont
  {F.}~\bibnamefont {Carbone}},\ }\href {\doibase 10.1038/s41467-018-05021-x}
  {\bibfield  {journal} {\bibinfo  {journal} {Nat. Commun.}\ }\textbf {\bibinfo
  {volume} {9}},\ \bibinfo {pages} {2694} (\bibinfo {year} {2018})}\BibitemShut
  {NoStop}%
\bibitem [{\citenamefont {Priebe}\ \emph {et~al.}(2017)\citenamefont {Priebe},
  \citenamefont {Rathje}, \citenamefont {Yalunin}, \citenamefont {Hohage},
  \citenamefont {Feist}, \citenamefont {Schäfer},\ and\ \citenamefont
  {Ropers}}]{priebe_attosecond_2017}%
  \BibitemOpen
  \bibfield  {author} {\bibinfo {author} {\bibfnamefont {K.~E.}\ \bibnamefont
  {Priebe}}, \bibinfo {author} {\bibfnamefont {C.}~\bibnamefont {Rathje}},
  \bibinfo {author} {\bibfnamefont {S.~V.}\ \bibnamefont {Yalunin}}, \bibinfo
  {author} {\bibfnamefont {T.}~\bibnamefont {Hohage}}, \bibinfo {author}
  {\bibfnamefont {A.}~\bibnamefont {Feist}}, \bibinfo {author} {\bibfnamefont
  {S.}~\bibnamefont {Schäfer}}, \ and\ \bibinfo {author} {\bibfnamefont
  {C.}~\bibnamefont {Ropers}},\ }\href {\doibase 10.1038/s41566-017-0045-8}
  {\bibfield  {journal} {\bibinfo  {journal} {Nat. Photonics}\ }\textbf
  {\bibinfo {volume} {11}},\ \bibinfo {pages} {793} (\bibinfo {year}
  {2017})}\BibitemShut {NoStop}%
\bibitem [{\citenamefont {Pellegrini}\ \emph {et~al.}(2016)\citenamefont
  {Pellegrini}, \citenamefont {Marinelli},\ and\ \citenamefont
  {Reiche}}]{pellegrini_physics_2016}%
  \BibitemOpen
  \bibfield  {author} {\bibinfo {author} {\bibfnamefont {C.}~\bibnamefont
  {Pellegrini}}, \bibinfo {author} {\bibfnamefont {A.}~\bibnamefont
  {Marinelli}}, \ and\ \bibinfo {author} {\bibfnamefont {S.}~\bibnamefont
  {Reiche}},\ }\href {\doibase 10.1103/RevModPhys.88.015006} {\bibfield
  {journal} {\bibinfo  {journal} {Rev. Mod. Phys.}\ }\textbf {\bibinfo {volume}
  {88}},\ \bibinfo {pages} {015006} (\bibinfo {year} {2016})}\BibitemShut
  {NoStop}%
\bibitem [{\citenamefont {Sargent~III}\ \emph {et~al.}(1974)\citenamefont
  {Sargent~III}, \citenamefont {Scully},\ and\ \citenamefont
  {Lamb~Jr.}}]{scullylamb}%
  \BibitemOpen
  \bibfield  {author} {\bibinfo {author} {\bibfnamefont {M.}~\bibnamefont
  {Sargent~III}}, \bibinfo {author} {\bibfnamefont {M.~O.}\ \bibnamefont
  {Scully}}, \ and\ \bibinfo {author} {\bibfnamefont {W.~E.}\ \bibnamefont
  {Lamb~Jr.}},\ }\href@noop {} {\emph {\bibinfo {title} {Laser Physics}}}\
  (\bibinfo  {publisher} {Addison-Wesley Publishing Company},\ \bibinfo {year}
  {1974})\BibitemShut {NoStop}%
\bibitem [{\citenamefont {Bonifacio}\ \emph {et~al.}(1990)\citenamefont
  {Bonifacio}, \citenamefont {Casagrande}, \citenamefont {Cerchioni},
  \citenamefont {de~Salvo~Souza}, \citenamefont {Pierini},\ and\ \citenamefont
  {Piovella}}]{boni90}%
  \BibitemOpen
  \bibfield  {author} {\bibinfo {author} {\bibfnamefont {R.}~\bibnamefont
  {Bonifacio}}, \bibinfo {author} {\bibfnamefont {F.}~\bibnamefont
  {Casagrande}}, \bibinfo {author} {\bibfnamefont {G.}~\bibnamefont
  {Cerchioni}}, \bibinfo {author} {\bibfnamefont {L.}~\bibnamefont
  {de~Salvo~Souza}}, \bibinfo {author} {\bibfnamefont {P.}~\bibnamefont
  {Pierini}}, \ and\ \bibinfo {author} {\bibfnamefont {N.}~\bibnamefont
  {Piovella}},\ }\href {\doibase 10.1007/BF02770850} {\bibfield  {journal}
  {\bibinfo  {journal} {La Rivista del Nuovo Cimento (1978-1999)}\ }\textbf
  {\bibinfo {volume} {13}},\ \bibinfo {pages} {1} (\bibinfo {year}
  {1990})}\BibitemShut {NoStop}%
\bibitem [{\citenamefont {Koekoek}\ \emph {et~al.}(2010)\citenamefont
  {Koekoek}, \citenamefont {Lesky},\ and\ \citenamefont {Swarttouw}}]{koekoek}%
  \BibitemOpen
  \bibfield  {author} {\bibinfo {author} {\bibfnamefont {R.}~\bibnamefont
  {Koekoek}}, \bibinfo {author} {\bibfnamefont {P.~A.}\ \bibnamefont {Lesky}},
  \ and\ \bibinfo {author} {\bibfnamefont {R.~F.}\ \bibnamefont {Swarttouw}},\
  }\href@noop {} {\emph {\bibinfo {title} {Hypergeometric Orthogonal
  Polynomials and Their \textit{q}-Analogues}}}\ (\bibinfo  {publisher}
  {Springer, Heidelberg},\ \bibinfo {year} {2010})\BibitemShut {NoStop}%
\bibitem [{Note6()}]{Note6}%
  \BibitemOpen
  \bibinfo {note} {Indeed, the small parameter for the perturbative expansion
  of $f_\protect \text {cl}$ (or $\protect \mathcal {W}$ close to the classical
  regime) is not $\tau $ but rather $\tau /\Delta \wp $. Hence, we require
  $\tau \ll \Delta \wp $ for this asymptotic expansion to converge. We
  emphasize that $\tau /\Delta \wp $ is a purely classical quantity, where no
  term with $\hbar $ enters.}\BibitemShut {Stop}%
\bibitem [{\citenamefont {{Meixner}}\ and\ \citenamefont
  {Schäfke}(1954)}]{meixner_mathieusche_1954}%
  \BibitemOpen
  \bibfield  {author} {\bibinfo {author} {\bibfnamefont {J.}~\bibnamefont
  {{Meixner}}}\ and\ \bibinfo {author} {\bibfnamefont {F.~W.}\ \bibnamefont
  {Schäfke}},\ }\href {\doibase 10.1007/978-3-662-00941-3} {\emph {\bibinfo
  {title} {Mathieusche Funktionen und Sphäroidfunktionen: Mit Anwendungen auf
  Physikalische und Technische Probleme}}}\ (\bibinfo  {publisher} {Springer
  Berlin Heidelberg},\ \bibinfo {year} {1954})\BibitemShut {NoStop}%
\bibitem [{\citenamefont {Olver}\ \emph {et~al.}(2018)\citenamefont {Olver},
  \citenamefont {{Olde Daalhuis}}, \citenamefont {Lozier}, \citenamefont
  {Schneider}, \citenamefont {Boisvert}, \citenamefont {Clark}, \citenamefont
  {Miller},\ and\ \citenamefont {Saunders}}]{NIST:DLMF}%
  \BibitemOpen
  \bibinfo {editor} {\bibfnamefont {F.~W.~J.}\ \bibnamefont {Olver}}, \bibinfo
  {editor} {\bibfnamefont {A.~B.}\ \bibnamefont {{Olde Daalhuis}}}, \bibinfo
  {editor} {\bibfnamefont {D.~W.}\ \bibnamefont {Lozier}}, \bibinfo {editor}
  {\bibfnamefont {B.~I.}\ \bibnamefont {Schneider}}, \bibinfo {editor}
  {\bibfnamefont {R.~F.}\ \bibnamefont {Boisvert}}, \bibinfo {editor}
  {\bibfnamefont {C.~W.}\ \bibnamefont {Clark}}, \bibinfo {editor}
  {\bibfnamefont {B.~R.}\ \bibnamefont {Miller}}, \ and\ \bibinfo {editor}
  {\bibfnamefont {B.~V.}\ \bibnamefont {Saunders}},\ eds.,\ \href
  {http://dlmf.nist.gov/} {\emph {\bibinfo {title} {{NIST Digital Library of
  Mathematical Functions}}}},\ \bibinfo {edition} {release 1.0.21}\ ed.\
  (\bibinfo {year} {2018})\BibitemShut {NoStop}%
\bibitem [{Note7()}]{Note7}%
  \BibitemOpen
  \bibinfo {note} {There is still an ambiguity in the choice of $\nu $, as $\nu
  '=\nu +n2\pi $, $n\in \protect \mathbb {N}$ yields the same periodic boundary
  condition. We circumvent this by demanding $\protect \qopname \relax
  m{lim}_{\alpha \to 0} \protect \mathrm {me}_{\nu }(\theta )=\protect \mathrm
  e ^{\protect \mathrm {i}\nu \theta }$ for $\nu \protect \not \in \protect
  \mathbb {Z}$. This approach is different form the convention in solid state
  physics, where $\nu $ is reduced to the first Brillouin zone, i.\protect
  \tmspace +\thinmuskip {.1667em}e. the interval $[-\pi ,\pi ]$, and an
  additional band index $n$ is introduced.}\BibitemShut {Stop}%
\bibitem [{\citenamefont {{Coïsson}}\ \emph {et~al.}(2009)\citenamefont
  {{Coïsson}}, \citenamefont {{Vernizzi}},\ and\ \citenamefont
  {{Yang}}}]{coisson_mathieu_2009}%
  \BibitemOpen
  \bibfield  {author} {\bibinfo {author} {\bibfnamefont {R.}~\bibnamefont
  {{Coïsson}}}, \bibinfo {author} {\bibfnamefont {G.}~\bibnamefont
  {{Vernizzi}}}, \ and\ \bibinfo {author} {\bibfnamefont {X.}~\bibnamefont
  {{Yang}}},\ }in\ \href {\doibase 10.1109/OSSC.2009.5416839} {\emph {\bibinfo
  {booktitle} {2009 IEEE International Workshop on Open-source Software for
  Scientific Computation (OSSC)}}}\ (\bibinfo  {publisher} {IEEE},\ \bibinfo
  {year} {2009})\BibitemShut {NoStop}%
\bibitem [{\citenamefont {Shirts}(1993)}]{shirts_computation_1993}%
  \BibitemOpen
  \bibfield  {author} {\bibinfo {author} {\bibfnamefont {R.~B.}\ \bibnamefont
  {Shirts}},\ }\href {\doibase 10.1145/155743.155796} {\bibfield  {journal}
  {\bibinfo  {journal} {ACM Trans. Math. Softw.}\ }\textbf {\bibinfo {volume}
  {19}},\ \bibinfo {pages} {377} (\bibinfo {year} {1993})}\BibitemShut
  {NoStop}%
\bibitem [{\citenamefont {Kandemir}(1998)}]{kandemir_wigner_1998}%
  \BibitemOpen
  \bibfield  {author} {\bibinfo {author} {\bibfnamefont {B.}~\bibnamefont
  {Kandemir}},\ }\href {\doibase 10.1016/S0375-9601(98)00354-5} {\bibfield
  {journal} {\bibinfo  {journal} {Phys. Lett. A}\ }\textbf {\bibinfo {volume}
  {245}},\ \bibinfo {pages} {209} (\bibinfo {year} {1998})}\BibitemShut
  {NoStop}%
\bibitem [{\citenamefont {Courant}\ and\ \citenamefont
  {Hilbert}(2008)}]{courant_methods_2008}%
  \BibitemOpen
  \bibfield  {author} {\bibinfo {author} {\bibfnamefont {R.}~\bibnamefont
  {Courant}}\ and\ \bibinfo {author} {\bibfnamefont {D.}~\bibnamefont
  {Hilbert}},\ }\href {\doibase 10.1002/9783527617210} {\emph {\bibinfo {title}
  {Methods of Mathematical Physics: Partial Differential Equations}}}\
  (\bibinfo  {publisher} {Wiley VCH},\ \bibinfo {year} {2008})\BibitemShut
  {NoStop}%
\bibitem [{\citenamefont {Louisell}\ \emph {et~al.}(1979)\citenamefont
  {Louisell}, \citenamefont {Lam}, \citenamefont {Copeland},\ and\
  \citenamefont {Colson}}]{louisell_exact_1979}%
  \BibitemOpen
  \bibfield  {author} {\bibinfo {author} {\bibfnamefont {W.~H.}\ \bibnamefont
  {Louisell}}, \bibinfo {author} {\bibfnamefont {J.~F.}\ \bibnamefont {Lam}},
  \bibinfo {author} {\bibfnamefont {D.~A.}\ \bibnamefont {Copeland}}, \ and\
  \bibinfo {author} {\bibfnamefont {W.~B.}\ \bibnamefont {Colson}},\ }\href
  {\doibase 10.1103/PhysRevA.19.288} {\bibfield  {journal} {\bibinfo  {journal}
  {Phys. Rev. A}\ }\textbf {\bibinfo {volume} {19}},\ \bibinfo {pages} {288}
  (\bibinfo {year} {1979})}\BibitemShut {NoStop}%
\bibitem [{\citenamefont {Lawden}(1989)}]{Lawden}%
  \BibitemOpen
  \bibfield  {author} {\bibinfo {author} {\bibfnamefont {D.~F.}\ \bibnamefont
  {Lawden}},\ }\href {\doibase 10.1007/978-1-4757-3980-0} {\emph {\bibinfo
  {title} {Elliptic Functions and Applications}}},\ \bibinfo {series} {Applied
  Mathematical Sciences}, Vol.~\bibinfo {volume} {80}\ (\bibinfo  {publisher}
  {Springer},\ \bibinfo {address} {New York},\ \bibinfo {year}
  {1989})\BibitemShut {NoStop}%
\bibitem [{\citenamefont {Bužek}\ and\ \citenamefont
  {Knight}(1995)}]{BUZEK19951}%
  \BibitemOpen
  \bibfield  {author} {\bibinfo {author} {\bibfnamefont {V.}~\bibnamefont
  {Bužek}}\ and\ \bibinfo {author} {\bibfnamefont {P.~L.}\ \bibnamefont
  {Knight}},\ }in\ \href {\doibase 10.1016/S0079-6638(08)70324-X} {\emph
  {\bibinfo {booktitle} {Prog. Opt.}}},\ Vol.~\bibinfo {volume} {34},\ \bibinfo
  {editor} {edited by\ \bibinfo {editor} {\bibfnamefont {E.}~\bibnamefont
  {Wolf}}}\ (\bibinfo  {publisher} {North Holland},\ \bibinfo {year}
  {1995})\BibitemShut {NoStop}%
\bibitem [{\citenamefont {Jackson}(1998)}]{jackson_classical_1998}%
  \BibitemOpen
  \bibfield  {author} {\bibinfo {author} {\bibfnamefont {J.~D.}\ \bibnamefont
  {Jackson}},\ }\href@noop {} {\emph {\bibinfo {title} {Classical
  {Electrodynamics}}}},\ \bibinfo {edition} {3rd}\ ed.\ (\bibinfo  {publisher}
  {Wiley},\ \bibinfo {address} {New York},\ \bibinfo {year} {1998})\BibitemShut
  {NoStop}%
\end{thebibliography}%

\end{document}